\UseRawInputEncoding
\documentclass[fleqn,usenatbib]{mnras}
\usepackage{newtxtext,newtxmath}

\usepackage[T1]{fontenc}
\usepackage{ae,aecompl}

\usepackage{graphicx}	
\usepackage{amsmath}	
\usepackage{natbib} 
\usepackage{upgreek}
\usepackage{xcolor}
\usepackage{amsfonts}
\usepackage{caption}
\usepackage{ulem}
\usepackage{comment}

\newcommand{\kms}{km~s$^{-1}$}
\newcommand{\unitflux}{erg~cm$^{-2}$~s$^{-1}$}
\newcommand{\unitlum}{erg~s$^{-1}$}
\newcommand{\Msun}{M$_{\odot}$}
\newcommand{\ha}{H$\upalpha$}
\newcommand{\hb}{H$\upbeta$}

\title[The nuclear transient AT\,2017gge]{The nuclear transient AT\,2017gge: a tidal disruption event in a dusty and gas-rich environment and the awakening of a dormant SMBH}
\author[F. Onori et al.]{F. Onori$^{1}$,\thanks{E-mail: francesca.onori@inaf.it}
G. Cannizzaro$^{2,3}$,
P.G. Jonker$^{3,2}$,
M. Kim$^{4}$,
M. Nicholl$^{5,6}$,
S. Mattila$^{7}$,
\newauthor
T. M. Reynolds$^{8}$,
M. Fraser$^{9}$,
T. Wevers$^{10}$,
E. Brocato$^{1,11}$,
J.~P. Anderson$^{10}$,
R. Carini$^{11}$,
\newauthor
P. Charalampopoulos$^{12}$,
P. Clark$^{13}$,
M. Gromadzki$^{14}$,
C. P. Guti\'errez$^{15,7}$
N. Ihanec$^{14,10}$,
\newauthor
C. Inserra$^{16}$,
A. Lawrence$^{17}$,
G. Leloudas$^{12}$,
P. Lundqvist$^{18}$,
T. E. M\"uller-Bravo$^{19}$,
\newauthor
S. Piranomonte$^{11}$,
M. Pursiainen$^{12}$,
K. A. Rybicki$^{14,20}$,
A. Somero$^{7}$,
D. R. Young$^{21}$,
\newauthor
K. C. Chambers$^{22}$,
H. Gao$^{22}$,
T. J.L. de Boer$^{22}$,
E. A. Magnier$^{22}$
\\
$^{1}$INAF-Osservatorio Astronomico d'Abruzzo,
via M. Maggini snc, I-64100 Teramo, Italy\\
$^{2}$ SRON, Netherlands Institute for Space Research, Niels Bohrweg 4, 2333 CA Leiden, The Netherlands\\
$^{3}$Department of Astrophysics/IMAPP, Radboud University, P.O. Box 9010, 6500 GL Nijmegen, the Netherlands\\
$^{4}$Department of Astronomy and Atmospheric Sciences, Kyungpook National University, Daegu 702-701, Korea;\\
$^{5}$ School of Physics and Astronomy, University of Birmingham, Birmingham B15 2TT, UK\\
$^{6}$Institute for Gravitational Wave Astronomy, University of Birmingham, Birmingham B15 2TT, UK\\
$^{7}$Tuorla Observatory, Department of Physics and Astronomy, FI-20014 University of Turku, Finland.\\
$^{8}$Niels Bohr Institute, University of Copenhagen, Jagtvej 128, 2200 København N, Denmark\\
$^{9}$School of Physics, University College Dublin, Belfield, Dublin 4, Ireland\\
$^{10}$ European Southern Observatory, Alonso de Córdova 3107, Casilla 19, Santiago, Chile\\
$^{11}$INAF-Osservatorio Astronomico di Roma, Via Frascati, 33, 00078 Monte Porzio Catone RM \\
$^{12}$DTU Space, National Space Institute, Technical University of Denmark, Elektrovej 327, 2800 Kgs. Lyngby, Denmark\\
$^{13}$Institute of Cosmology and Gravitation, University of Portsmouth, Portsmouth, PO1 3FX, UK\\
$^{14}$Astronomical Observatory, University of Warsaw, Al. Ujazdowskie 4, 00-478 Warszawa, Poland\\
$^{15}$Finnish Centre for Astronomy with ESO (FINCA), FI-20014 University of Turku, Finland\\
$^{16}$Cardiff Hub for Astrophysics Research and Technology, School of Physics \& Astronomy, Cardiff University, Queens Buildings,\\ The Parade, Cardiff, CF24 3AA, UK\\
$^{17}$Institute for Astronomy, SUPA, University of Edinburgh Royal Observatory Edinburgh, Blackford Hill, Edinburgh EH9 3HJ, UK\\
$^{18}$Department of Astronomy, Stockholm University, The Oskar Klein Centre, AlbaNova, SE-106 91 Stockholm, Sweden\\
$^{19}$Institute of Space Sciences (ICE, CSIC), Campus UAB, Carrer de Can Magrans, s/n, E-08193 Barcelona, Spain.\\
$^{20}$Department of Particle Physics and Astrophysics, Weizmann Institute of Science, Rehovot 76100, Israel\\
$^{21}$Astrophysics Research Centre, School of Mathematics and Physics, Queen’s University Belfast, Belfast BT7 1NN, UK\\
$^{22}$Institute of Astronomy, University of Hawaii, 2680 Woodlawn Drive, Honolulu, HI 96822, USA\\
}

\date{Accepted 2022 September 8. Received 2022 August 1; in original form 2022 June 1}

\pubyear{2022}

\begin{document}
\label{firstpage}
\pagerange{\pageref{firstpage}--\pageref{lastpage}}
\maketitle

\begin{abstract}
\small
We present the results from a dense multi-wavelength (optical/UV, near-infrared (IR), and X-ray) follow-up campaign of the nuclear transient AT\,2017gge, covering a total of 1698 days from the transient's discovery. The bolometric light-curve, the black body temperature and radius, the broad H and \ion{He}{I} $\lambda$5876 emission lines and their evolution with time, are all consistent with a tidal disruption event (TDE) nature. A soft X-ray flare is detected with a delay of $\sim$200 days with respect to the optical/UV peak and it is rapidly followed by the emergence of a broad \ion{He}{II} $\lambda$4686 and by a number of long-lasting high ionization coronal emission lines. This indicate a clear connection between a TDE flare and the appearance of extreme coronal line emission (ECLEs). An IR echo, resulting from dust re-radiation of the optical/UV TDE light is observed after the X-ray flare and the associated near-IR spectra show a transient broad feature in correspondence of the \ion{He}{I} $\lambda$10830 and, for the first time in a TDE, a transient high-ionization coronal NIR line (the [\ion{Fe}{XIII}] $\lambda$10798) is also detected. The data are well explained by a scenario in which a TDE occurs in a gas and dust rich environment and its optical/UV, soft X-ray, and IR emission have different origins and locations. The optical emission may be produced by stellar debris stream collisions prior to the accretion disk formation, which is instead responsible for the soft X-ray flare, emitted after the end of the circularization process.
\end{abstract}




\section{Introduction}
When an unlucky star wanders too close to a supermassive black hole (SMBH) it is ripped apart by the strong tidal forces. 
In this process, approximately half of the stellar material is expelled in unbound orbits, while the rest streams back to the SMBH and circularize to form a new accretion disk \cite[][]{strubbe09, lodato11}. 
These tidal disruption events (TDEs) manifest themselves as a luminous, short-lived, flares coming from the nuclei of otherwise quiescent galaxies \citep{hills75,rees88, phinney89, evans89} and represent an important tool to study the properties of dormant SMBHs. The emission usually peaks in the UV/optical or in soft X-rays and the bolometric luminosity is expected to follow the bound debris fallback rate, with a powerlaw decline $\propto t^{-5/3}$ on the timescale of months to years (e.g., \citealt{evans89, cannizzo90,rees90, lodato09}). 

The fist TDE candidates were discovered in the X-rays \cite[][]{komossa99}, but thanks to the development of wide-field optical surveys dedicated to the search of transients, the population of observed TDEs has quickly grown and the optical band has become the main discovery channel.
Multi-wavelength monitoring campaigns represents a crucial instrument to identify TDEs among nuclear flares and have revealed an intriguing and puzzling diversity in the observational properties \cite[see the reviews from][]{vanVelzen20, saxton20, gezari21}. The optical spectra are characterized by a strong blue continuum at early times and are dominated by broad ($\sim$10$^{4}$ km s$^{-1}$) H and/or He emission lines with different strengths and relative ratios \cite[][]{arcavi14, leloudas19, charalampopoulos22}. A fraction of TDEs have shown broad Bowen fluorescence emission lines  \cite[][]{blagorodnova19, leloudas19, onori19} and evidence for \ion{Fe}{II} emission lines have been found in a small subset of TDEs \cite[][]{wevers19, cannizzaro21}.
Following these discoveries, the TDE population has been divided into three main spectral classes, depending on the appearance or lack of the different broad spectral features \cite[][]{arcavi14, leloudas19, vanVelzen20}.
In some TDE candidates, high ionization coronal emission lines have been identified \cite[][]{komossa08, komossa09, wang11, yang13, palaversa16}, suggesting the presence of a gas-rich environment surrounding these sources. Furthermore, infrared (IR) echos, resulting from the reradiation of the TDE emission by dust, have been detected thanks to dedicated observations \cite[][]{mattila18}, follow-up campaigns or {\it WISE} archival searches \cite[see][for a recent review]{jiang21,vanVelzen21b}. Recent findings suggest that some TDEs are expected to be so highly dust enshrouded that they could have remained out of the reach of optical or X-ray surveys due to the large column densities of obscuring dust and gas \cite[][]{reynolds22}. From photometric analysis, surprisingly low and constant black body temperatures of $\sim$10$^{4}$ K, which cannot be explained through traditional accretion process, have been derived \cite[][]{hinkle20,vanVelzen21}. Despite what is expected in the case of emission from a newly formed accretion disk,
TDEs selected in the optical are typically not detected in the X-rays.
Only few exceptions have been discovered, with some events showing also soft X-ray emission, sometimes delayed with respect to the optical peak emission \cite[e.g ASASSN-14li, ASASSN-15oi, AT\,2019dsg, AT\,2018fyk, AT\,2019qiz, AT\,2019azh;][respectively]{holoien16a, gezari17,cannizzaro21,wevers19,nicholl20,liu22}. 

Despite the recent progress in this field, there are many aspects that remain unclear, such as the emission mechanism behind all the observed features, the geometry of the emitting region and the X-ray non-detection of the optically-selected TDEs. Different scenarios have been proposed, including the presence of an optically thick atmosphere which reprocess the high energy radiation emitted during the accretion process \citep{guillochon14, roth16, roth18}, or emission by shocks from intersecting stellar debris streams during the disk-formation phase \cite[][]{piran15, shiokawa15, bonnerot17}.
However, in these models, the X-rays are expected to show up eventually, once the wind calm down or the circularization process ends. \cite{dai18}, instead, have  explained the X-ray/optical dichotomy in the framework of a TDE unified model in which optically thick winds are produced following the formation of the accretion disk and are responsible for the X-ray obscuration. Inclination effect together with the physical properties of the mass outflow determine the detection of the X-ray emission.\\

AT\,2017gge was first detected on 2017 August 03 (MJD~57\,968.35) by the Asteroid Terrestrial-impact Last Alert System survey \cite[ATLAS,][]{tonry18}, which reported  a discovery magnitude in the ATLAS orange filter of $o = 18.70\pm0.17$ mag \cite[in the AB system,][]{atlas17gge}. The transient is located within 0\arcsec.1 (0.1 kpc) from the center of the host galaxy, specifically at the coordinates RA(J2000) = 16:20:34.99 and DEC(J2000)= +24:07:26.5. The host galaxy has been identified to be  SDSS\,J162034.99+240726.5, a spiral galaxy at redshift $z=0.0665$, classified as a star-forming galaxy from the Sloan Digital Sky Survey (SDSS) pre-transient optical spectrum. The field has been observed in the 0.1-2.4 keV energy band by the ROentgen SATellite ({\it ROSAT}) within the {\it ROSAT} all Sky Survey (RASS) on the 1990 July 30, and no signs of X-ray emission have been detected (only an upper limit of F$_{0.3-10keV}$=1.0$\times$10$^{-12}$ \unitflux\/ can be derived). A mid-infrared (MIR) flare in the WISE light curve of AT\,2017gge has been recently reported by \cite{jiang21}.
Due to its proximity to the host galaxy core, AT\,2017gge is classified as a nuclear transient. The spectroscopic classification reported an uncertain nature for AT\,2017gge  \cite[][]{fraser17}, with a spectrum taken three weeks after the transient's discovery characterized by a blue continuum and by a broad component at the base of both the narrow \ha\/ and \hb\/ emission lines, which are not consistent with a SN scenario. \cite{fraser17} suggest the possibility of a TDE nature for AT\,2017gge, as the smooth evolution of the lightcurve is atypical for AGN variability. Recently, based on the X-ray and optical analysis, \cite{wang22} claimed a TDE nature for this transient.

In this paper we present the results of our dense multi-wavelength follow-up campaign of AT\,2017gge, which covers a total of 1698 days from the transient's discovery. The data-set includes optical photometric and spectroscopic data obtained with a number of ground based facilities, soft X-ray observations delivered by {\it Swift} and a series of near-infrared (NIR) spectra obtained after an enhanced MIR emission.
Thanks to this dense, long-lasting and multi-band data-set we have been able to further investigate the nature of this transient, to accurately study the spectroscopic evolution of the broad emission lines and the connection between a delayed soft X-ray flare and the emergence of a broad \ion{He}{II} $\lambda$4686 emission line and of a number of high ionization coronal emission lines. Finally, thanks to late-time, high resolution spectroscopy, the activity status of the host galaxy, after the occurrence of the nuclear transient, has been inspected and the black hole (BH) mass has been derived, using two independent methods. 
In this manuscript, all the times are reported with respect to MJD~57\,968.35, which corresponds to the discovery date of AT\,2017gge as reported by the ATLAS survey. Throughout the paper we use a luminosity distance of $d_{\rm L}$ = 297.6 Mpc, based on a WMAP9 cosmology with $H_{\rm 0}$ = 69.32 km s$^{-1}$ Mpc$^{-1}$ , $\Omega_{\rm M}$ = 0.29, $\Omega_{\Lambda}$ = 0.71 \cite[][]{hinshaw2013}.

\section{Observations and Data reduction}
\label{sec:observations}

Our monitoring campaign of the nuclear transient AT\,2017gge started on 2017 September 14. In order to provide good coverage of the transient's evolution, we have followed its emission by using different instruments with a multi-wavelength approach. In particular, while the X-ray data has been obtained through the XRT instrument on board the Neil Gehrels {\it Swift} observatory \cite[][]{gehrels04}, the optical/NIR photometric and spectroscopic follow-up has been carried out by using a variety of ground-based facilities together with the UV-optical photometric observations delivered by {\it Swift}/UVOT instrument. In the following section we describe the observational set-up and data reduction for all the instruments used.


\subsection{Ground-based observations}

A dense photometric coverage of the first $\sim$40 days of AT\,2017gge have been performed in the $o$ and $c$ filters by the ATLAS survey \cite[][]{smith20} and in the $i^\prime$ filter by the Panoramic Survey Telescope and Rapid Response System \cite[Pan-STARRS,][]{huber15, chambers16}. Additional $w$ images have been obtained by Pan-STARRS between 600 and 700 days from the transient discovery.    
The spectroscopic monitoring was carried out mainly by using the ESO Faint Object Spectrograph and Camera \cite[EFOSC2,][]{buzzoni84}, mounted on the 3.58 m New Technology Telescope (NTT) as part of the ePESSTO ESO public survey \cite[][]{smartt15} and the Andalucia Faint Object Spectrograph and Camera (ALFOSC), mounted on the 2.56 m Nordic Optical Telescope (NOT), as part of the NOT Unbiased Transient Survey (NUTS\footnote{http://csp2.lco.cl/not/}).
Additional optical observations and late time images have been obtained with the Device Optimized for the LOw RESolution (DOLORES) installed at the 3.58 m Telescopio Nazionale Galileo (TNG), the Auxiliary-port CAMera (ACAM) mounted on the 4.2 m William Herschel Telescope (WHT) and the Optical Wide Field Camera (IO:O) at the Liverpool Telescope \cite[LT,][]{steele04}, respectively. Higher resolution spectra were obtained with the Gemini North Multi-Object Spectrographs (GMOS-N), mounted on the 8.1 m Frederick C.~Gillett Gemini North telescope, with the Optical System for Imaging, the low-Intermediate-Resolution Integrated Spectroscopy (OSIRIS), mounted at the 10.4 m Gran Telescopio CANARIAS (GTC) and the X-shooter spectrograph \cite[][]{Vernet11}, mounted on the UT3 the Very Large Telescope (VLT), located at the ESO Paranal observatory. NIR spectra and images in the $J$, $H$ and $Ks$ bands have been obtained with the infrared spectrograph and imaging camera Son of ISAAC \cite[SofI][]{morvood98}, mounted on the NTT. 
After the transient light had faded below the detection levels, host galaxy spectra and images were taken for each ground-based facility and with the same observational set-up used during the follow-up.

All the spectroscopic observations have been carried out with the slit oriented at the parallactic angle. 
In Table~\ref{tbl:obsSpec} the main observing information, such as the observing date, the grism used, the exposure time, the slit width, the airmass and seeing condition, are reported for each instrument. The sequence of the optical spectra is shown in Figure \ref{fig:spec}, the SofI NIR spectra are shown in the lower panel of Figure \ref{fig:wise}, while we show the whole X-shooter spectrum (UVB, VIS and NIR arms) in Figure \ref{fig:Xsh_spec}.

\begin{table*}
\centering
 \begin{minipage}{140mm}
 \caption{Spectroscopic observations of AT\,2017gge.} 
 \label{tbl:obsSpec}
 \begin{center}
 \begin{tabular}{@{}lccllllll}
 \hline
\hline
MJD      & Phase   & UT date & Instrument & Grism  & Exp. time   & Slit    & Airmass  & Seeing  \\  
         & [days]  &         &            &        & [s]             &[\arcsec]&[\arcsec] & [\arcsec]\\
(1)      & (2)     &  (3)    & (4)        & (5)    & (6)             &(7)      & (8)      & (9)        \\
\hline
58\,010.01  & \phantom{1}41.7    & 2017 Sep 14 & NTT/EFOSC2 & Gr\#13 & 1x900\phantom{0}  & 1.5 & 2.4 & 1.3\\
58\,011.88  & \phantom{1}43.5  & 2017 Sep 15 & WHT/ACAM   & V400   & 1x1800            & 1.0 & 1.5 & 1.0\\
58\,012.88  & \phantom{1}44.5  & 2017 Sep 16 & WHT/ACAM   & V400   & 2x1800            & 1.0 & 1.5 & 0.5\\
58\,162.21  &193.9 & 2018 Feb 13 & NOT/ALFOSC  & Gr\#4  & 1x2400            & 1.0 & 1.4 & 1.1 \\
58\,186.35  &218.0 & 2018 Mar 09 & NTT/EFOSC2 & Gr\#13 & 1x2700            & 1.0 & 1.9 & 0.9\\
58\,187.36  &219.0 & 2018 Mar 09 & NTT/EFOSC2 & Gr\#18 & 1x3600            & 1.0 & 1.8 & 0.8\\
58\,198.14  &229.8 & 2018 Mar 21 & TNG/DOLORES & LR--B & 1x2700            & 1.0 & 1.3 & 1.2 \\
58\,200.12  &231.8 & 2018 Mar 23 & NOT/ALFOSC   & Gr\#19 & 1x3200           & 1.3 & 1.2 & 1.2\\
58\,201.17  &232.8 & 2018 Mar 24 & NOT/ALFOSC   & Gr\#4  & 1x3600           & 1.0 & 1.1 & 0.9 \\
58\,202.33  &234.0 & 2018 Mar 25 & NTT/EFOSC2  & Gr\#13 & 1x3600           & 1.0 & 1.8 & 1.0 \\
58\,203.36  &235.0 & 2018 Mar 25 & NTT/EFOSC2  & Gr\#18 & 1x3000           & 1.0 & 1.7 & 0.7 \\
58\,207.16  &238.8 & 2018 Mar 30 & TNG/DOLORES & LR--B  & 1x3600           & 1.0 & 1.0 & 0.5 \\
58\,207.18  &238.8 & 2018 Mar 30 & WHT/ACAM    & V400   & 6x600\phantom{0} & 1.0 & 1.0 & 0.8 \\
58\,216.35  &248.0 & 2018 Apr 08 & NTT/EFOSC2  & Gr\#18 & 1x3600           & 1.0 & 1.7 & 0.7\\
58\,217.34  &249.0 & 2018 Apr 09 & NTT/EFOSC2  & Gr\#13 & 1x3000           & 1.0 & 1.7 &1.0\\
58\,229.28  &260.9 & 2018 Apr 21 & NTT/EFOSC2  & Gr\#13 & 1x3000           & 1.0 & 1.7 &0.6 \\ 
58\,229.34  &261.0 & 2018 Apr 21 & NTT/EFOSC2  & Gr\#18 & 2x1500           & 1.0 & 1.7 & 0.7 \\
58\,243.26  &274.9 & 2018 May 05 & NTT/EFOSC2  & Gr\#18 & 1x3600           & 1.0 & 1.7 & 0.7 \\
58\,244.24  &275.9 & 2018 May 06 & NTT/EFOSC2  & Gr\#13 & 1x3600           & 1.0 & 1.7 &1.1 \\
58\,260.15  &291.8 & 2018 May 22 & NTT/EFOSC2  & Gr\#13 & 2x3000           & 1.0 & 1.8 &0.7 \\
58\,261.13  &291.8 & 2018 May 23 & NTT/EFOSC2  & Gr\#18 & 2x3000           & 1.0 & 2.0 &0.7 \\
58\,279.02  &310.7 & 2018 Jun 10 & NOT/ALFOSC   & Gr\#19 & 2x2800           & 1.0 & 1.0 &1.1 \\  
58\,299.06  & 320.7 & 2018 Jun 30 & NTT/EFOSC2   & Gr\#13 & 2x3600           & 1.0 & 1.7 &0.9 \\  
58\,307.00  &338.6 & 2018 Jul 07 & GTC/OSIRIS  & R500B  & 2x1800           & 1.0 & 1.1 &0.9 \\ 
58\,347.07  &378.7 & 2018 Aug 17 & NTT/SofI    & BG     & 1x2160           & 1.0 & 1.7 &0.5 \\
58\,376.22  &407.9 & 2018 Sep 15 & Gemini/GMOS-N      & B600   & 4x940\phantom{0} & 1.0 & 1.2 & 0.9 \\
58\,525.21  &556.9 & 2019 Feb 11 & NOT/ALFOSC  & Gr\#4  & 1x2800           & 1.3 & 1.3 &0.7 \\
58\,540.57  &572.2 & 2019 Feb 26 & Gemini/GMOS-N      & B600   & 4x900\phantom{0} & 1.0 & 1.4 & 0.8 \\
58\,552.16  &583.8 & 2019 Mar 10 & TNG/DOLORES & LR--B  & 2x2000           & 1.5 & 1.2 &2.5 \\
59\,314.29  &1346.0 & 2021 Apr 10 & NTT/SofI   & BG     & 1x3240          & 1.0 & 1.7 & 1.0 \\
59\,316.29  &1348.0 & 2021 Apr 12 & NTT/EFOSC2 & Gr\#13 & 1x1800           & 1.0 & 1.7 & 0.9 \\
59\,316.31  &1348.0 & 2021 Apr 12 & NTT/EFOSC2 & Gr\#18 & 1x1800           & 1.0 & 1.7 & 0.8 \\
59\,666.29  &1697.9 & 2022 Mar 28 & VLT/Xshooter & UVB/VIS/NIR & 2x1360 & 1.0/0.9/0.9 & 1.7 & 0.6 \\
\hline
\end{tabular}
\end{center}
Notes: (1) Modified Julian dates of observations; (2) Phase with respect to the discovery date MJD 57\,968.35 (3) UT date, (4) Telescope and instrument; (5) Grism; (6) Exposure time; (7) Slit width;  (8) Airmass; (9) Seeing. 
\noindent
\end{minipage}
\end{table*}

\subsubsection{NTT/EFOSC2+SofI}
As shown in Table \ref{tbl:obsSpec}, a total of 16 spectra have been taken with EFOSC2 instrument by using the grism Gr $\#$13, which covers a wide wavelength range (3685-9315 \AA\/) with a resolution of  $R=\lambda/\Delta\lambda \sim 300$, calculated for a 1\arcsec.0 slit (a slightly higher resolution is achieved in case of seeing better than 1.0\arcsec). 
In order to perform more detailed observations in the \ion{He}{II} $\lambda$4686 region, we also used the grism Gr$\#$18, which provides a resolution of $R=\lambda/\Delta\lambda \sim 750$ (calculated for a 1.0\arcsec slit and for a 1.0\arcsec\/ seeing) in the wavelength range 4700-6770 \AA. Two NIR spectra have been obtained with the SofI instrument in nodding mode with the BG grism, which covers the wavelength range 9500-16400 \AA\/ and provides a resolution $R=\lambda/\Delta\lambda \sim 1000$, calculated for a 0.6\arcsec slit. We use the two instruments also in imaging mode in order to obtain images of the field in the $g^\prime$ and $r^\prime$ filters (with EFOSC2) and in the $J$, $H$ and $K$ filters (with SofI).

All these spectra and images have been reduced using the \texttt{ePESSTO NTT Pipeline v.2.4.0}\footnote{https://github.com/svalenti/pessto} (in spectroscopic and imaging mode, respectively), which is based on standard IRAF tasks, such as bias, flat-field and cosmic rays correction. The spectroscopic wavelength and flux calibration are performed using arc lamps and standard stars, respectively. Multiple spectra taken on the same night are averaged in order to increase the SNR. The photometric zero points of the images are calculated using SDSS stars for the Sloan filters, while the stars in the 2MASS catalog are used for the SofI images.

\subsubsection{NOT/ALFOSC}
A total of 5 spectra of AT\,2017gge were obtained with the ALFOSC instrument mounted on the NOT at days 193.9, 231.8, 232.8, 310.7 and 556.9. For these observations we first used the grism Gr$\#$4, which covers the wavelength range 3200-9600 \AA\/ with a resolution $R=\lambda/\Delta\lambda \sim 360$, calculated for a 1.0\arcsec slit under seeing conditions of 1\arcsec~or larger. We performed more detailed observations in the \ion{He}{II} $\lambda$4686 region by using the grism Gr$\#$19, which provides a resolution of $R=\lambda/\Delta\lambda \sim 970$ (for a 1.0\arcsec slit under seeing conditions of 1\arcsec~or larger) over the 4400-6950 \AA\ wavelength range.

We reduced the spectra by using the \texttt{foscgui 1.7} pipeline\footnote{\texttt{foscgui} is a graphical user interface aimed at extracting spectroscopy and photometry obtained with FOSC-like instruments. It was developed by E. Cappellaro. A package description can be found at http://sngroup.oapd.inaf.it/foscgui.html.}, which is based on standard IRAF reduction tasks \citep{tody86}, including bias and flat field correction, cosmic ray cleaning, wavelength calibration by using arc lamps and flux calibration with a standard star.

We used ALFOSC also to obtain images of the transient field in $g^\prime$ and $r^\prime$ filters. These images have been reduced with the \texttt{foscgui} 1.7 pipeline in imaging mode, which is based on standard IRAF reduction tasks and includes bias, flat fields and cosmic rays correction. It also provides the World Coordinate System calibration using SDSS stars. The photometric zero points are derived using SDSS stars in the field of view.

\subsubsection{WHT/ACAM}

We observed AT\,2017gge with the low-resolution spectrograph of ACAM on three nights. In particular, we used the V400 grating which provides a nominal wavelength coverage of 3950-9400 \AA\/ and a resolution $R=\lambda/\Delta\lambda \sim 430$ for a 1.0\arcsec slit, under seeing conditions of 1\arcsec~or larger. We reduced the data by using a the standard IRAF tasks including bias, flat field and cosmic ray correction, wavelength and flux calibration with arc lamps and standard stars, respectively. 

\subsubsection{TNG/DOLORES}
A total of three spectra of AT\,2017gge were obtained with the low resolution spectrograph and camera DOLORES.
In the spectroscopic mode we used the the LR-B grating which provides a wavelength coverage of 3000-8430 \AA\/ and a resolution $R=\lambda/\Delta\lambda \sim 580$ for a 1.0\arcsec slit and under seeing condition of 1.0\arcsec or larger. 
We used DOLORES also in imaging mode in order to obtain  $u^\prime$, $g^\prime$ and $r^\prime$ images of the field of the transient. The data have been reduced by using the standard IRAF tasks which includes bias, flat-field and cosmic rays correction. The spectroscopic wavelength and flux calibration are performed by using arc lamps and standard stars, respectively.    

\subsubsection{LT/IO:O}

We observed AT\,2017gge with the IO:O at the Liverpool Telescope in order to obtain images of the transient field in the $u^\prime$, $g^\prime$ and $r^\prime$ filters. The images were reduced with the IO:O pipeline\footnote{https://telescope.livjm.ac.uk/TelInst/Pipelines/\#ioo} and the zeropoints calculated using stars in the the American Association of Variable Star Observers (AAVSO) Photometric All-Sky Survey (APASS) catalogue \citep{henden19}. 

\subsubsection{GTC/OSIRIS}
Two spectra of AT\,2017gge have been taken 338.6 days after the transient discovery with the OSIRIS instrument, located at the Nasmyth-B focus of GTC. We use the spectrograph in the long slit mode and with the R500B grism, which covers the 3600-7200 \AA\/ wavelength range with a resolution $R=\lambda/\Delta\lambda \sim 540$, calculated for a 0.6\arcsec\/ slit under seeing condition of 0.6\arcsec\/ or larger. The data has been reduced by using the dedicated \texttt{GTCMOS} pipeline \cite[][]{gomezgonzales16}, which is an IRAF-based script and performs bias, flat-field, cosmic rays correction. It also delivers wavelength calibrated 2D spectral images (calibrated by using arc lamps) and flux calibration is applied on the extracted spectra by using a standard star for reference. The spectra taken on the same night are then combined together, in order to increase the S/N.

\subsubsection{Gemini/GMOS-N}
Two additional late-time, higher resolution spectra of AT\,2017gge have been obtained with the Gemini Multi-Object Spectrograph (GMOS) mounted at Gemini North telescope, in single long slit mode (1.0\arcsec\/ slit) and using the B600 grism. This configuration allows to approximately cover the 4400-7500 wavelength range with a resolution $R=\lambda/\Delta\lambda = 844$, achieved with a slit width of 1.0\arcsec\/ under seeing condition of 1.0\arcsec\/ or larger. The science data was obtained through dithering in the spatial and spectral direction to avoid contamination from bad pixels and columns. Data have been reduced with the \texttt{Gemini iraf} package, including bias subtraction, flat-field, wavelength calibration. The flux calibration was performed with the spectra of the standard stars EG 131 and Hilt 600. Four individual exposures were combined into each final spectrum.    

\subsubsection{VLT/X-shooter}
One spectrum of the AT\,2017gge host galaxy has been taken at day 1697.9 with the X-shooter instrument which is an intermediate resolution spectrograph covering a wide wavelength range. In particular, it spans the 3000$-$25000 \AA\/ range in a single observation, thanks to the presence of three arms (UVB, VIS and NIR) working simultaneously. Specifically, the UVB arm covers the 3000$-$5595 \AA\/ wavelength range, the VIS arm spans the 5595$-$10240 \AA\/ and the NIR arm ranges from 10240$-$24800 \AA\ . We used the 1.0\arcsec\/, 0.9\arcsec\/ and 0.9\arcsec\/ slit widths configuration  for the UVB, VIS and NIR respectively, which deliver the respective nominal resolutions of $R=\lambda/\Delta\lambda$ = 5400, 8900 and 5600. Data have been reduced by using the X-shooter pipeline in the  \texttt{EsoReflex} environment \cite[][]{freudling13} and telluric corrections were performed by using the \texttt{molecfit} V. 4.2 software \cite[][]{smette15,kausch15}. 

\subsection{Swift observations}
AT\,2017gge was monitored by the Neil Gehrels {\it Swift} observatory over a period spanning $\sim$200 days, starting from $\sim$60 days after the transient discovery. A total of 19 XRT and UVOT observations have been obtained. In particular, we were awarded six epochs of \textit{Swift} ToO observations between 2018 March 09 and 2018 April 07 for late-time follow-up of the source. In Table~\ref{tab:swift_log} a list of the {\it Swift} observations for AT\,2017gge are reported, together with the main properties (date of observation, time after the transient discovery and instruments exposure times). All the XRT observation have been executed in photon counting mode.

\begin{table}
\caption{List of the {\it Swift} observations executed for the monitoring of AT\,2017gge. The XRT data have been obtained in photon counting mode. (1) Date of observation; (2) time after discovery; (3) XRT exposure time (4) UVOT exposure time.}
\label{tab:swift_log}
\begin{tabular}{llcc}
\hline
MJD			& Time & XRT exp. time & UVOT exp. time  \\
(days)      & (days)           &(s)             & (s)             \\
(1)         & (2)        & (3)                 & (4)             \\
\hline
58\,031.12    & \phantom{10}62.8 & 1631 & 1555\\
58\,041.75    & \phantom{10}73.4 & 1604 & 1532 \\
58\,046.29    & \phantom{10}77.9 & 1403 & 1338 \\
58\,051.78    & \phantom{10}83.4 & 2084 & 2020 \\
58\,055.04    & \phantom{10}86.7 & 2232 & 2180\\
58\,143.57    &\phantom{1}175.2 & 1532 & 1480\\
58\,156.93    &\phantom{1}188.6 & 2153 & 2104 \\
58\,162.98    &\phantom{1}194.6 & 1504 & 1480 \\
58\,168.35    &\phantom{1}200.0 & 2109 & 2058 \\
58\,174.28    &\phantom{1}205.9 & \phantom{1}564 &\phantom{1}536\\
58\,177.72    &\phantom{1}209.4 & 1432 & 1387 \\
58\,186.15    &\phantom{1}217.8 & \phantom{1}908 & \phantom{1}906 \\
58\,193.72    &\phantom{1}225.4 & \phantom{1}490 & \phantom{1}489 \\
58\,200.76    &\phantom{1}232.4 & \phantom{1}379 & \phantom{1}378\\
58\,211.60    &\phantom{1}243.3 & \phantom{1}560 & \phantom{1}559 \\
58\,213.85    &\phantom{1}245.5 & \phantom{1}535 & \phantom{1}534\\
58\,215.98    &\phantom{1}247.6 & \phantom{1}576 & \phantom{1}578\\
59\,651.84 & 1683.5 & 3574 & 3502\\
59\,655.29 & 1686.9 & 2283 & 2210\\
\hline
\end{tabular}
\end{table}

\subsubsection{Swift/UVOT data}
The \textit{Swift}/UVOT observations include images in the filters \textit{UVW2} (1928 \AA), \textit{UVM2} (2244 \AA), \textit {UVW1} (2600 \AA), \textit {U} (3465 \AA), \textit {B} (4392 \AA) and \textit {V} (5468 \AA) which have been reduced using the standard pipeline with the updated calibrations from the \texttt{HEAsoft}-6.25 \texttt{ftools} package. In order to derive the apparent magnitudes of the transient we have used the \texttt{HEAsoft} routine \texttt{uvotsource}. For each filter we have measured the aperture photometry using a 5\arcsec\/ aperture centered on the position of the transient and  a background region of 60\arcsec\/ radius placed in an area free of sources. We derive the Galactic extinction correction for the UVOT filters by assuming $R_V$=3.1 and $A_V$= 0.193 \cite[]{schlafly11} and a Cardelli extinction law \cite[]{cardelli89}.

In Table \ref{tbl:UVOTphot} we provide the \textit{Swift}/UVOT \textit {UVW2}, \textit {UVM2} and \textit {UVW1} apparent magnitudes (in the AB system) together with the \textit{Swift}/UVOT \textit{U}, \textit{B} and \textit{V} filters apparent magnitudes (in Vega system), not corrected for for foreground extinction.

\subsubsection{Swift/XRT data}
\label{sec:xrt}

\begin{figure}
\centering
\includegraphics[scale=0.25, angle=-90]{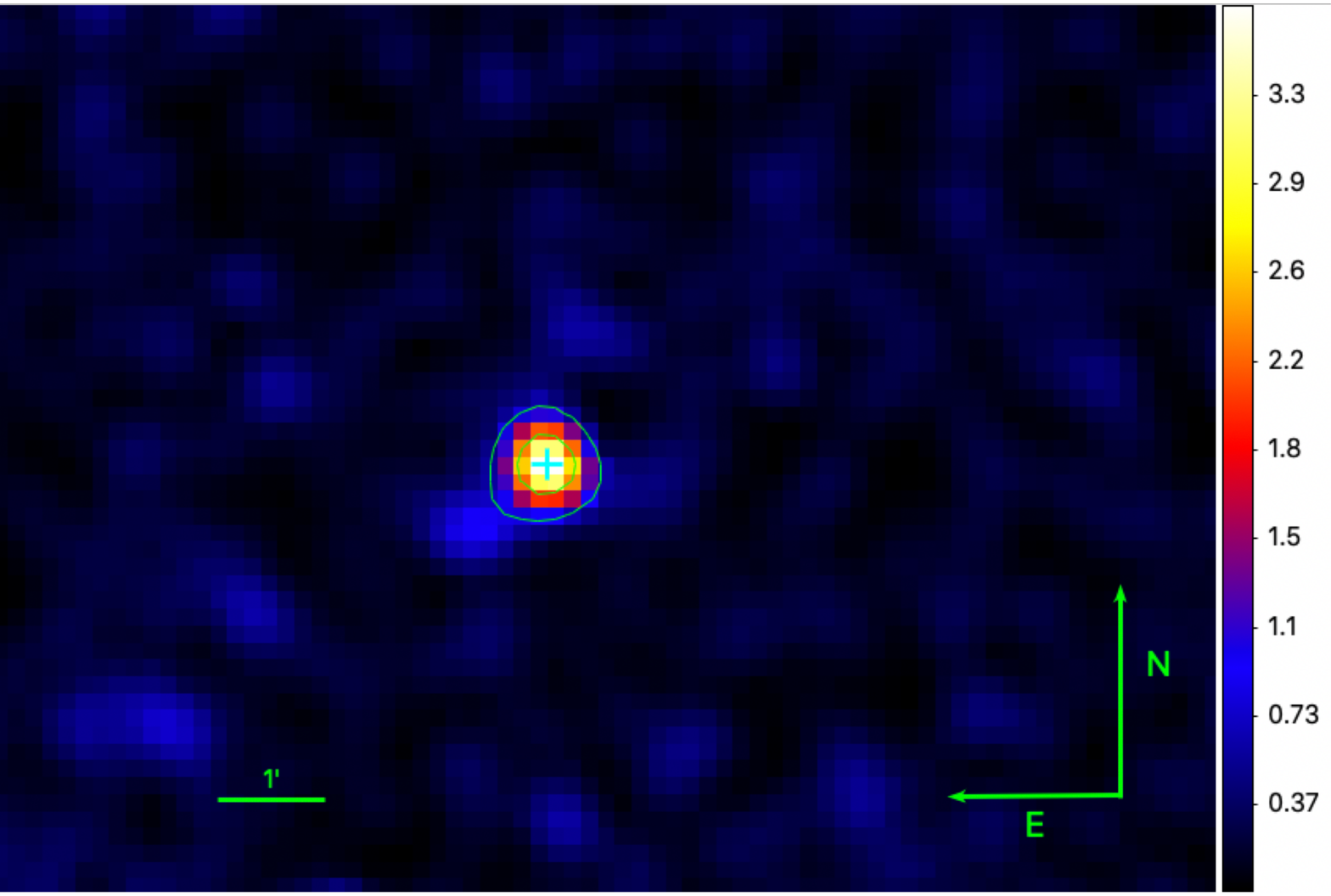}
\caption{Stacked {\it Swift}/XRT image in the 0.3-10 keV band of AT\,2017gge. The position of the transient is marked with a cyan cross. An X-ray source is clearly visible at the transient position.}
\label{fig:xray_image}
\end{figure}

We processed the XRT observations by using the online analysis tools provided
by the UK Swift Science Data Centre \cite[][]{evans07, evans09}. The data have been combined together in order to obtain a single deep stack of total exposure time $\sim$22 ks which has been analyzed locally. The stacked image is shown in Figure \ref{fig:xray_image}. An X-ray source at the position of AT\,2017gge is clearly visible. By using a 25\arcsec\/ wide aperture placed at the transient coordinates we measured an excess of 78$\pm$12 counts above the background (at 6.6$\sigma$). The X-ray emission was delayed with respect to the optical one (see below).
When active AT2017gge showed a 0.3-10 keV count rate of $\sim 10^{-2}$ ct s$^{-1}$. 

\section{Photometric analysis}
\label{phot}

\subsection{Differential photometry}
In order to analyse the photometric properties of AT\,2017gge, we first have carried out differential photometry procedures against the host galaxy for each images of our data-set. This has allowed us to subtract the host galaxy contribution and thus to obtain images where only the nuclear transient's emission is present. To this purpose we used the pre-transient host galaxy images available in the Pan-STARRS1 data archive (DR2) in the $g^\prime$ and $r^\prime$ filters, and the host cataloged $u^\prime$ image available on the SDSS archive \cite[DR16,][]{ahumada20, doi10}. For the data in the NIR taken with NTT/SofI, we have used the host galaxy images obtained with the same instrument at very late times, 1346 days from the AT\,2017gge discovery. 

\begin{figure}
\centering
\includegraphics[scale=0.25, angle=0]{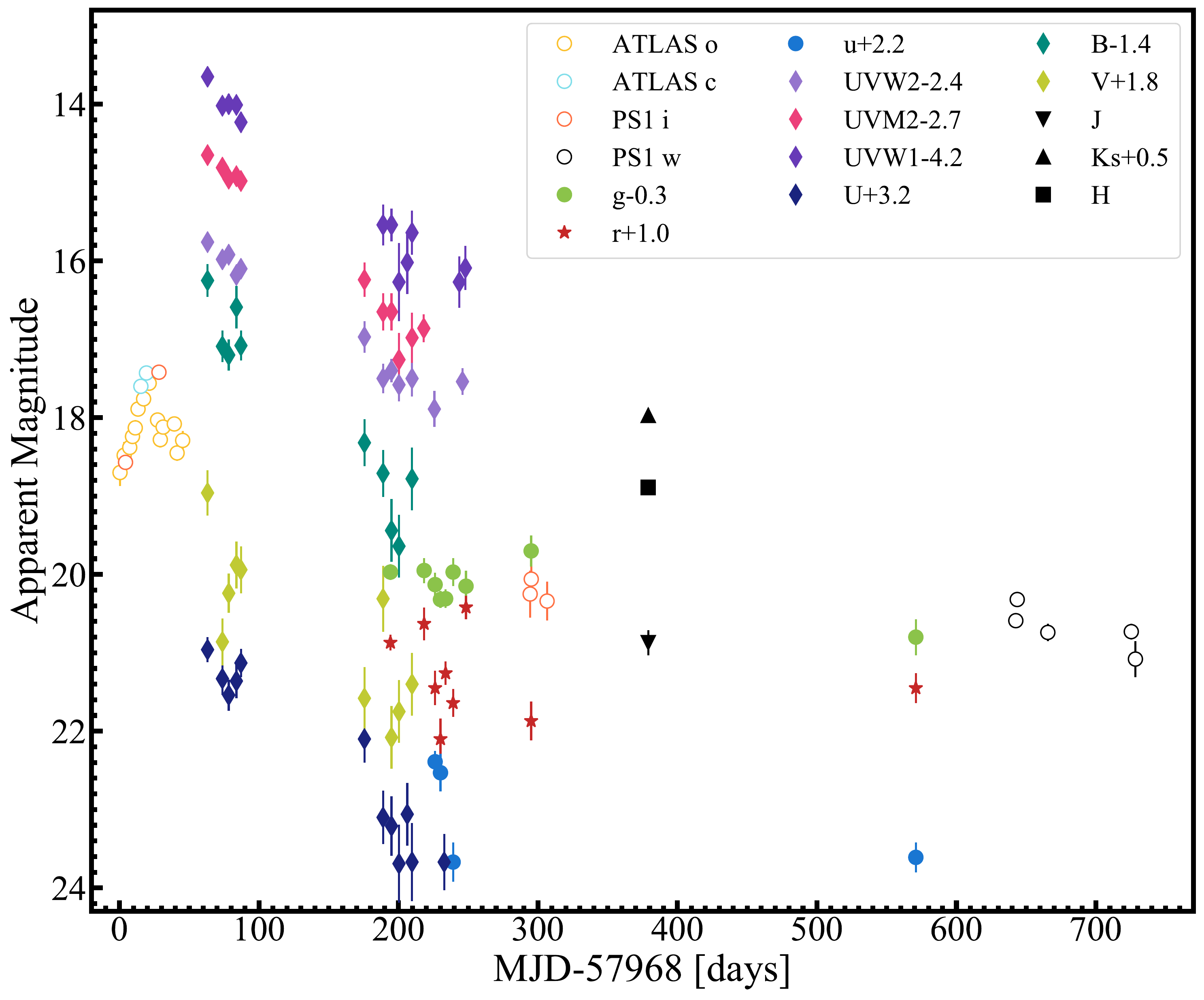}
\caption{Lightcurves for AT\,2017gge. Open circles indicate the optical photometric data taken from the ATLAS and PS1 database, filled circles and filled stars indicate the NOT/ALFOSC, NTT/EFOSC2, TNG/DOLORES and LT/IO:O data from this work, while the \textit {Swift}/UVOT data are shown as colored filled diamonds. The black filled points (square and triangles) indicate the NIR photometry from NTT/SofI photometry. All the magnitudes are in AB system and not corrected for foreground Galactic extinction. All magnitudes are host-subtracted.
The day zero on the x-axis corresponds to the transient's discovery date as reported by the ATLAS survey \citep{atlas17gge}.
}
\label{fig:LC}
\end{figure}

For each images we subtracted the host galaxy contribution by using \texttt{hotpants} V5.1.11 software \citep{hotpants}. We have then applied aperture photometry on the resulting images by using the \texttt{sextractor} software with apertures of variable size, depending on the seeing conditions. In Table \ref{tbl:Optphot} we report the measured host-subtracted optical and IR magnitudes of AT\,2017gge, not corrected for foreground extinction and in their common systems (AB for Sloan filters and Vega for Johnson filters). In the case of the UVOT filters, we have used the host contribution as determined from the photometric analysis performed on the very late-time observation (taken at $\sim$1680 days from the transient discovery). The resulting light curves, including also the host-subtracted UVOT, ATLAS and PS1 data, are shown in Figure \ref{fig:LC}.

\subsection{Bolometric light curve}
\begin{figure}
\centering
\includegraphics[scale=0.3, angle=0]{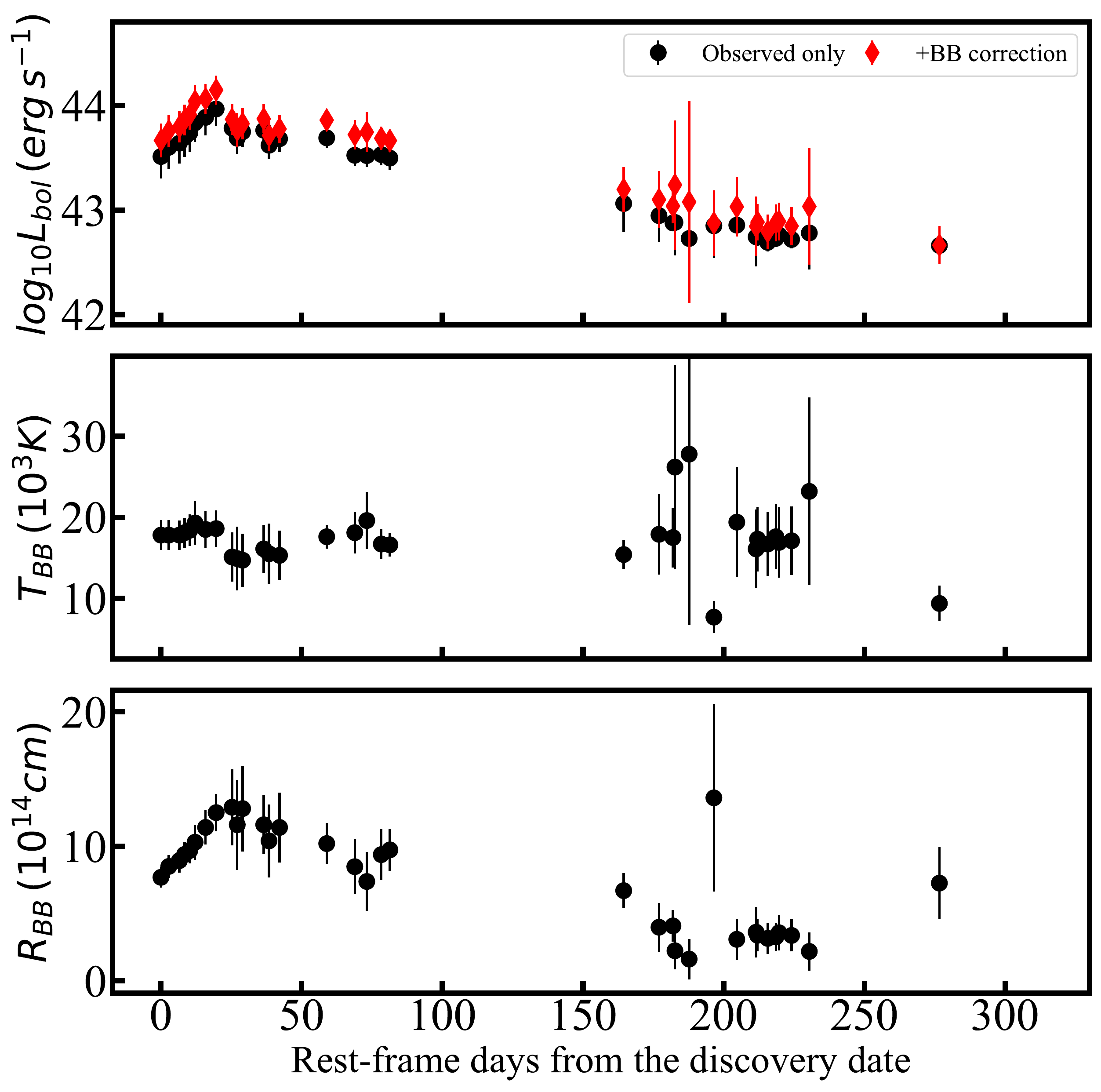}
\caption{{\it Top}: Evolution of the bolometric luminosity during the first 300 days from the AT\,2017gge discovery. The black filled circles show the pseudo-bolometric luminosity calculated by integrating over the observed fluxes, while with red filled diamonds we show the luminosity derived by fitting the SED with a single component black body model. {\it Middle}: Black body temperature evolution. {\it Bottom}: Black body radius evolution.}
\label{fig:bbfit}
\end{figure}

From the multi-color photometry derived from our analysis, together with the values from the ATLAS and PS1 surveys, we computed the bolometric luminosities for AT\,2017gge. We used the \texttt{python} routine \texttt{superbol} \cite[]{nicholl18}, all the input magnitudes have been corrected for the Galactic extinction from \cite{schlafly11}, which assumes a reddening law with $R_{V}=3.1$, and K-correction \cite[][]{oke68} has been applied. When needed, we extrapolated the photometry assuming a constant color evolution for the light curve. Subsequently, we have integrated over the spectral energy distribution (SED) inferred from the multi-band data (for each epoch) and, finally, we fitted the SED with a single black body function. The best-fit black body at each epoch was then used to compute the additional flux bluewards of the UVW2 band and redwards of $K$ band. In Figure \ref{fig:bbfit} we show the results for the first $\sim$300 days from the transient's discovery: the bolometric luminosity evolution (upper panel, black circles for the luminosity calculated by integrating over observed fluxes only, the red diamonds show the luminosity derived using a single black body model to fit the SED); the black body temperature and radius evolution are shown in the central and lower panel, respectively. While the black body temperature is characterized by no significant evolution over time, being consistent with a constant value of $T_{\rm BB} \sim$1.8$\times$10$^{4}$ K, the black body radius show a first phase of expansion from $R_{\rm BB}$=(7.7$\pm$0.8)$\times$10$^{14}$ cm to a maximum value of $R_{\rm BB}\sim$(12.9$\pm$2.8)$\times$10$^{14}$ cm, reached at $\sim$25 days from the transient's discovery, followed by a phase of decline finally reaching a value of $R_{\rm BB}$=(7.3$\pm$2.7)$\times$10$^{14}$ after $\sim$300 days form the transient's discovery.  

\begin{figure}
\centering
\includegraphics[scale=0.3, angle=0]{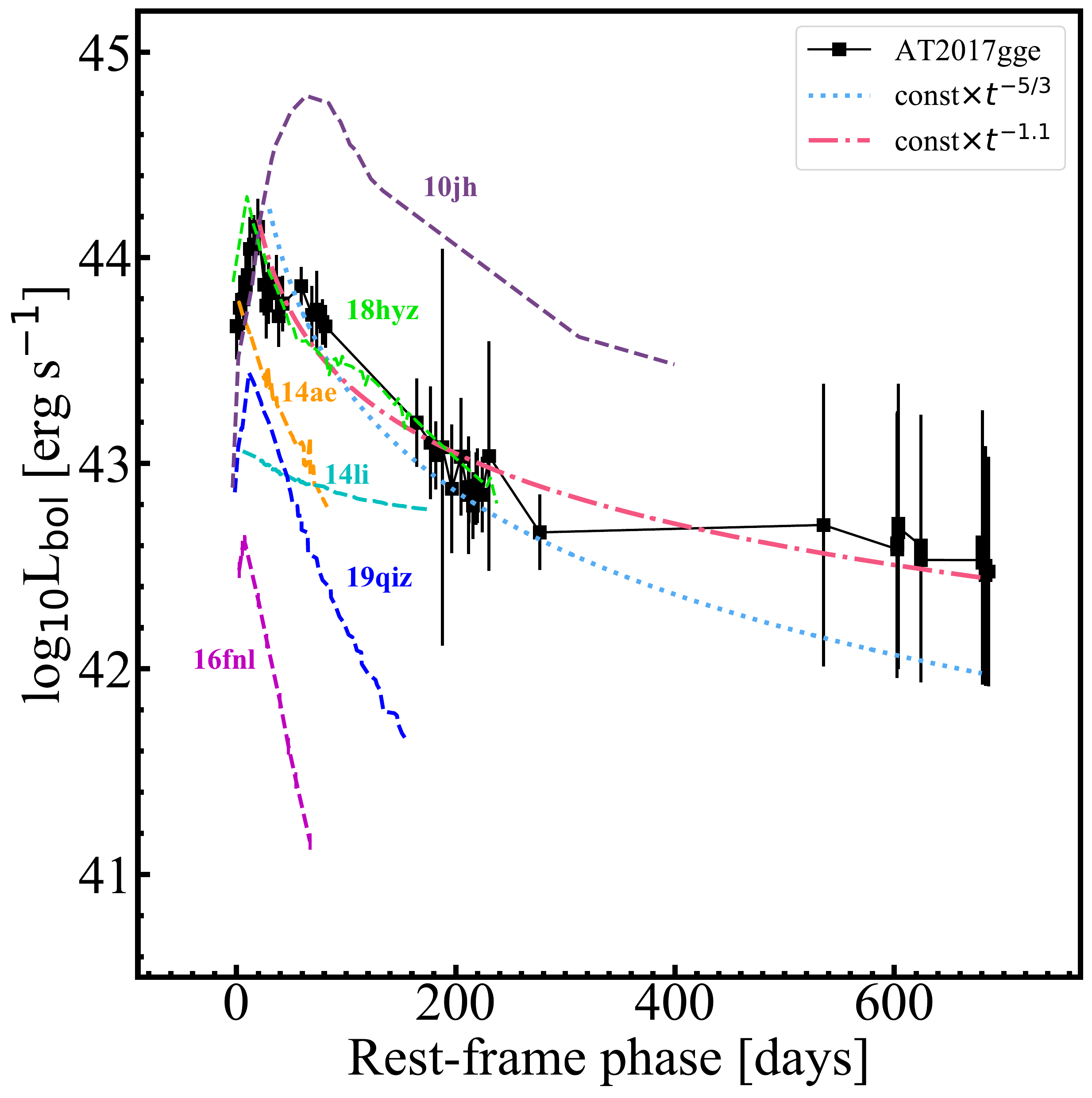}
\caption{AT\,2017gge bolometric light curve (filled black squares), with a powerlaw decline of $t^{-5/3}$ and $t^{-1.1}$ indicated by a light blue dotted line and a red dot-dashed line, respectively. We show, for comparison, the bolometric light curves of some other TDEs with colored dashed lines: PS1-10jh \citep{gezari12}; ASASSN-14ae \citep{holoien14}; ASASSN-14li \citep{holoien16a}; iPTF16fnl \citep{blagorodnova17, onori19}; AT\,2018hyz \citep{short20} and AT\,2019qiz \citep{nicholl20}. In all cases the phase is reported with respect to the discovery date.
}
\label{fig:Lbol}
\end{figure}

In Figure \ref{fig:Lbol} we show the bolometric luminosity of AT\,2017gge (derived using the black body correction, filled black square) in comparison with the bolometric light curves derived for other TDEs\footnote{Photometric data retrieved from The Open TDE Catalog https://tde.space} by applying the same method used for AT\,2017gge data (colored dashed lines). After PS1-10jh \cite[][]{gezari12}, AT\,2017gge is the second brightest object in the plot, with a luminosity and an evolution notably similar to what observed for AT\,2018hyz. Moreover, for the first $\sim$100 days it shows a light curve behaviour comparable to the one observed for AT\,2019qiz. From the bolometric luminosity evolution we estimate that the emission reached its peak value of $L_{\rm bol}$=(1.4$\pm$0.5)$\times$10$^{44}$ \unitlum\/ at MJD\,57\,989.26, which corresponds to $\sim$20 days from the transient's discovery. The rise of the light curve is consistent with a $t^{1/2}$ powerlaw, while in decline it is better represented by a $t^{-1.1}$ powerlaw (red dot-dashed line) rather than the $t^{-5/3}$ trend (light-blue dot-dashed line), when considering $\sim$600 days of emission. We note that during the first $\sim$200 days the light curve decline can be well represented also by the $t^{-5/3}$ power-law. The total energy radiated ($E_{\rm rad}$) is derived by integrating the bolometric luminosity over time and it results $E_{\rm rad}$= (1.0$\pm$0.1)$\times 10^{51}$ erg. 
In their recent letter on AT\,2017gge, \cite{wang22} independently derived the bolometric lightcurve, together with the black body temperature and radius by using all the UVOT data but V filter and only the ATLAS $o$ band data. Although their values are consistent with our results, they found a declining trend for the bolometric lightcurve luminosity compatible with the $t^{-5/3}$ powerlaw. This discrepancy can be explained with the use of a more complete and well sampled photometric dataset in our analysis.

The photometric properties, such as the steep rise to the peak luminosity, reached after $\sim$20 days from the transient's discovery, the powerlaw decay of the bolometric luminosity 
the constant black body temperature at $T_{\rm BB} \sim$1.8$\times$10$^{4}$ K and the black body radius $R_{\rm BB}$ $\sim$10$^{15}$ cm evolving with time, are all commonly observed in TDEs \cite[][and reference therein]{hinkle20, vanVelzen20,vanVelzen21,zabludoff21}, and thus are indicative of the TDE nature of AT\,2017gge. 

\subsection{IR emission}
\begin{figure}
\includegraphics[width=0.47\columnwidth, angle=0]{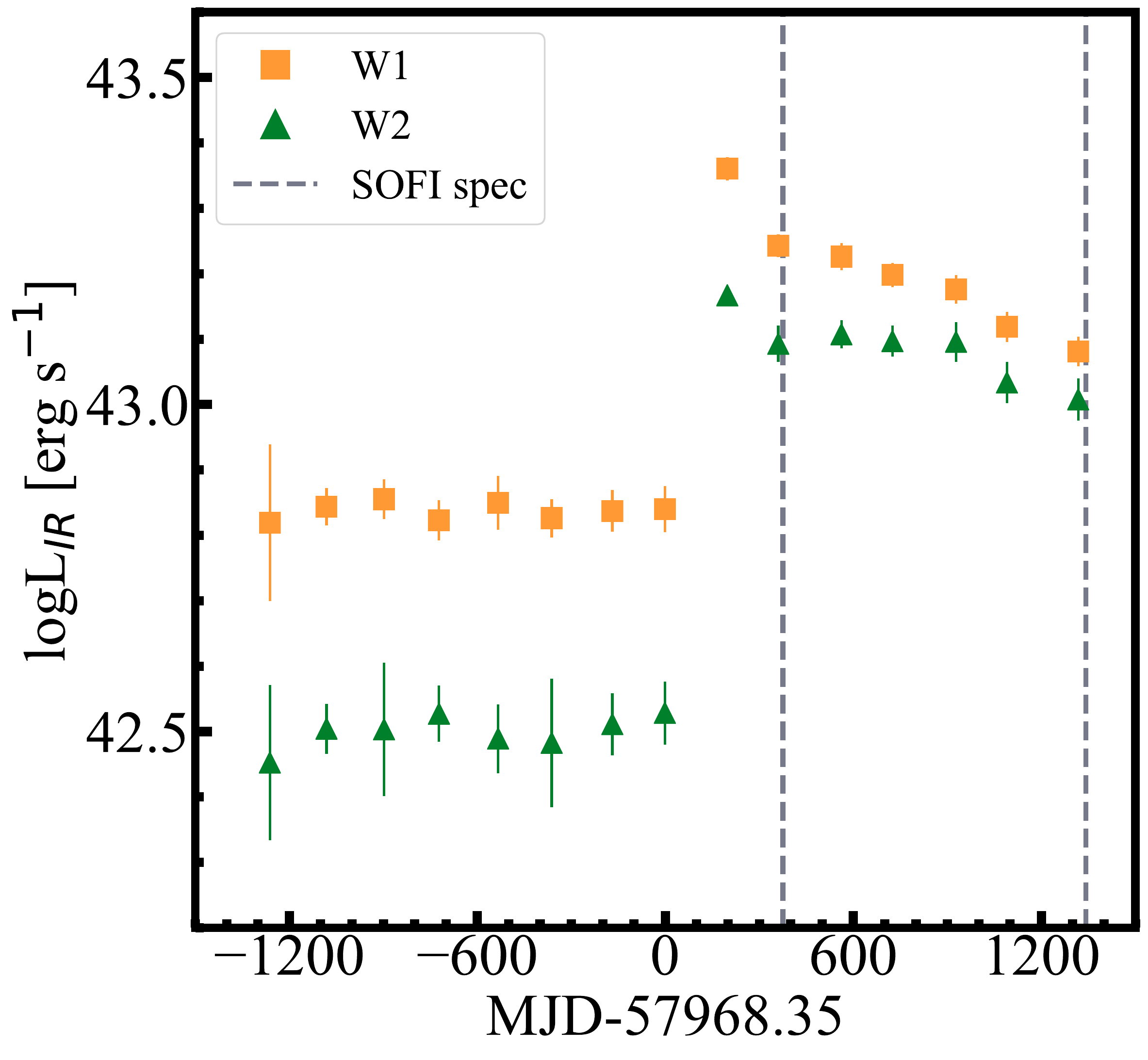}
\includegraphics[width=0.52\columnwidth, angle=0]{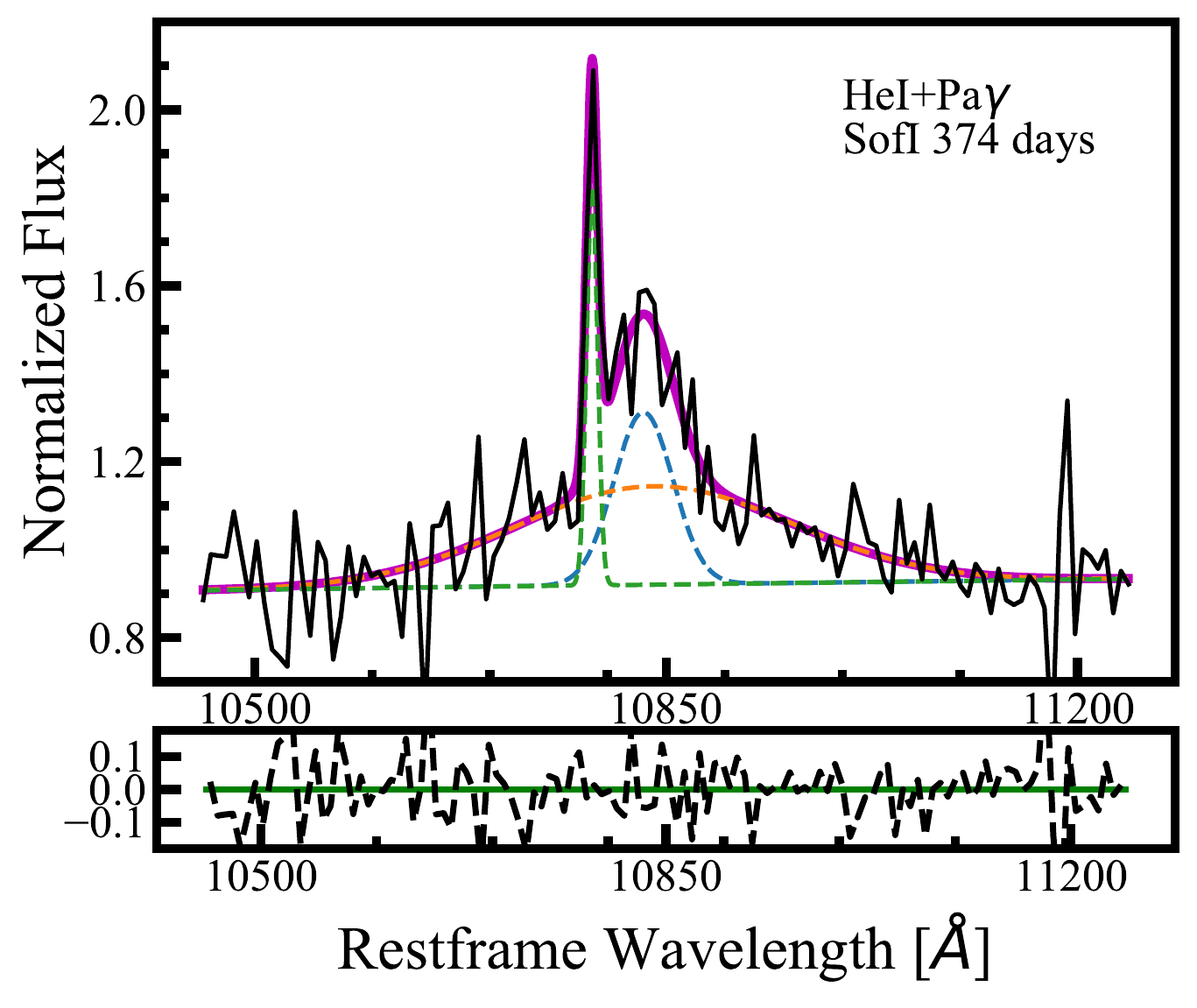}
\includegraphics[width=1\columnwidth, angle=0]{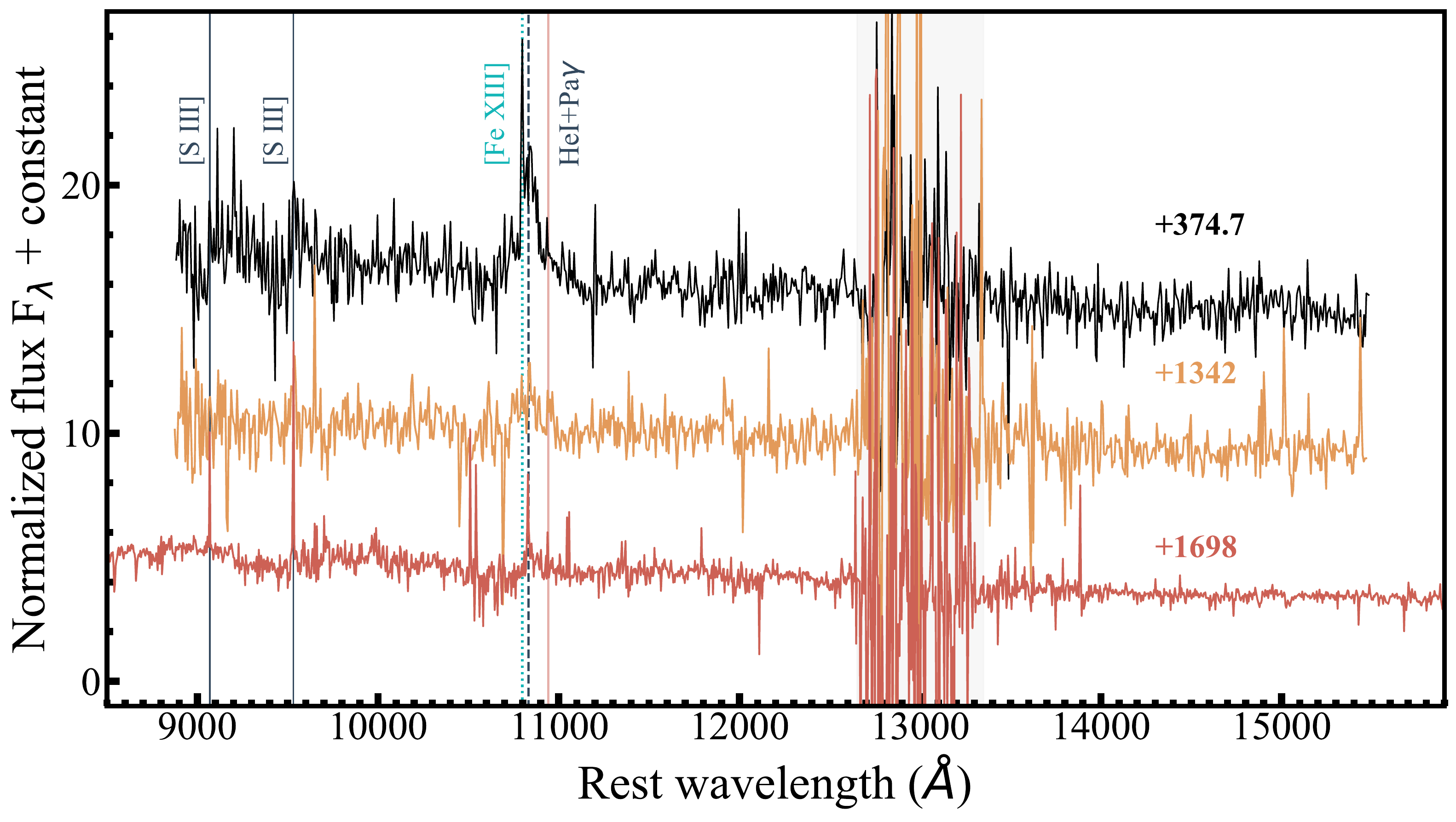}
\caption{ {\it Upper left:} IR light curves for AT\,2017gge, covering a range of $\sim$ 2400 days around the transient's discovery. The luminosities were derived by using the WISE $W1$ (3.4 $\upmu$m, orange squares) and W2 (4.6 $\upmu$m, green triangles) magnitudes, corrected for foreground extinction. Grey dashed vertical lines indicate the phase at which IR SofI spectra have been taken. {\it Upper right:} Multi-component Gaussian fit of the \ion{He}{I} $\lambda$10830 region of the SofI spectrum taken at 374 days. Dashed colored lines indicate each component, while the total model is shown with the magenta solid line. Residuals with respect to the fitting model are shown at the bottom. {\it Lower:} Rest-frame NIR spectra of AT\,2017gge taken with SofI at two different epochs from the transient's discovery (374.7 days and 1342 days, respectively) in comparison with the NIR X-shooter spectrum taken at 1698 days. The spectra have been corrected for the foreground reddening. The detected emission lines are indicated by vertical lines. The regions affected by telluric absorption lines is indicated by grey bands.}
\label{fig:wise}
\end{figure}

Interestingly, AT\,2017gge is listed in the Mid-InfraRed Outburst in Nearby Galaxies (MIRONG) sample \cite[][]{jiang21}, which includes a total of 137 low-redshift SDSS galaxies that have experienced
recent MIR flares of at least an amplitude of 0.5 mag in their WISE light curves. In Figure \ref{fig:wise} we show the AT\,2017gge WISE light curves for the $W1$ and $W2$ filters (with half a year sampling) and the epochs of the two SofI spectra. The MIR flare is detected at $\sim$197 days from the transient's discovery in both filters and the first SofI spectrum has been obtained 178 days after the MIR flare, when the MIR emission is still $\sim$0.6 dex higher than during the quiescent phase. 

Notably, in their work, \cite{jiang21} propose IR echoes (i.e. optical/UV light absorbed and reradiated by dust) of emission originating from transient accretion onto SMBHs as the main source of the detected MIR outbursts and they derive some physical properties of the dust responsible for the detected emission, such as the temperature, luminosity, mass and its distance from the central heating source. Specifically, for the case of AT\,2017gge they infer a distance of the dust of $\log R_{\rm dust}$=-1.51$\pm$0.12 pc or $\log R_{dust}$=-1.01$\pm$0.16 pc, depending on the considered absorption coefficient, which correspond to $R_{\rm dust}\sim$10$^{17}$cm. 
We independently computed the time lags ($\tau$) between the optical and MIR data
using two different methods, the interpolated cross correlation function \cite[ICCF,][]{white94} and the \texttt{javelin} tool 
\cite[][]{zu13,zu16}. From the ICCF method, the time lag $\tau$ for $W1$ band is estimated to be $\tau=204.6^{+77.6}_{-43.5}$ days, which is consistent with that computed applying the \texttt{javelin} method ($\tau \approx 207.6^{+24.7}_{-24.7}$ days). These methods are commonly used in active galactic nuclei (AGN) reverberation mapping studies and are based on the tight correlation between the IR luminosity and the dust radius of ordinary AGNs (size--luminosity relation). Thus, it is possible to infer the luminosity of a possible AGN from the estimated time lag. In particular, by using the size--luminosity relation from \cite{lyu19}, we found $L_{\rm AGN}\approx 10^{45}$ erg s$^{-1}$. 

Our estimated peak bolometric luminosity for AT 2017gge from the observed optical and UV fluxes was (1.4$\pm$0.5)$\times$10$^{44}$~erg s$^{-1}$, but we can infer that the true peak was larger than this due to the presence of the IR echo. Therefore it is not unreasonable to assume that the intrinsic TDE luminosity was large enough to be consistent with this estimate. The time lag measurements above also provide values for the distance to the dust producing the IR echo, which are $0.161^{+0.061}_{-0.034}$~pc for CCF, and $0.163\pm0.019$~pc for \texttt{javelin}. These values are similar to the dust sublimation radius of 0.15~pc found for the TDE PTF-09ge by \citet{vanvelzen16}, implying that is is reasonable for AT~2017gge to have also sublimated dust out to this radius. The estimates for the dust radius in \citet{jiang21} are somewhat smaller than this, and precise measurements require a more involved model. 
It is possible to use the infrared light echo to estimate the required luminosity of the flare needed to sublimate the dust to the sublimation radius. In particular, we have obtained a flare peak luminosity of $L_{\rm abs}$=1.14$\times$10$^{45}$\unitlum by using the the Equation 1 reported in \cite{vanvelzen16}, the (independently derived) sublimation radius of 0.16 pc and assuming a sublimation temperature of 1850 K for the dust. Notably, this value is consistent with the luminosity inferred by using the AGN size--luminosity relation of \cite{lyu19} ($L_{\rm AGN}\approx 10^{45}$ erg s$^{-1}$) and thus further support the hypothesis of an intrinsic TDE luminosity larger than what derived from the observed optical and UV fluxes.
Interestingly, from the the MIR luminosity analysis, \cite{wang22} derived a covering factor of $\sim$0.2 for the dusty environment. As already discussed by the authors, such a high value is comparable with what is usually observed for the AGN torus.
However, prior to the transient occurrence, an AGN classification has been excluded given: a) the X-ray non-detection in the RASS observations, b) the quiescent WISE color \cite[W1-W2$\sim$0.6,][]{stern12} and c) the the SDSS spectrum based location in the Baldwin, Phillips \& Terlevich (BPT) diagrams. We however note that the pre-transient SDSS values placed the host galaxy on the (theoretical) extreme starburst line and this may indicate the presence of a weak AGN contribution, see Section \ref{subsec:BPT} for a detailed discussion).

\citet{wang22} estimated the total radiated energy of AT 2017gge in the IR (until July 2021) of 1.3$\times$10$^{51}$ erg. Combined with our estimated radiated energy in the optical the total radiated energy detected is $\sim$2$\times$10$^{51}$ erg and this is a lower limit for the total radiated energy from the TDE. This is still significantly lower than the total energy of 10$^{53}$ erg expected to be released by the accretion of a solar mass star \citep[see e.g.][]{Giannios2011}. However, the covering factor of $\sim$0.2 implies that a large proportion of radiation in the UV could still be unobserved in both the optical and IR bands and make up for this discrepancy. 

\section{The delayed X-ray emission}
In this section we describe the spectral analysis performed on the X-ray source detected at the location of AT\,2017gge in the stacked {\it Swift}/XRT image, shown in Figure \ref{fig:xray_image}. We extracted the spectrum of the source by using a 40\arcsec\/ wide region centered on the position of the transient and a background region of 140\arcsec\/ radius placed in an area free of sources. The data fitting procedure has been performed by using the software \texttt{XSPEC} v12.11.1 \cite[]{Arnaud96} and the Cash statistics. The extracted spectrum has been grouped to at least four counts per bin. We used a fixed Galactic column density of $N_{\rm H}$=5.0$\times$10$^{20}$cm$^{-2}$, as derived for the source position by using the Heasarch \texttt{nH} \texttt{ftools} and the HI4PI Collaboration map \cite[][]{HI4PI}. 
\subsection{The X-ray light-curve}
\begin{figure}
\centering
\includegraphics[scale=0.30, angle=0]{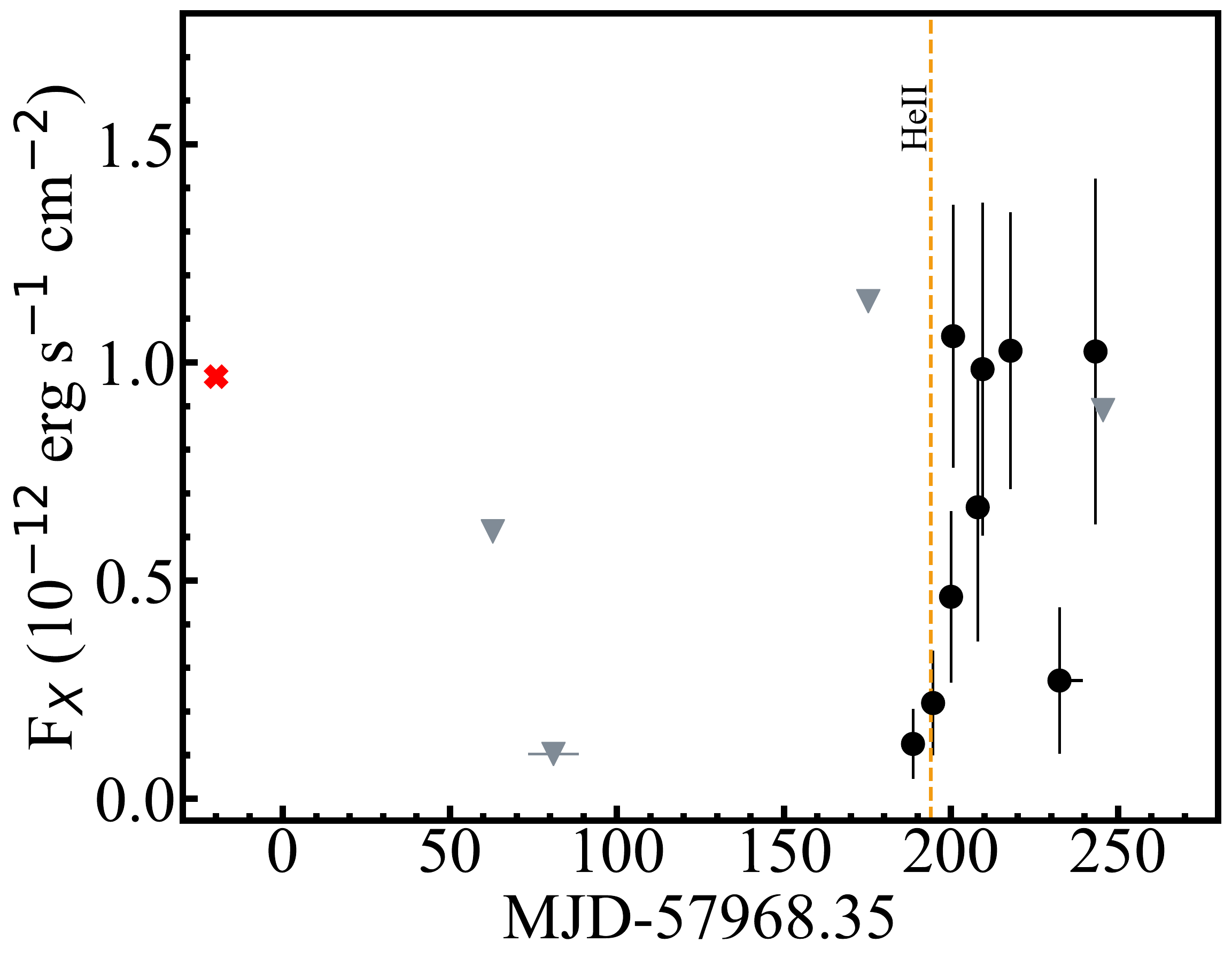}
\caption{The X-ray light curve in the 0.3-10 keV band from the XRT observations. The grey triangles represent upper limits. The upper limit at $\sim$80 days from the transient discovery are derived by stacking 4 nearby observation (from MJD 58041.75 to MJD 58055.04). The orange vertical dashed line indicates the  first detection of the broad \ion{He}{II} $\lambda$4686 emission lines in the optical spectra. The pre-transient upper limits from the RASS observation of the AT\,2017gge field, taken at MJD=48182, are shown with a red cross.}
\label{fig:xray_lc}
\end{figure}

In Figure \ref{fig:xray_lc} the XRT light curve in the 0.3-10 keV band, produced by specifying a minimum of 3 counts per bin, is shown. In order to convert the count rates to unabsorbed fluxes, we used the results obtained from the spectral fitting with an absorbed powerlaw model (see subsection \ref{subsec:xray_spec_fit} for a detailed discussion on the spectral analysis), which gives a conversion factor of 0.07$\times$10$^{−9}$ erg cm$^{-2}$ cts$^{-1}$. The RASS pre-transient upper limit of F$_{0.3-10 keV}$=1.0$\times$10$^{-12}$ \unitflux, corresponding to observation taken at MJD 48102, is also shown before the transient detection for comparison (red cross in Figure \ref{fig:xray_lc}).   
The detected X-ray emission has a transient nature, with only upper limits (grey triangles in the Figure) for the first 170 days from the discovery of AT\,2017gge and showing the peak around 200 days. Furthermore, no X-rays have been detected in the recent XRT observations taken at $\sim$1684 days \cite[see][]{wang22}.
Such a delayed X-ray brightening with respect to the optical/UV peak has already been observed in some TDEs, specifically, ASASSN-14li \cite[][]{pasham17, gezari17}, ASASSN-15oi \cite[][]{gezari17, holoien18}, AT\,2019azh \cite[][]{vanVelzen21, liu22} and OGLE16aaa \cite[][]{kajava20,shu20}, which have shown $\sim$30 days, 1 year, 200 and 140 days of delay in the X-ray emission, respectively. In all these cases the authors investigate the different scenarios for the origin and location of the optical/UV emission and the delayed X-ray flare, including the stream-stream collision and reprocessing photosphere/winds hypothesis. Among these, a two-process scenario, where the UV/optical emission is produced by the stellar debris streams during the circularization phase, whereas a delayed formation of an accretion disk is responsible for the X-rays, has been suggested and, in some cases, has been even preferred \cite[i.e. ASASSN-15oi, ASASSN-14li, OGLE16aaa, AT\,2019azh][]{gezari17, pasham17, kajava20, liu22}.

\subsection{The X-ray spectral analysis} 
\label{subsec:xray_spec_fit}
\begin{figure*}
\centering
\includegraphics[scale=0.25, angle=-90]{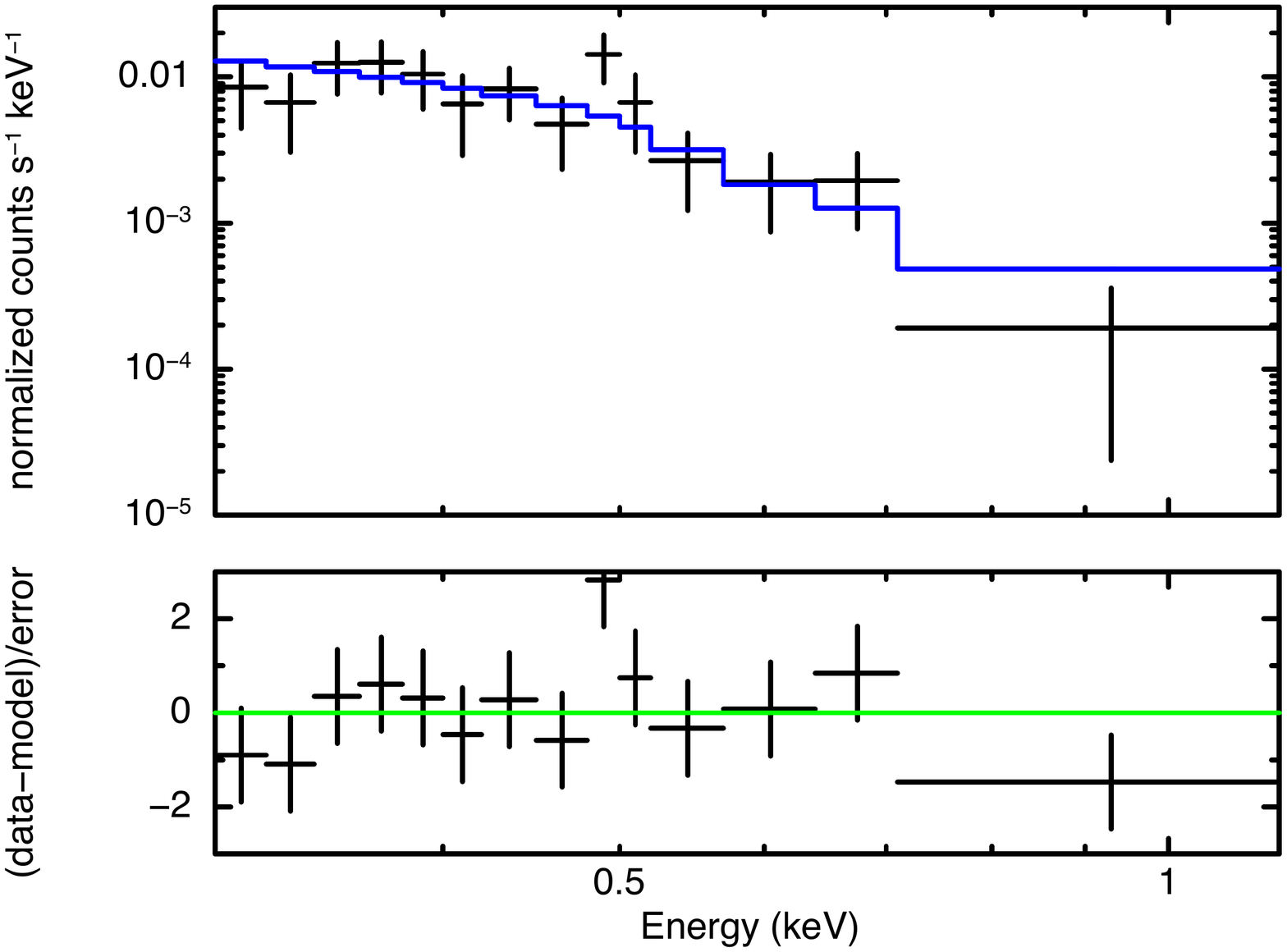}
\includegraphics[scale=0.25, angle=-90]{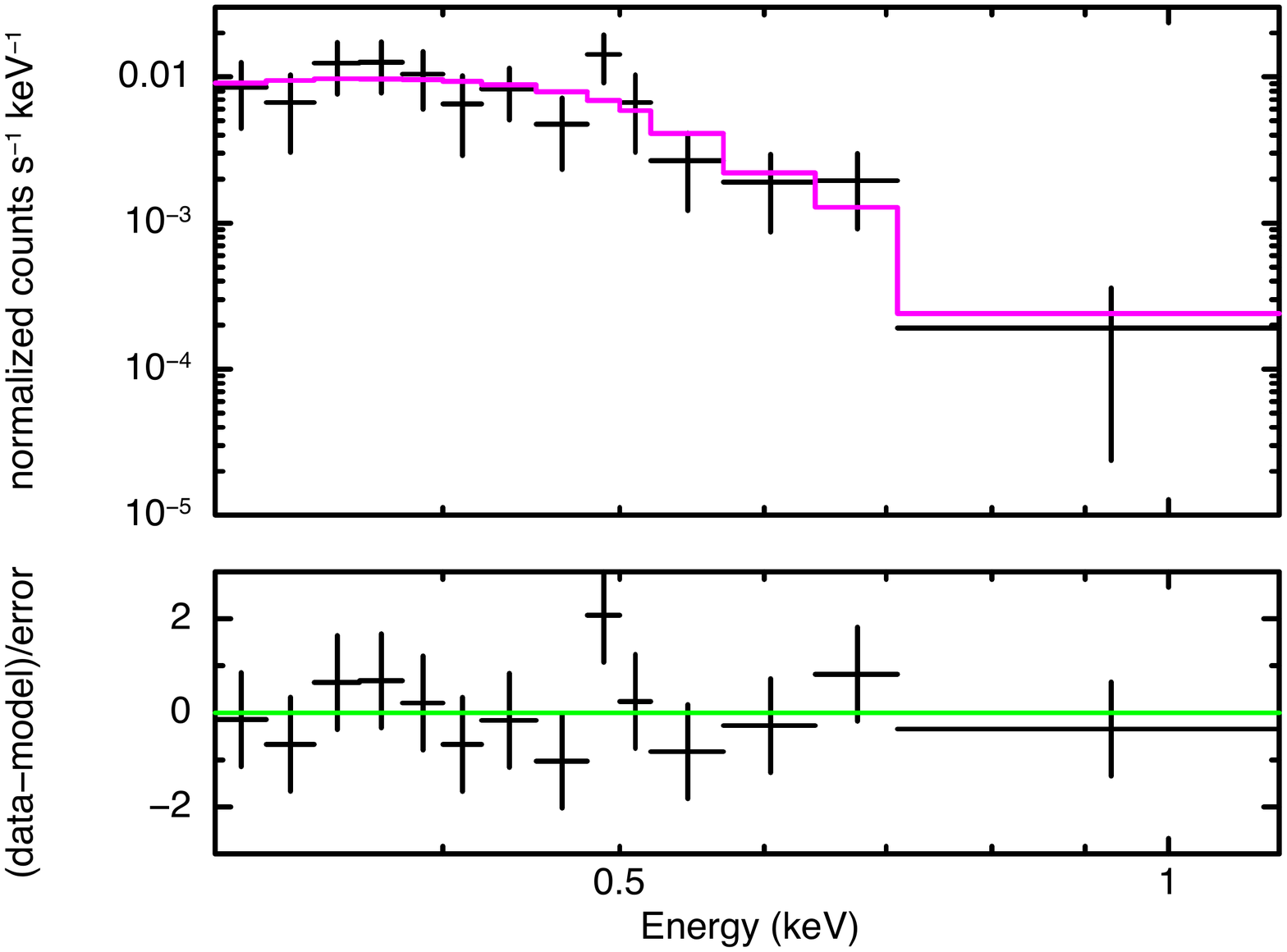}
\includegraphics[scale=0.25, angle=-90]{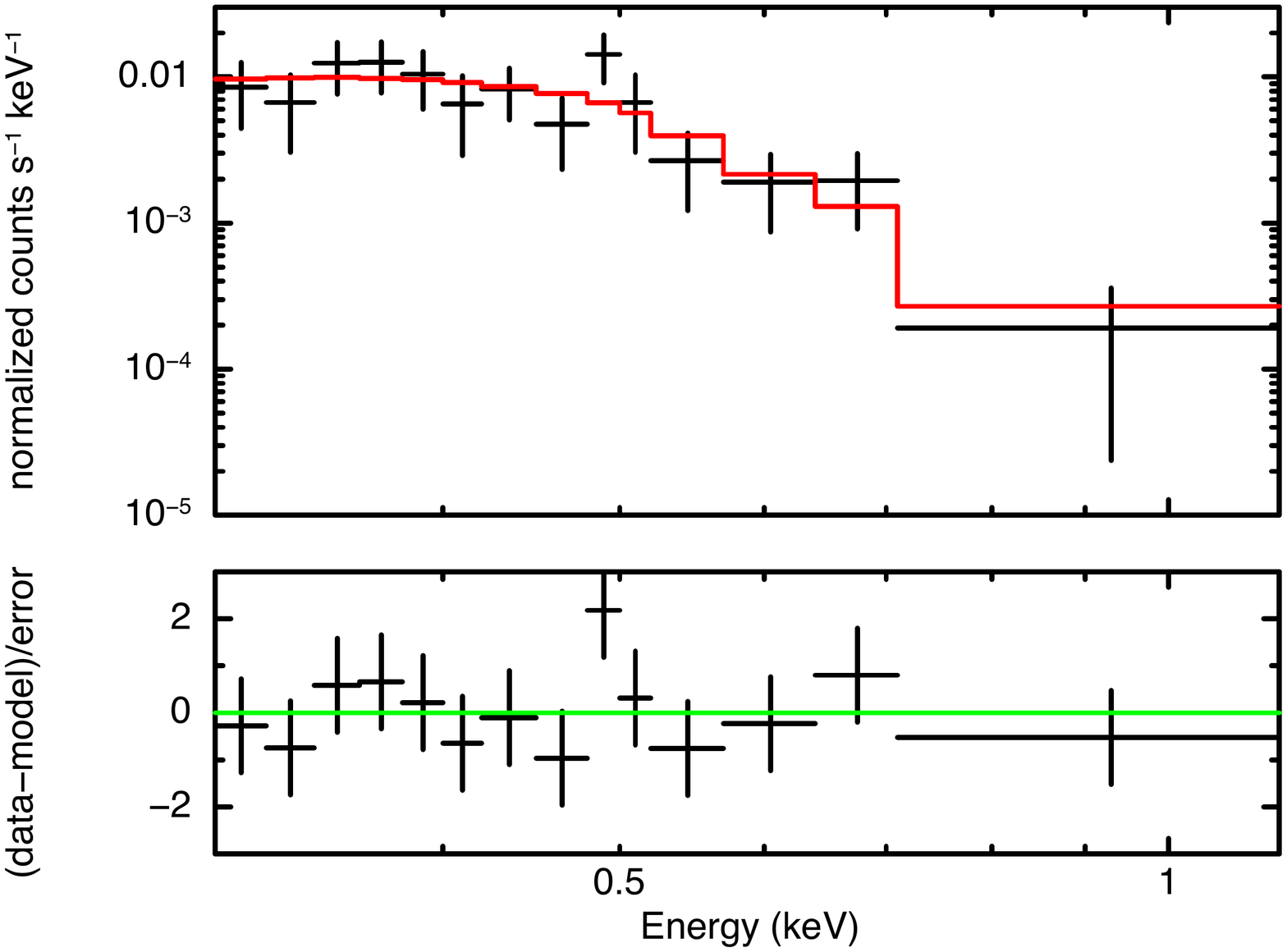}
\caption{Spectral analysis of the AT\,2017gge {\it Swift}/XRT spectrum extracted from deep stacked data (total exposure time $\sim$22 ks) obtained by combining 17 observations taken between MJD 58\,031.12 and MJD 58\,215.98. The results obtained by using an absorbed powerlaw model (\texttt{TBABS*ZASHIFT*POWERLAW}, the \texttt{TBABS*ZASHIFT*BBODYRAD} model and the 
the \texttt{TBABS*ZASHIFT*DISKBB} model are shown in the upper left, upper right and lower panels, respectively. The residuals with respect to the fitting models are shown at the bottom of each figure. Colored solid lines refer to the total model.}
\label{fig:xray_spec}
\end{figure*}

The spectrum of the source is very soft, with the emission detected only between the 0.3-1 keV energy range. We modelled the data with an absorbed powerlaw by using the model \texttt{TBABS*ZASHIFT*POWERLAW} (solid blue line, left panel) and we obtain a photon index of $\Gamma$=6.0$\pm$0.8 and a c-stat/d.o.f.=11.74/12 
The unabsorbed flux in the 0.3-10 keV band results to be $F_{0.3-10}$=(0.18$\pm$0.05)$\times$10$^{-12}$ \unitflux, which corresponds to an X-ray luminosity of $L_{\rm X}$=(2.4$\pm$0.7)$\times$10$^{42}$ \unitlum, calculated for the distance of AT\,2017gge. These spectral features are usually observed in the X-ray emission of non-jetted TDEs, which show very soft spectra well represented either by a single powerlaw with a photon index value $\Gamma$>4 or by a black body model with temperatures between 10-100 eV \cite[][]{auchettl17, saxton20}. Therefore, we also modelled the X-ray spectrum using a black body model \texttt{TBABS*ZASHIFT*BBODYRAD}  (middle panel of Figure \ref{fig:xray_spec}, magenta solid line). In this case we obtain c-stat/d.o.f.=7.24/12. The inferred temperature and radius are k$T$=70$\pm$10 eV and $R$=(1.4$^{+0.7}_{-0.4}$)$\times$10$^{11}$ cm. These values are consistent with a TDE nature for this transient. The unabsorbed flux in the 0.3-10 keV band is $F_{0.3-10}$=(0.15$\pm$0.04)$\times$10$^{-12}$ \unitflux, which corresponds to a luminosity of $L_{\rm X}$=(1.8$\pm$0.5$)\times$10$^{42}$ \unitlum\/. The results of the spectral analysis are reported in Table \ref{tab:xrt_fit} and are shown if Figure \ref{fig:xray_spec}. These results are all consistent with the X-ray analysis already presented in the recent discovery paper of AT\,2017gge \cite[][]{wang22}.
Additionally, in order to investigate the origin of the delayed X-ray flare as produced by a newly formed accretion disk, we have modelled the data with the \texttt{TBABS*ZASHIFT*DISKBB} model (right panel of Figure \ref{fig:xray_spec}, red solid line). In this case we obtain a c-stat/d.o.f.=7.76/12, a temperature at the inner disk radius of T$_{in}$=83$\pm$10 eV, an unabsorbed flux of $F_{0.3-10}$=(0.15$\pm$0.04)$\times$10$^{-12}$ \unitflux, which corresponds to a luminosity of $L_{\rm X}$=(1.9$\pm$0.5)$\times$10$^{42}$ \unitlum\/. We note that although it is difficult to discriminate between these two models given the very similar c-stats values, the accretion disk model is able to represent well the data. Thus, given the derived X-ray properties, we can conclude that the AT\,2017gge X-ray emission resembles what is usually seen in non-jetted TDEs, as it is characterized by a very soft emission and a steep spectrum well modelled either by an absorbed powerlaw, by a black body or by an accretion disk model. 

\begin{table}
\caption{Results from the spectral analysis on AT\,2017gge XRT stacked data. 
The reported fluxes are unabsorbed and calculated in the 0.3-10 keV band.}
\label{tab:xrt_fit}
\centering
\begin{tabular}{lll}
\hline\hline
Model				&Parameter		   & Value \\
\hline			
\texttt{TBabs}      &$N_{\rm H}$($\times$10$^{22}$ cm$^{-2}$)  & 0.05 (frozen) \\
\texttt{zashift}    & Redshift         & 0.0655113 (frozen)\\
\texttt{POWERLAW}   &$\Gamma$ 		   &6.00$\pm$0.80 \\
                    & norm ($\times$10$^{-6}$)             &5.28$^{+4.0}_{-2.6}$ \\
                    &C-stat/d.o.f.                  &11.74/12\\
\hline
\multicolumn{2}{l}{$F_{0.3-10keV}$=(0.18$\pm$0.05)$\times$10$^{-12}$ \unitflux}\\
\multicolumn{2}{l}{$L_{0.3-10keV}$=(2.4$\pm$0.7)$\times$10$^{42}$ \unitlum}\\
\hline
\texttt{TBabs}      &$N_{\rm H}$($\times$10$^{22}$ cm$^{-2}$)     & 0.05 (frozen) \\
\texttt{zashift}    & Redshift         & 0.0655113 (frozen)\\
\texttt{BBODYRAD}   & k$T$ (keV) 		   & 0.070$\pm$0.01 \\
                    & norm             & 2072$^{+4416}_{-1324}$ \\
                    &C-stat/d.o.f.                  &7.24/12\\
\hline
\multicolumn{2}{l}{$F_{0.3-10}$=(0.15$\pm$0.04)$\times$10$^{-12}$ \unitflux}\\
\multicolumn{2}{l}{$L_{0.3-10keV}$=(1.8$\pm$0.5)$\times$10$^{42}$ \unitlum}\\
\hline
\texttt{TBabs}      &$N_{\rm H}$($\times$10$^{22}$ cm$^{-2}$)     & 0.05 (frozen) \\
\texttt{zashift}    & Redshift         & 0.0655113 (frozen)\\
\texttt{DISKBB}     & $T_{in}$ (keV)   & 0.083$\pm$0.01 \\
                    & norm             & $<$4187 \\
                    & C-stat/d.o.f.    & 7.76/12\\
\hline
\multicolumn{2}{l}{$F_{0.3-10}$=(0.15$\pm$0.04)$\times$10$^{-12}$ \unitflux}\\
\multicolumn{2}{l}{$L_{0.3-10keV}$=(1.9$\pm$0.5)$\times$10$^{42}$ \unitlum}\\
\hline
\end{tabular}
\end{table}

\section{Optical and NIR spectroscopic analysis}
\label{sec:spectroscopy}

The rest-frame optical spectra of AT\,2017gge taken with the different ground-based facilities described in section \ref{sec:observations} are shown in Figure \ref{fig:spec}, while the rest-frame SofI NIR spectra and the UVB, VIS and NIR X-shooter spectra are shown in the lower panel of Figure \ref{fig:wise} and in Figure \ref{fig:Xsh_spec}, respectively. We corrected all the reduced spectra for the
foreground extinction by using the Cardelli function \cite[][]{cardelli89} with $A_V=0.193$ mag and $R_V=3.1$. In this section we discuss the spectroscopic features detected in the optical and NIR spectral sequence.

\subsection{The optical spectra and the detection of high ionization coronal emission lines}

\begin{figure*}
\includegraphics[scale=0.28, angle=0]{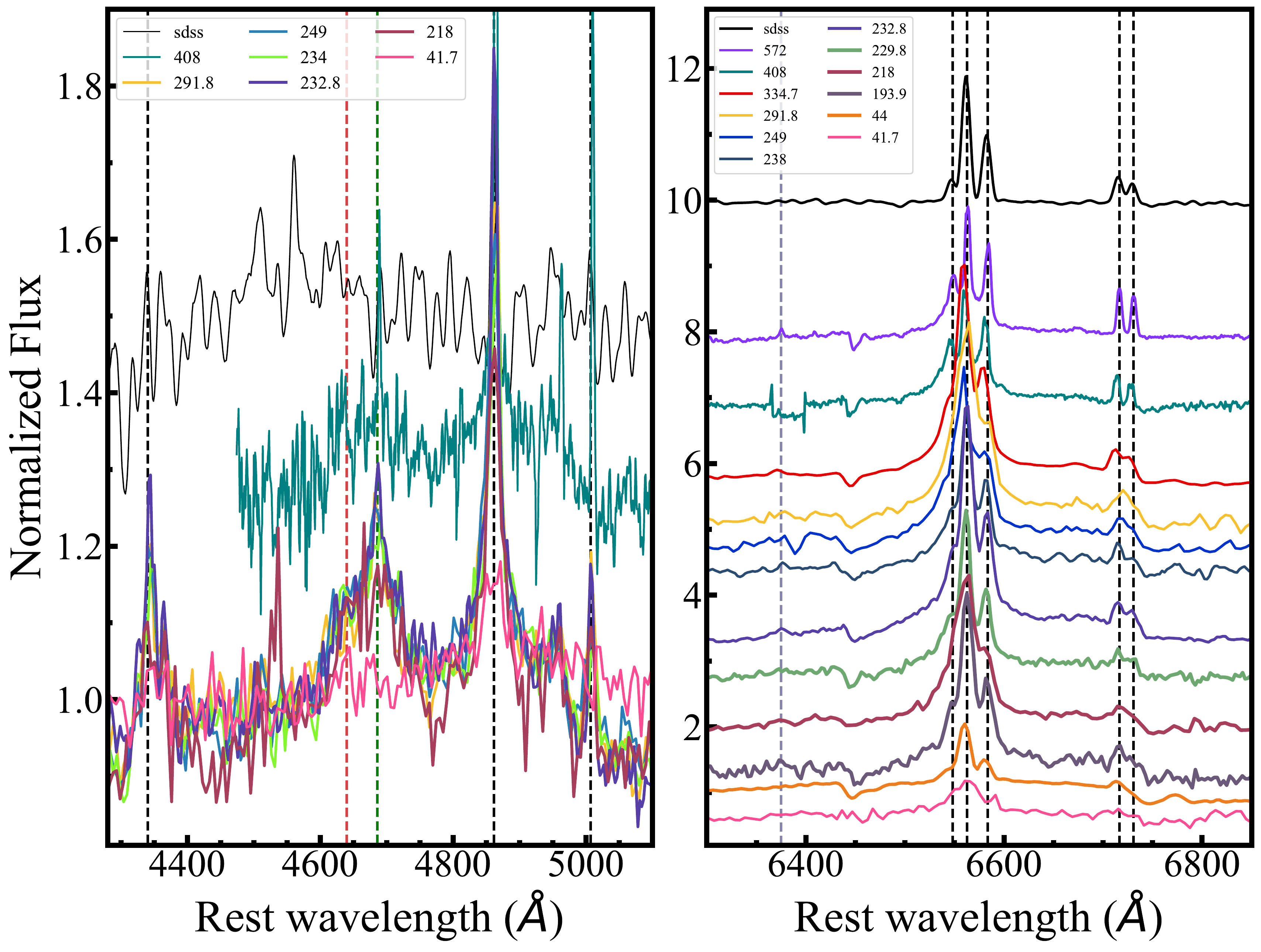}
\caption{Sequence of a selection of normalized spectra showing the lines profile evolution starting from 42 days from the transient's discovery for the \hb\/ and \ha\/ region (left and right panels, respectively). The SDSS pre-transient spectrum is shown for comparison (upper black spectrum in both panels). Vertical black dashed lines show the main emission lines (O, H, [\ion{N}{II}] and [\ion{S}{II}]). Red and green dashed lines indicate the position of the Bowen lines at 4640 \AA\/ and the \ion{He}{II} $\lambda$4686, respectively. Grey dashed line in the right panel shows the position of the high ionization coronal emission line [\ion{Fe}{X}] $\lambda$6374.}
\label{fig:spec_zoom}
\end{figure*}

The first spectra of our data-set (taken $\sim$42-45 days from the transient's discovery) are characterized by broad \hb\/, \ion{He}{I}$  \lambda$5876 and \ha\/ emission lines superimposed on a blue continuum. However, from the spectral sequence shown in Figure \ref{fig:spec}, a clear evolution emerges with time of the broad emission line profiles and, more in general, of the spectral properties, with a number of features, consistent with high ionization coronal emission lines, appearing later in time, following the X-ray flare. 
In Figure \ref{fig:spec_zoom}, the evolution of the lines profile from a selection of normalised spectra is shown for the \hb\/ and the \ha\/ regions (left and right panel, respectively). 
The transition of the \ha\/ from a broad-component-dominated profile toward a narrow-component-dominated profile (pink and violet spectra in Figure \ref{fig:spec_zoom}, respectively) is clearly visible, as well as the emergence of the narrow emission lines of [\ion{N}{II}] and [\ion{S}{II}]. A similar behaviour is observed for the emission lines in the \hb\/ region.

Of particular interest is the emergence of the broad \ion{He}{II} $\lambda$4686, which is not present in the first spectrum of the sequence nor in the pre-transient SDSS spectrum \cite[][]{smee13}, as shown in Figure \ref{fig:spec_zoom} (pink and black spectra, respectively). This feature is first detected in the spectrum taken after 193.9 days from the transient's discovery and it is characterized by a line profile time evolution, with the gradual narrowing of the broad component until only a narrow core remains, which is well visible in the Gemini spectrum taken at day 407.9 (dark green in Figure \ref{fig:spec_zoom}). Notably, the phase in which the broad \ion{He}{II} $\lambda$4686 first appears corresponds to the detection epoch of the soft X-ray flare, as shown in the X-ray light-curve in Figure \ref{fig:xray_lc}. 

\begin{figure*}
\includegraphics[scale=0.30, angle=0]{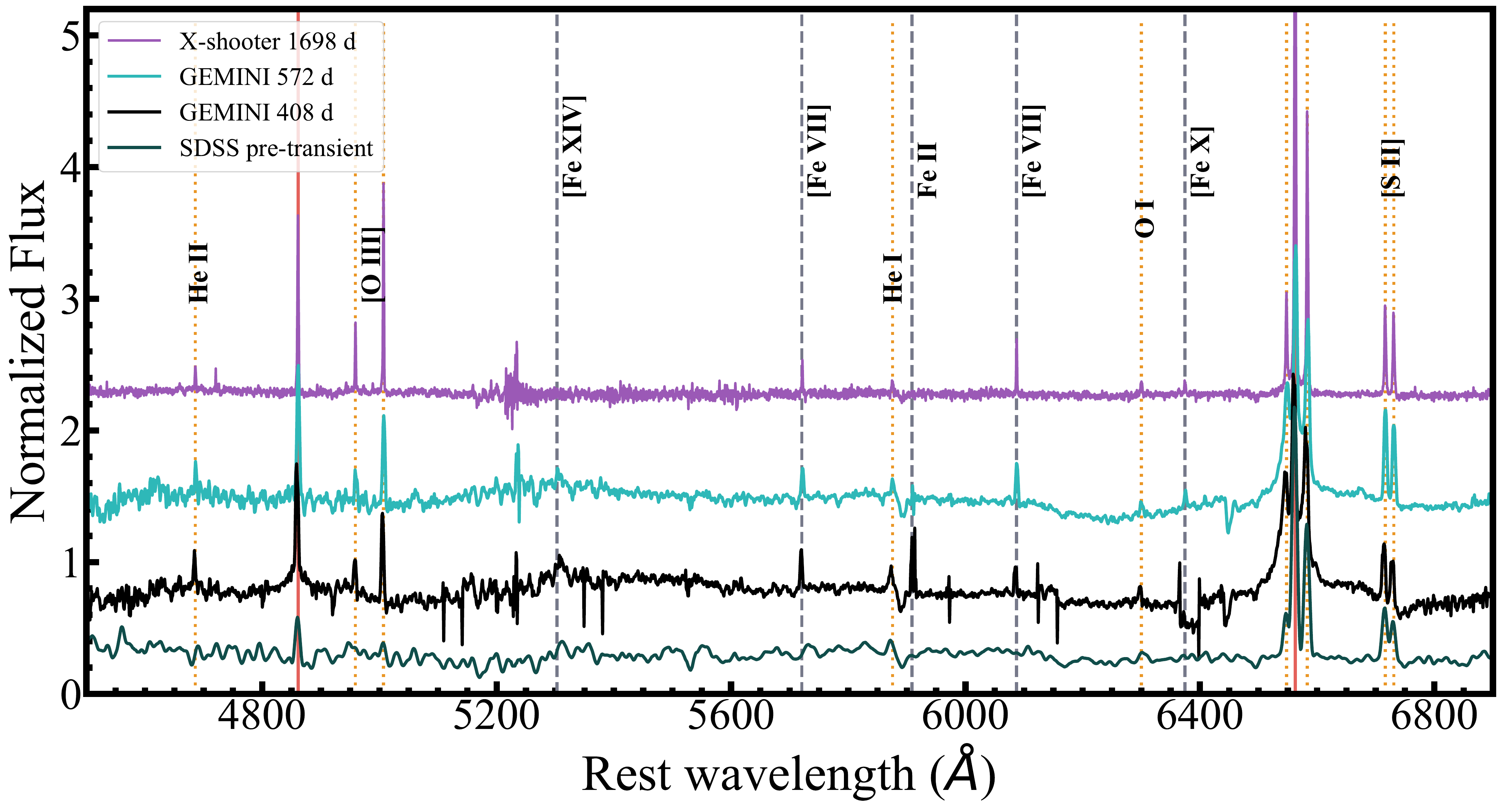}
\caption{Comparison between the SDSS spectrum of AT\,2017gge host galaxy taken before the transient first detection (in grey) and the late-times Gemini and X-shooter spectra taken at days 408, 572 and 1698 (black, cyano and purple, respectively). The positions of the main emission lines are indicated. 
}
\label{fig:sdss_gemini}
\end{figure*}

In Figure \ref{fig:sdss_gemini} we show a comparison between the late-time Gemini and X-shooter spectra (taken at days 408, 572 and 1698, respectively) and the pre-transient SDSS spectrum. A number of high ionization coronal narrow lines compatible with some iron transitions (i.e. ([\ion{Fe}{XIV}] $\lambda$5303, [\ion{Fe}{VII}] $\lambda\lambda$5720,6087, [\ion{Fe}{X}] $\lambda$6374) are well visible in the Gemini and X-shooter spectra, with the exception of the [\ion{Fe}{XIV}] $\lambda$5303, which is not detected anymore in the X-shooter spectrum. We note that the derived X-ray temperature of $\sim$70 eV, imply that these lines arise from photoionization, rather than from collisional ionization equilibrium \cite[][]{lundqvist22,arnaud92}.
As most of these coronal lines are first detected 218 days from the transient's discovery and are not detected at the resolution of the pre-transient SDSS spectrum, we conclude that these features have a transient nature. These findings are remarkable as, together with the TDE AT\,2019qiz \cite[][]{nicholl20}, they represent the first time in which high ionization coronal emission lines are detected in the optical spectra of a TDE soon after the occurrence of the X-ray flare (see Short et al., in prep., for an accurate study on the detection of the high ionization coronal emission lines in the optical spectra of AT\,2019qiz). This strongly suggests a close connection between the TDE X-ray outburst and the presence of high ionization coronal emission lines in the TDE late-time spectra and host galaxy. 

\subsection{The NIR spectra}
In the lower panel of Figure \ref{fig:wise} we show the NIR spectra taken with SofI at two different epochs from the transient's discovery, specifically at 374.7 days (in black) and 1342 days (in orange) in comparison with a part of the NIR X-shooter spectrum taken after 1698 days (in red). 
Although our NIR data-set is composed of only three spectra, a spectral time evolution is visible also in this wavelength range. Indeed, the first SofI spectrum shows a prominent and composite emission line in the \ion{He}{I} $\lambda$10830 region, with a very broad component and a narrower one (FWHM=7660$\pm$1500 \kms\/ and FWHM=1639$\pm$330 \kms\/, respectively), blended with the Pa$\upgamma$ in which we also detect an intense narrow emission line consistent with the high ionization coronal line [\ion{Fe}{XIII}] $\lambda$10798 (see the upper right panel of Figure \ref{fig:wise}). This finding is particularly notably as it represents the first detection of a transient high-ionization coronal NIR line in a TDE. Indeed, such a broad and complex feature becomes less prominent in the second SofI spectrum, where the broad component becomes fainter, the Pa$\upgamma$ starts to emerge and the narrow [\ion{Fe}{XIII}] $\lambda$10798 is not detected anymore. Finally, this broad feature has completely disappeared in the X-shooter spectrum, the \ion{He}{I} $\lambda$10830 and the Pa$\upgamma$ are well detected and show only narrow components. Also in this case, the narrow [\ion{Fe} {XIII}] $\lambda$10798 is not detected.

\section{The subtraction of the host galaxy contribution}
In order to isolate the nuclear transient spectral features, we have first subtracted the host galaxy contribution from each spectrum of our data-set by using the penalized pixel fitting (\texttt{ppxf}) method  \cite[][]{cappellari04,cappellari17}. In this way it has been possible to model the host galaxy spectral features present in the AT\,2017gge spectra through convolution with a host-galaxy template spectrum. Subsequently, the best fit host-galaxy model is subtracted from the observed transient+host spectrum, resulting in the final transient-only spectrum. During the fitting procedure, we excluded all the spectral regions where intense emission lines were present (with particular care for the \ha\/ and \hb\/ regions) and the areas affected by telluric absorption \cite[see also][for an application of this method]{onori19}. 

\begin{figure}
\includegraphics[width=1\columnwidth, angle=0]{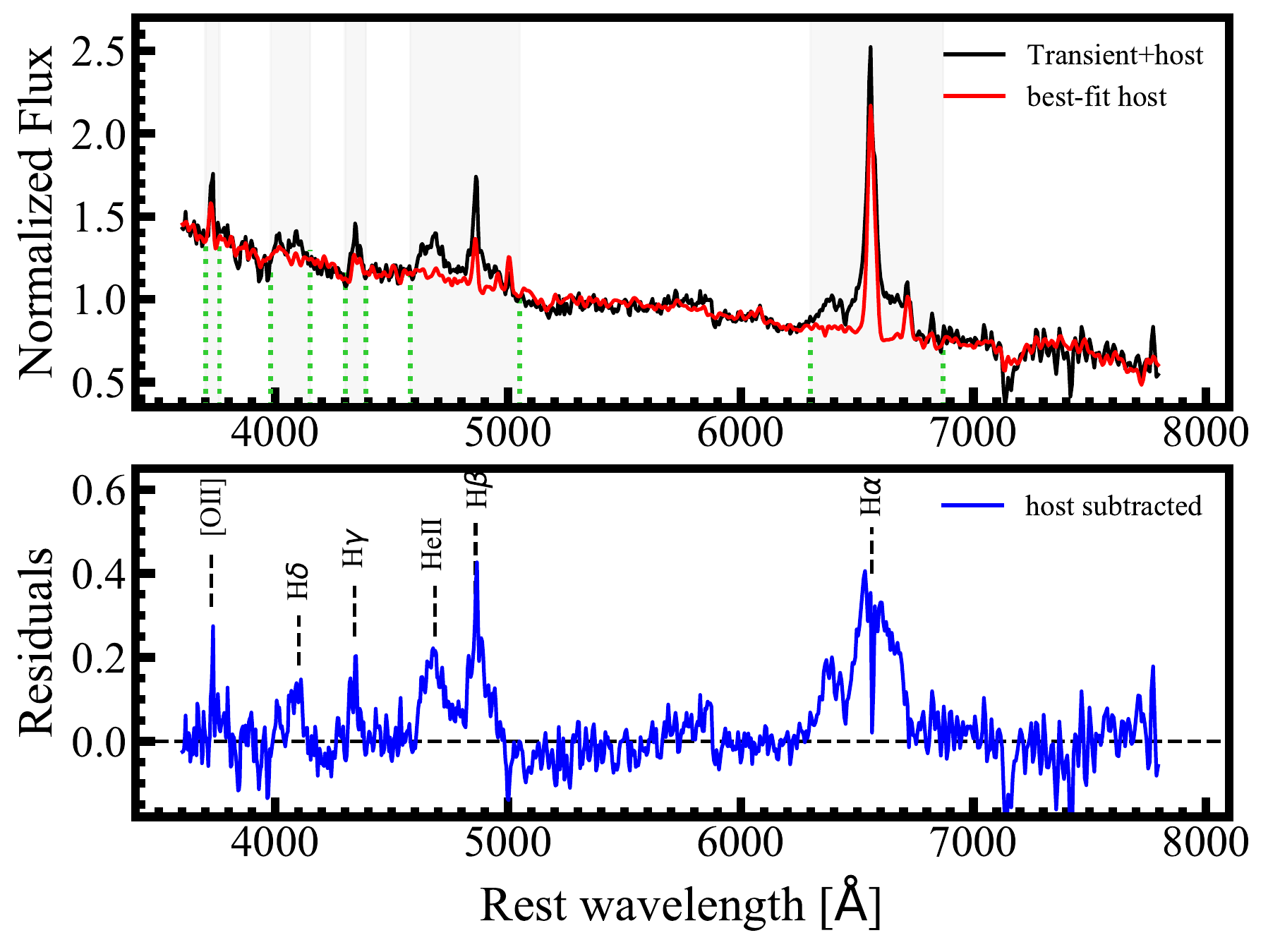}
\caption{Subtraction of the host galaxy contribution from the AT\,2017gge EFOSC2/NTT Gr$\#$13 spectrum, taken on 2018 March 24. In the upper panel, the best-fit host galaxy (in red) is plotted over the transient+host spectrum (in black). Gray areas delimited by the dotted green vertical lines show the spectral regions excluded during the fitting procedure. The resulting host-subtracted spectrum is shown in the lower panel (in blue), together with the identification of the main emission lines.}
\label{fig:ppxf}
\end{figure}

Based on our photometric analysis, the transient's emission can be considered exhausted (or at least faded beyond the detection in the optical bands) after $\sim$500 days from its discovery. Thus we used the spectra taken after this phase as the host galaxy templates in our host-subtraction procedure. In particular, in order to have a host galaxy spectrum with the same observational set-up as the one used for the transient+host spectra taken during our follow-up campaign, we have collected late--time spectra with different instruments: ALFOSC/NOT (Gr$\#$4, at day 557), DOLORES/TNG (LR-B, at day 584) and EFOSC2/NTT (Gr$\#$13 and $\#$18 at day 1348). For each spectrum of our data-set, we then have applied the \texttt{ppxf} method by using as host-galaxy template the aforementioned late-time spectra. 
Due to the lack of late--time spectra obtained with ACAM/WHT, ALFOSC/NOT in the Gr$\#$19 configuration and OSIRIS/GTC, we have used as host galaxy template the late--time spectra taken with DOLORES/TNG, EFOSC2/NTT in the Gr$\#$18 and in the Gr$\#$13 configuration, respectively. In Figure \ref{fig:ppxf} we show the application of this host subtraction method in the case of the AT\,2017gge spectrum taken with EFOSC2/NTT (Gr$\#$13 grism) on 2018 March 25 (at 234 days from the transient's discovery), where we used as host galaxy template the spectrum taken with the same observational set-up (EFOSC2/NTT, Gr$\#$13 grism) on 2021 April 12 (1348 days). In Figure \ref{fig:host_sub_spec}, the full sequence of the host-subtracted AT\,2017gge spectra is shown.

\section{The TDE broad emission lines}
\label{sec:broad_lines}
In this section we describe the method used for the emission line fitting and the results from the spectral analysis of the AT\,2017gge host-subtracted spectra. 

\subsection{Fitting of the emission lines}
\label{subsec:gaussian_fit}
\begin{figure}
\includegraphics[width=0.49\columnwidth, angle=0]{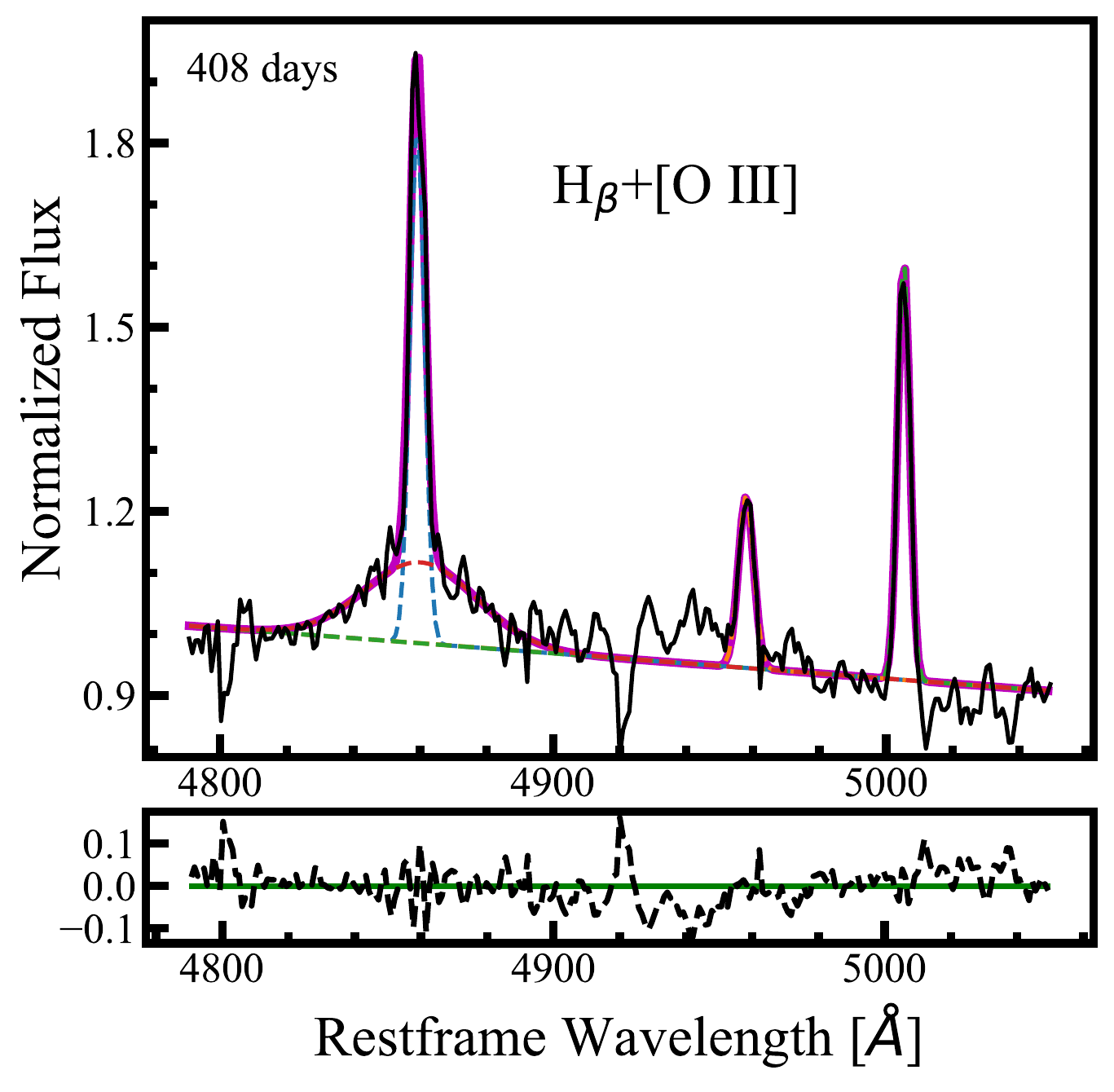}
\includegraphics[width=0.49\columnwidth, angle=0]{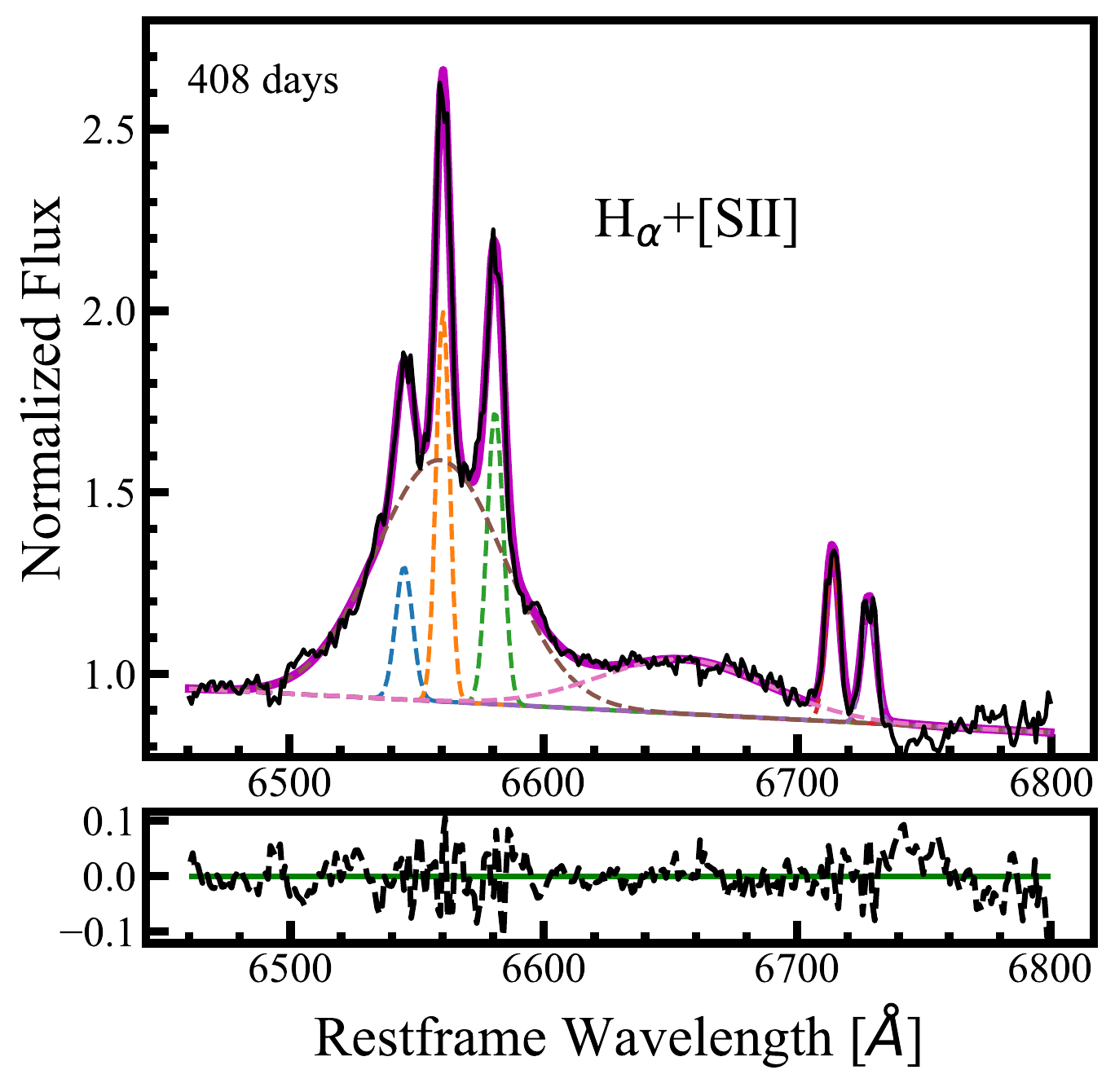}
\includegraphics[width=0.49\columnwidth, angle=0]{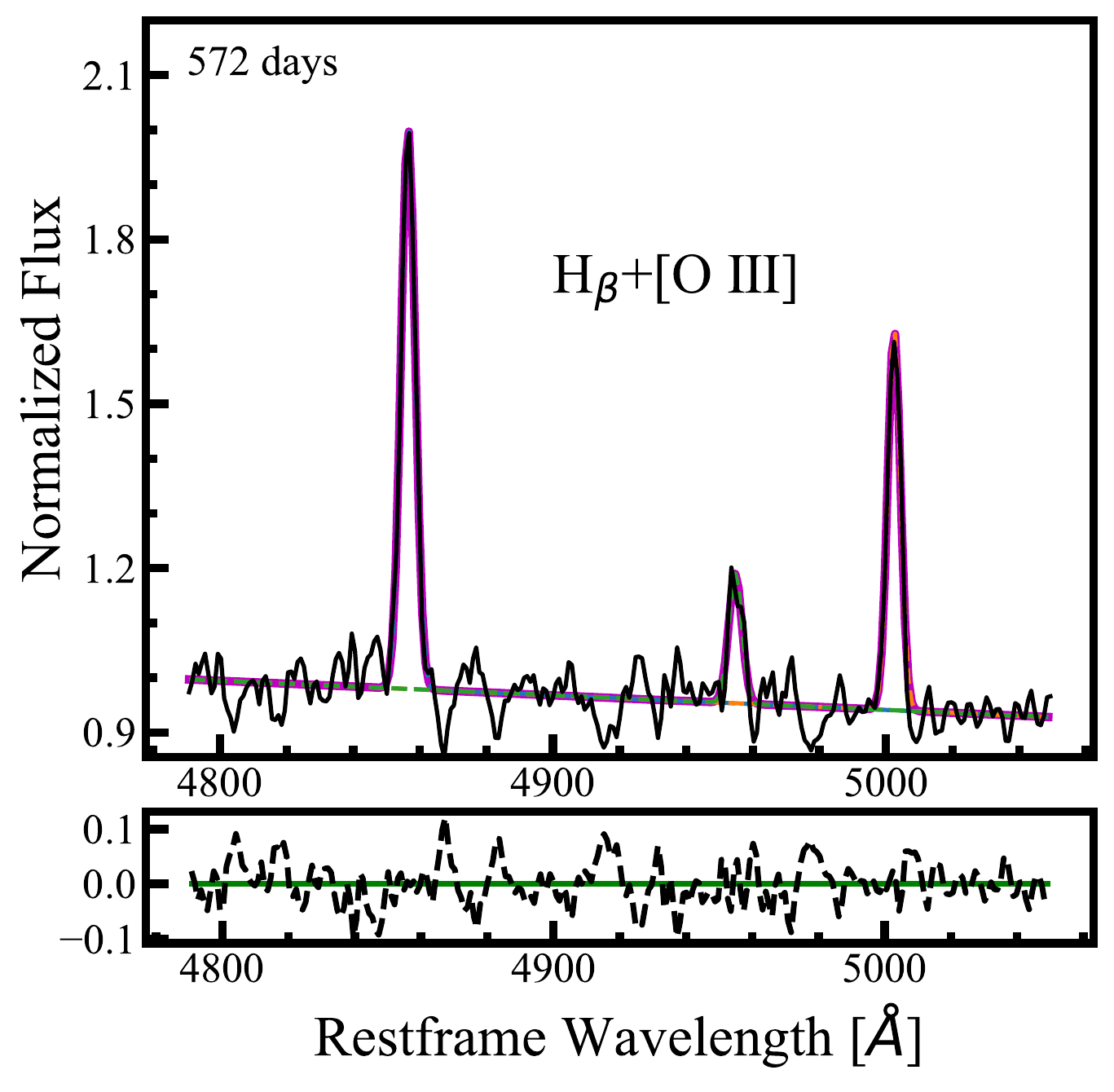}
\includegraphics[width=0.49\columnwidth, angle=0]{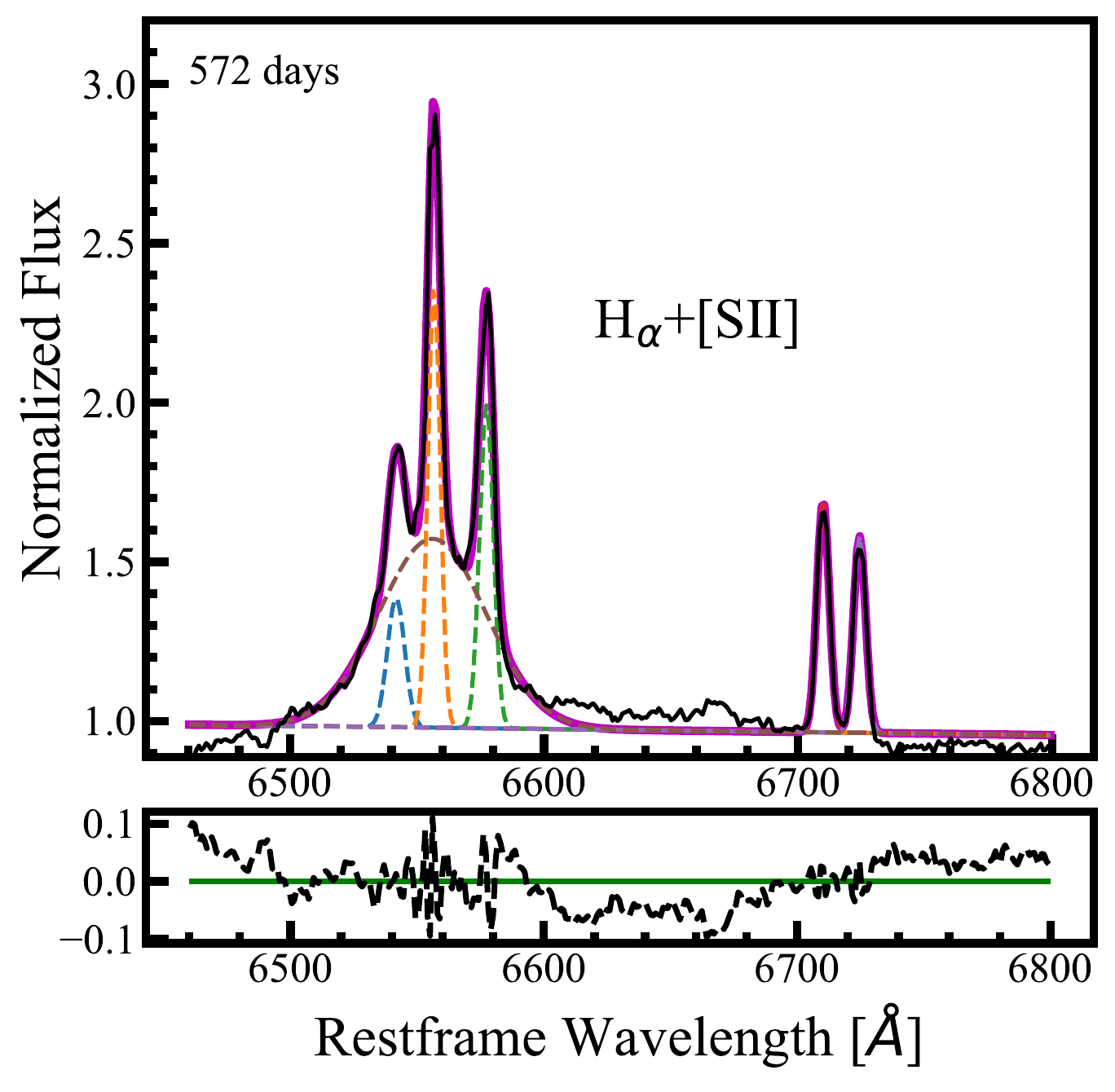}
\includegraphics[width=0.49\columnwidth, angle=0]{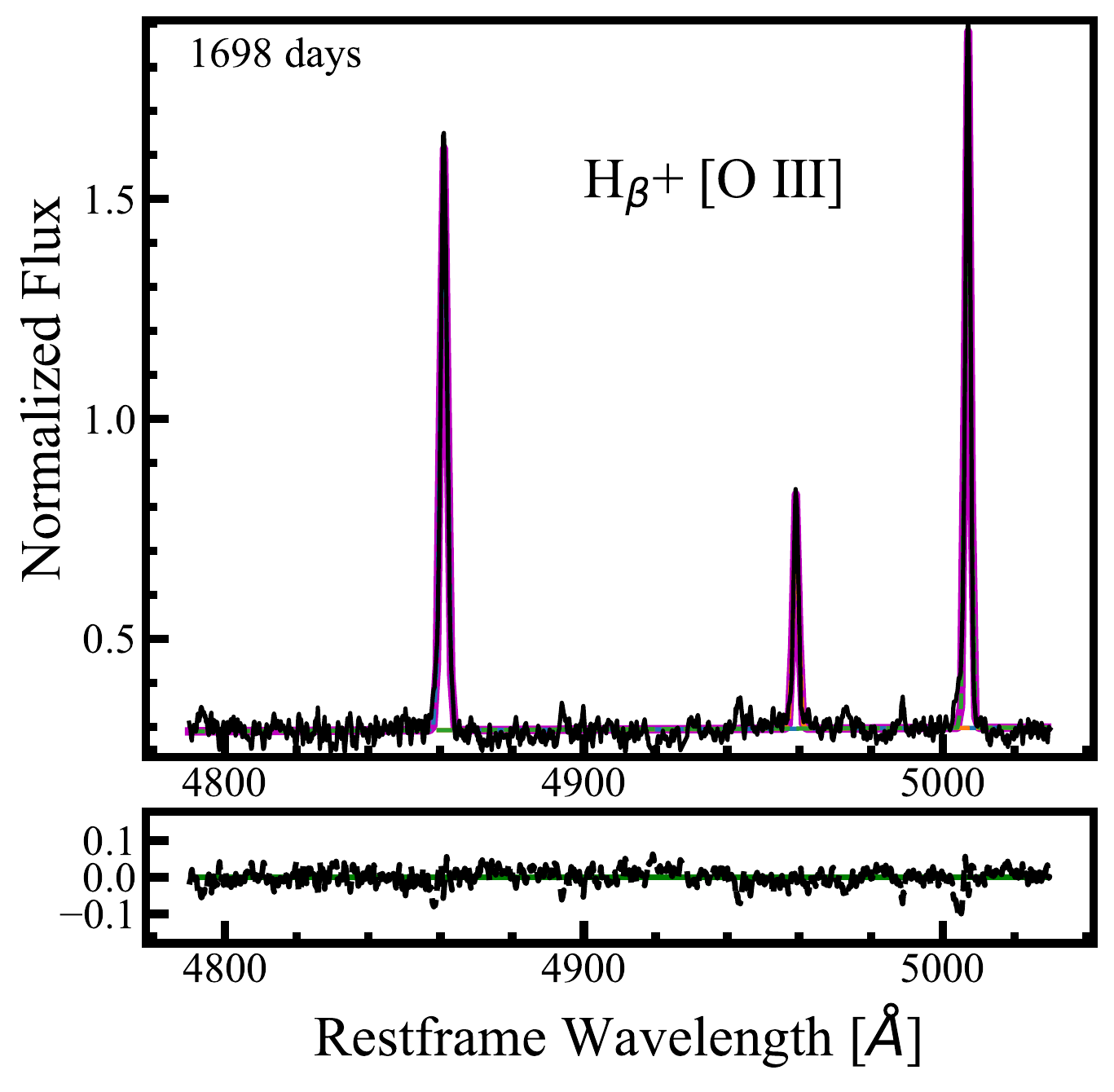}
\includegraphics[width=0.49\columnwidth, angle=0]{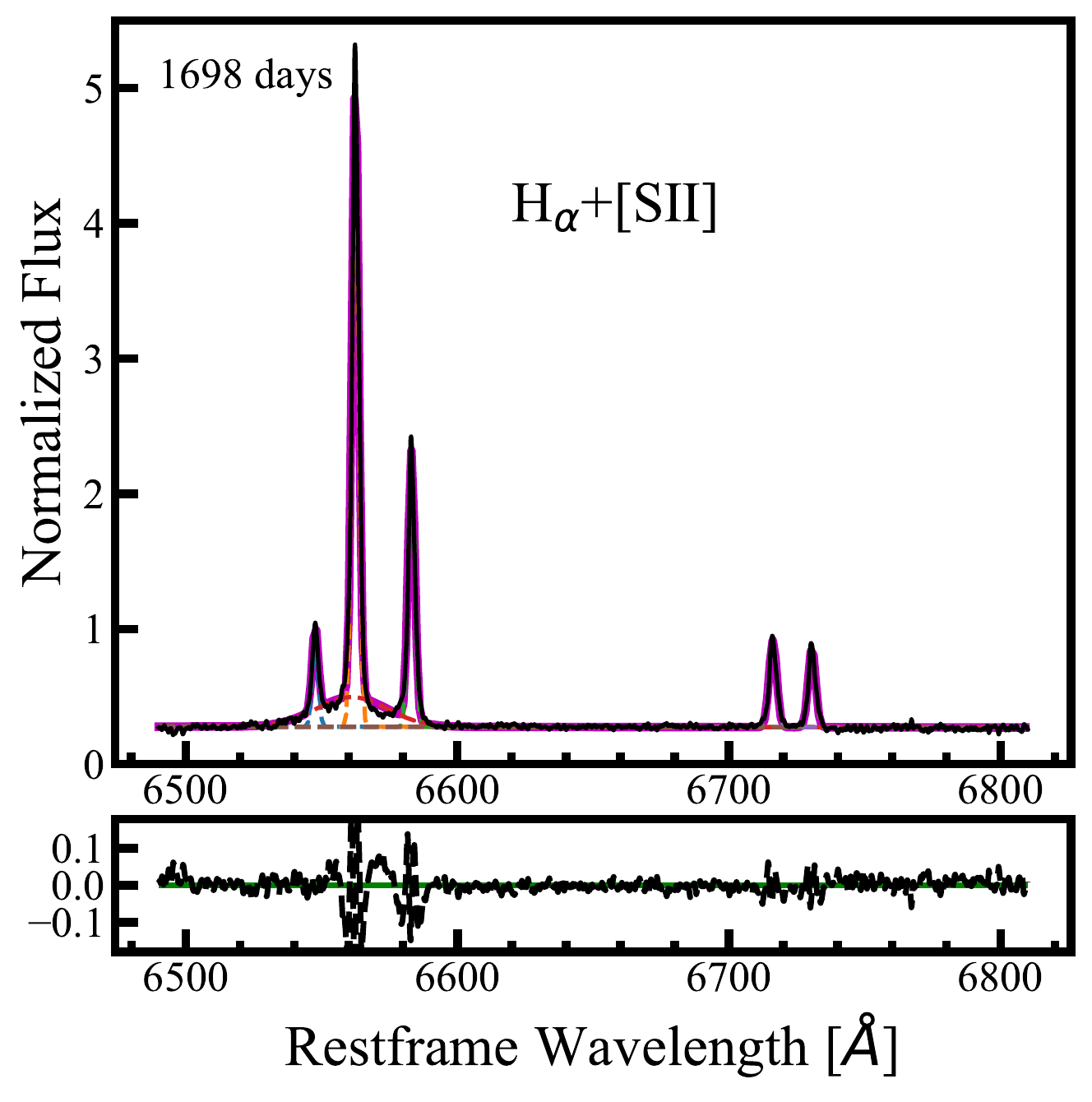}
\caption{Multi-components Gaussian fit of the \hb\/(left panel) and \ha\/ (right panel) emission lines in the AT\,2017gge Gemini spectra obtained 408 and 572 days from the transient's discovery (upper and central panels, respectively) and in the X-Shooter spectra taken at 1698 days. Each single Gaussian component is shown with dashed colored lines, while the total model is shown with a solid magenta line. Residuals with respect to the fitting models are shown at the bottom of each panel. A broad component in the \hb\/ is detected only in the Gemini spectrum of day 408, while a broad \ha\/ component is clearly visible in all the three spectra.}
\label{fig:geminifit}
\end{figure}

In order to investigate the properties and evolution of the main spectral features observed in the host-subtracted spectra of AT\,2017gge (see Figure \ref{fig:host_sub_spec} for the entire sequence), we modelled them with Gaussian functions by using the \texttt{python} packages \texttt{curvefit} and \texttt{leastsq}. In the case of simultaneous presence of broad and narrow components in the line profiles, a multi-component Gaussian fit has been applied. During the fitting procedure, a wavelength window of $\sim$1100 \AA\/ has been selected, in order to include both the spectral features of interest and the local continuum. A narrower fitting window has been selected in late time spectra, when the width of the broad emission lines became smaller.
The same fitting procedure has been applied also in the case of not host-subtracted spectra, such as the SDSS, SofI, Gemini and X-shooter spectra. In Figure \ref{fig:geminifit} the application of the multi-component Gaussian fit for the \hb\/ and \ha\/ region in the Gemini and X-shooter spectra is shown, while in the upper right panel of Figure \ref{fig:wise} it is shown for the \ion{He}{I} $\lambda$10830 region of the SofI spectrum of day 374.   

\subsection{The time evolution of the optical broad emission lines}

We analyzed the evolution of the emission line properties, such as the full width at half maximum (FWHM), the equivalent width (EW) and the EW ratio of the main broad emission lines detected in the host-subtracted spectra of AT\,2017gge. Specifically, we focus on the properties of the \ion{He}{II} $\lambda$4686, H$\upbeta$, \ion{He}{I} $\lambda$5876 and H$\upalpha$ emission lines, as derived from the fitting procedure described in section \ref{subsec:gaussian_fit}. In Figure \ref{fig:line_evolution} the results from this analysis are shown. Broad components in the H$\upbeta$, \ion{He}{I} $\lambda$5876 and \ha\/ are detected starting from the first spectra of our data-set, taken 41.7 and 44 days from the transient's discovery, and they last until phase 291 (for \ion{He}{I}), 408 (for H$\upbeta$)  and 572 (for H$\upalpha$). As shown in the upper panels of Figure \ref{fig:line_evolution}, all these three features have high values of the FWHM in the initial phases of the transient emission with FWHM$\sim$(1.3-2.0)$\times$10$^{4}$ \kms\/, which are consistent with the broad emission lines widths typically observed in TDEs \cite[][]{arcavi14, vanVelzen20, charalampopoulos22}. In this picture, the \ion{He}{II} $\lambda$4686 broad emission line is an exception: it is characterized by a late-time development, a FWHM one order of magnitude smaller (FWHM$\sim$10$^{3}$ km s$^{-1}$) than what measured for the H and \ion{He}{I} $\lambda$5876 lines and it shows a completely different time evolution.

\subsubsection{The FWHM time evolution}
In the upper panels of Figure  \ref{fig:line_evolution} we show the FWHM time evolution for the \ha, \hb, \ion{He}{I} $\lambda$5876 and \ion{He}{II} $\lambda$4686 broad emission lines. The evolution of \ha\/ and \ion{He}{I} is quite similar, with high values measured over longer times compared to what is observed in H$\upbeta$, which instead shows a more rapid decline. However, although the width of the \ion{He}{I} is kept at an almost constant value of FWHM$\sim$1.3$\times$10$^{4}$ \kms\/ for 250 days, its decline became quite fast between 250 and 291.8 days, where this feature rapidly changes its width from FWHM=(1.0$\pm$0.3)$\times$10$^{4}$ \kms\/ to FWHM=(0.20$\pm$0.06)$\times$10$^{4}$ \kms\/ and completely disappears from the subsequent spectra. Instead, the \ha\/ shows a steady decline from the initial value of FWHM=(1.8$\pm$0.1)$\times$10$^{4}$ \kms\/, as measured at phase 41.7 days, to a FWHM=(0.23$\pm$0.01)$\times$10$^{4}$ \kms\/ as derived at 572.2 days. It is worth noting that a broad \ha\/ component of FWHM=(0.12$\pm$0.01)$\times$10$^{4}$ \kms\/ is still detected in the X-shooter spectrum (at day 1698), while any other broad feature is not present anymore. 

The \hb\/ evolution is quite fast, with a rapid change of the line width in $\sim$150 days, from a value similar to what is observed in the H$\upalpha$, FWHM=(2.0$\pm$0.2)$\times$10$^{4}$ \kms\/ at 41.7 days, to a narrower broad component of FWHM=(0.56$\pm$0.06)$\times$10$^{4}$ \kms\/ at 193.9 days. Finally, a broad \ion{He}{II} $\lambda$4686 is first detected only after $\sim$193.9 days from the transient's discovery, it is characterized by smaller value of FWHM with respect to H$\upbeta$, \ion{He}{I} and H$\upalpha$, with a nearly constant value at FWHM$\sim$0.6$\times$10$^{4}$ \kms\/ until its last detection at 338.6 days, in which a FWHM=(0.56$\pm$0.05)$\times$10$^{4}$ \kms\/ is measured. Later time spectra show only a narrow \ion{He}{II} emission line. 

\subsubsection{The EW time evolution}
The middle panels of Figure \ref{fig:line_evolution} show the EW time evolution of the \ion{He}{II} $\lambda$4686, \hb, \ion{He}{I} $\lambda$5876 and \ha\/ broad components. As for the FWHM, the EW of these features follows a declining trend, with the \ha\/ characterized by the higher values (from an initial EW$\sim$120 \AA\/ to EW$\sim$30 \AA\/ as measured at 572 days) with respect to the other lines during the whole monitoring campaign. 
Indeed, the \hb\/ shows a trend similar to what is observed in the \ha, but at EW values one order of magnitude lower, while the \ion{He}{I} is characterized by an initial growing phase until 193.9 days, followed by a very rapid declining trend which ends at 291.8 days. Also in this case the EW values are one order of magnitude lower than what measured for the \ha\/. The behaviour of the \ion{He}{II} is instead completely different from the aforementioned lines. In fact, it is characterized by a late-time development and by a rapid variation of the EW, which oscillates between different values (from EW$\sim$8 \AA\/  to EW$\sim$30 \AA\/) within a time interval of $\sim$20 days.   

\subsubsection{The broad line ratios}
\begin{figure*}
\includegraphics[scale=0.20, angle=0]{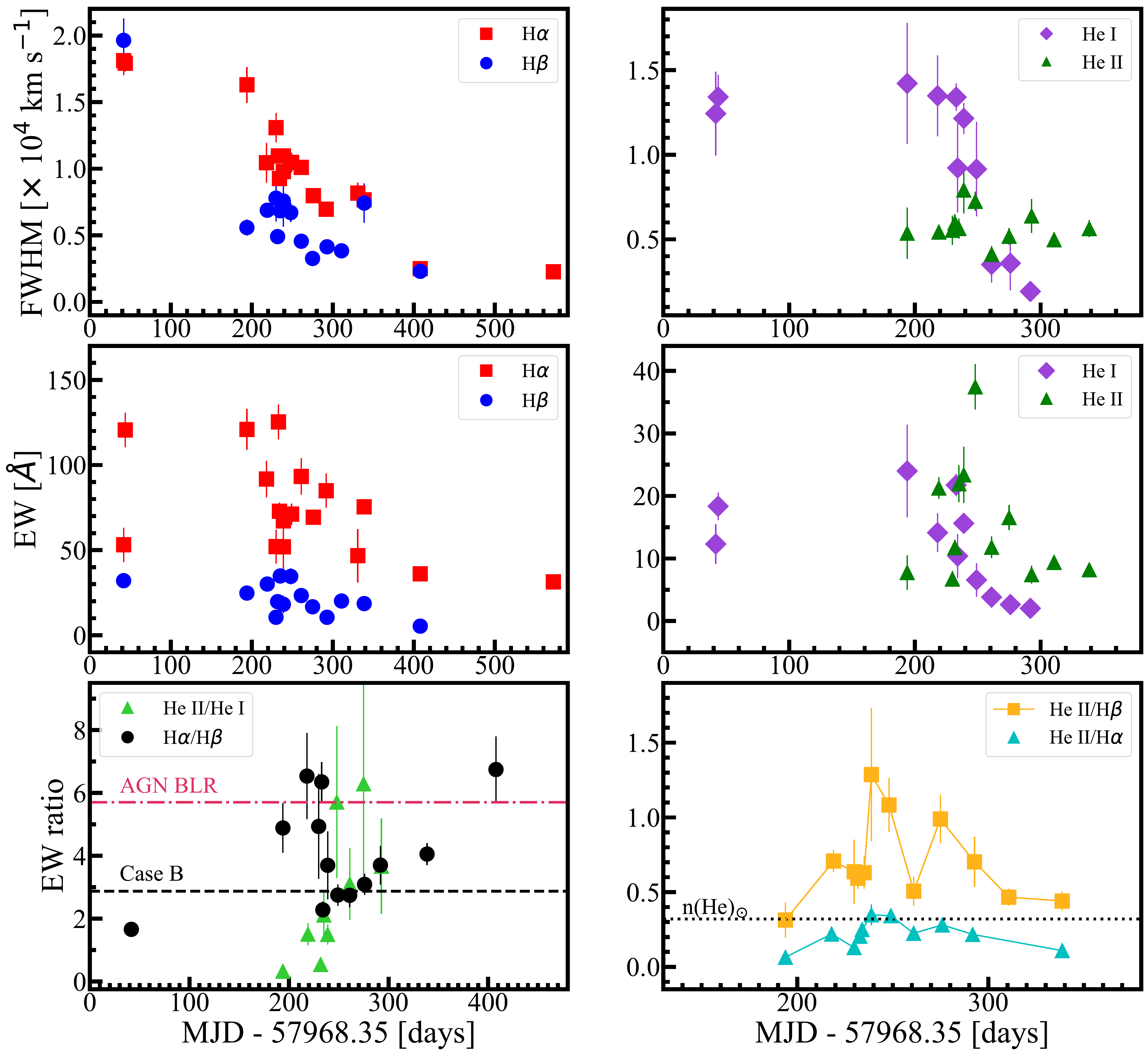}
\caption{{\it Upper:} Time evolution of the FWHM for the broad components in the \ha\/, \hb\/(left panel, red squares and blue circles, respectively), \ion{He}{II} $\lambda$4686 and \ion{He}{I} $\lambda$5876 (right panel, green triangles and purple diamonds, respectively). {\it Middle:} Time evolution of the the broad components EW for \ha\/, \hb\/(left panel, red squares and blue circles, respectively), \ion{He}{II} $\lambda$4686 and \ion{He}{I} $\lambda$5876 (right panel, green triangles and purple diamonds, respectively). {\it Bottom}: Time evolution of the H$\upalpha$/H$\upbeta$ and \ion{He}{II}/\ion{He}{I} EW ratios (left panel, black filled circles and green filled triangles, respectively) and of the \ion{He}{II}/H$\upbeta$ and \ion{He}{II}/H$\upalpha$ (right panel, yellow squares and cyan triangles, respectively). The location of the typical value for the H$\upalpha$/H$\upbeta$ ratio in the Case B recombination ($T=10^{4}$ K and $n_e$=10$^{4}$ cm$^{-3}$, \citealt{osterbrock06}) is shown by the black dashed line, while its value in homogeneous AGN BLR models (\citealt{osterbrock06}) is shown by the red dotted-dashed line. The black dotted line in the bottom right panel shows the expected He-to-H ratio in a nebular environment in the case of solar abundance \citealt{hung17}.}
\label{fig:line_evolution}
\end{figure*}

In the bottom left panel of Figure \ref{fig:line_evolution} we show the time evolution for the H$\upalpha$/H$\upbeta$ and of the \ion{He}{II}/\ion{He}{I} line ratios. In particular, the H$\upalpha$/H$\upbeta$  line ratio is consistent with the value of 2.86, expected for the theoretical Case B recombination \cite[black dashed line][]{osterbrock06} but with two remarkable exceptions. Indeed, the H$\upalpha$/H$\upbeta$ line ratio is greatly enhanced around day 200, when it changes from a value of 1.7$\pm$0.2 at 41.7 days to the values of 4.9$\pm$0.8 and 6.5$\pm$1.4 at 193.8 and 218 days, respectively. After $\sim$10 days, it comes back to values in agreement with the Case B recombination.
This consistency with the Case B recombination value lasts for $\sim$80 days, when the line ratio gradually increases again toward the value of 6.7$\pm$1.1 at 407.9 days. It is interesting to note that the high H$\upalpha$/H$\upbeta$ line ratio values detected at days 193.8, 218 and 407.9 are all consistent with the line ratio value expected if the emission is dominated by an AGN broad line region \cite[BLR, red dotted-dashed line,][]{osterbrock06}. This variation in the H$\upalpha$/H$\upbeta$ is mainly due to the broad \hb\/ declining more rapidly than the broad \ha, which started around 194 days.

The \ion{He}{II}/\ion{He}{I} line ratios, instead is characterized by a steep and rapid rising trend, from an initial values of 0.32$\pm$0.15, as measured at 193.9 days, to a final value of 3.67$\pm$1.52 as measured at 292.8 days. This behaviour has been already observed in the recent spectroscopic study on a sample of 16 TDEs performed by \cite{charalampopoulos22}. In their paper, the authors ascribe the observed temporal rising of the \ion{He}{II}/\ion{He}{I} line ratio with a photoionizazion increase in the Helium line emitting region. In AT\,2017gge, the development of the \ion{He}{II} soon after the X-ray flare and the simultaneous disappearance of the \ion{He}{I} suggests that they are emitted from the same region which is being further ionised probably as a result of the delayed X-rays. However, we stressed that although our results can be explain within this scenario, they are not strong enough to extend such a delayed X-ray photoionization of He II broad region model to the general TDE behaviour. Indeed, we outline that other events characterized by a multiwavelength behavior similar to that of AT\,2017gge, have shown broad \ion{He}{II} emission line in the early-time spectra, well before the X-ray detection \cite[i.e. ASASSN-15oi and AT2019azh][]{gezari17,vanVelzen21,hinkle21}

In the bottom right panel of Figure \ref{fig:line_evolution} the time evolution of the \ion{He}{II}/H$\upbeta$ and \ion{He}{II}/H$\upalpha$ line ratios in comparison with the value expected in the case of a nebular environment for a solar helium abundance \cite[black dotted line,][]{hung17} are shown. The \ion{He}{II}-to-\ha\/ line ratio is in agreement with the nebular argument for all the duration of the monitoring campaign. This behaviour has been also shown to hold for many others TDEs \cite[see Figure 9 of][]{charalampopoulos22}. 


\section{The host galaxy activity status and the black hole mass}
In Figure \ref{fig:sdss_gemini}, we show the late time high resolution spectra of AT\,2017gge taken with Gemini (after 408 and 572 days from the transient's discovery) and X-shooter (taken at 1698 days) in comparison with the pre-transient SDSS spectrum. It is clearly visible how the host galaxy spectrum has been modified after the nuclear transient took place. Indeed, in the SDSS spectrum only narrow \ha\/ and \hb\/ emission lines are present, with no signs of broad components or outflows, and the [\ion{O}{III}] $\lambda$5007 is barely detected. In contrast, the Gemini spectra are characterized by \hb\/ and \ha\/ line profiles similar to those usually observed in AGN \cite[][]{osterbrock06}, with broad components clearly visible in both lines in the spectrum taken at day 408, and only the broad component in the \ha\/ detected at 572 days. In the X-shooter spectrum, only a faint broad \ha\/ emission line is detected, together with a clear enhancement of the [\ion{O}{III}] $\lambda$5007 emission line. Furthermore, as already discussed in Section \ref{sec:spectroscopy}, a narrow \ion{He}{II} $\lambda$4686 and many high ionization emission lines compatible with iron transitions are present only in the Gemini and X-shooter spectra, but are not detected in the pre-transient SDSS spectrum.

\subsection{The host galaxy activity status}
\label{subsec:BPT}
We investigated on the activity status of the AT\,2017gge host galaxy by using the BPT diagrams \citep{baldwin81} and the EW of the narrow \ha, \hb, [\ion{O}{III}] $\lambda$5007, [\ion{N}{II}] $\lambda$6583 and [\ion{S}{II}] $\lambda \lambda$ 6716,6731 lines detected in these spectra. The lines profiles have been modelled following the method described in Section \ref{subsec:gaussian_fit}. 

The line ratios derived from the SDSS spectrum are 
log([\ion{N}{II}]/H$\upalpha$)=$-$0.27$\pm$0.0), log([\ion{O}{III}]/H$\upbeta$)=$-$0.40$\pm$0.33 
and log([\ion{S}{II}]/H$\upalpha$)=$-$0.45$\pm$0.10.
Despite the faint [\ion{O}{III}] $\lambda$5007 emission line, these values place the galaxy on the boundary between the HII and composite regions in the BPT diagram shown in the left panel of Figure \ref{fig:BPT} and well inside the star--forming region in the BPT diagram shown in the right panel of Figure \ref{fig:BPT}. We note that the location on the extreme starburst line may indicate that an additional ionising component is required to explain the line ratios. 
The line ratios obtained from the Gemini spectrum of day 408 are  
log([\ion{N}{II}]/H$\upalpha$)=$-$0.04$\pm$0.08, 
log([\ion{O}{III}]/H$\upbeta$)=$-$0.08$\pm$0.11 
and log([\ion{S}{II}]/H$\upalpha$)=$-$0.09$\pm$0.12. While the line ratios obtained from the Gemini spectrum of day 572 are
log([\ion{N}{II}]/H$\upalpha$)=$-$0.03$\pm$0.06, 
log([\ion{O}{III}]/H$\upbeta$)=$-$0.2$\pm$ 0.10 
and log([\ion{S}{II}]/H$\upalpha$)=0.01$\pm$ 0.07.
In this case, the derived line ratios values are consistent between each other and place the galaxy well inside the LINER region and on the boundary between the composite and the AGN regions (cyan square and magenta triangle in Figure \ref{fig:BPT}), indicating an increase in the activity of the galaxy nucleus after the nuclear transient took place. 
Finally, the line ratios obtained from the X-shooter spectrum (taken at 1698 days) are  
log([\ion{N}{II}]/H$\upalpha$)=$-$0.36$\pm$0.03, 
log([\ion{O}{III}]/H$\upbeta$)=$-$0.08$\pm$0.03, 
and log([\ion{S}{II}]/H$\upalpha$)=$-$0.54$\pm$0.04
and place the galaxy inside the composite and the starforming regions in the BPT diagrams of Figure \ref{fig:BPT} (green star in the left and right panels, respectively), indicating a further variation in the host galaxy activity, which come back to values in agreement with what derived from the pre-transient spectrum of SDSS. 
A similar variation in the activity status of a TDE host galaxy has been observed also in AT\,2019qiz (Short et al, in prep.). 

The discovery of AT\,2017gge in a starforming galaxy make this TDE particularly interesting as it probes a host galaxy population (the blue galaxies) poorly represented in most optically and X-ray selected TDE host galaxies samples which result to be dominated by green valley galaxies \cite[][]{hammerstein21,sazonov21}.

\subsection{The SMBH mass}
\begin{figure}
\includegraphics[width=1\columnwidth, angle=0]{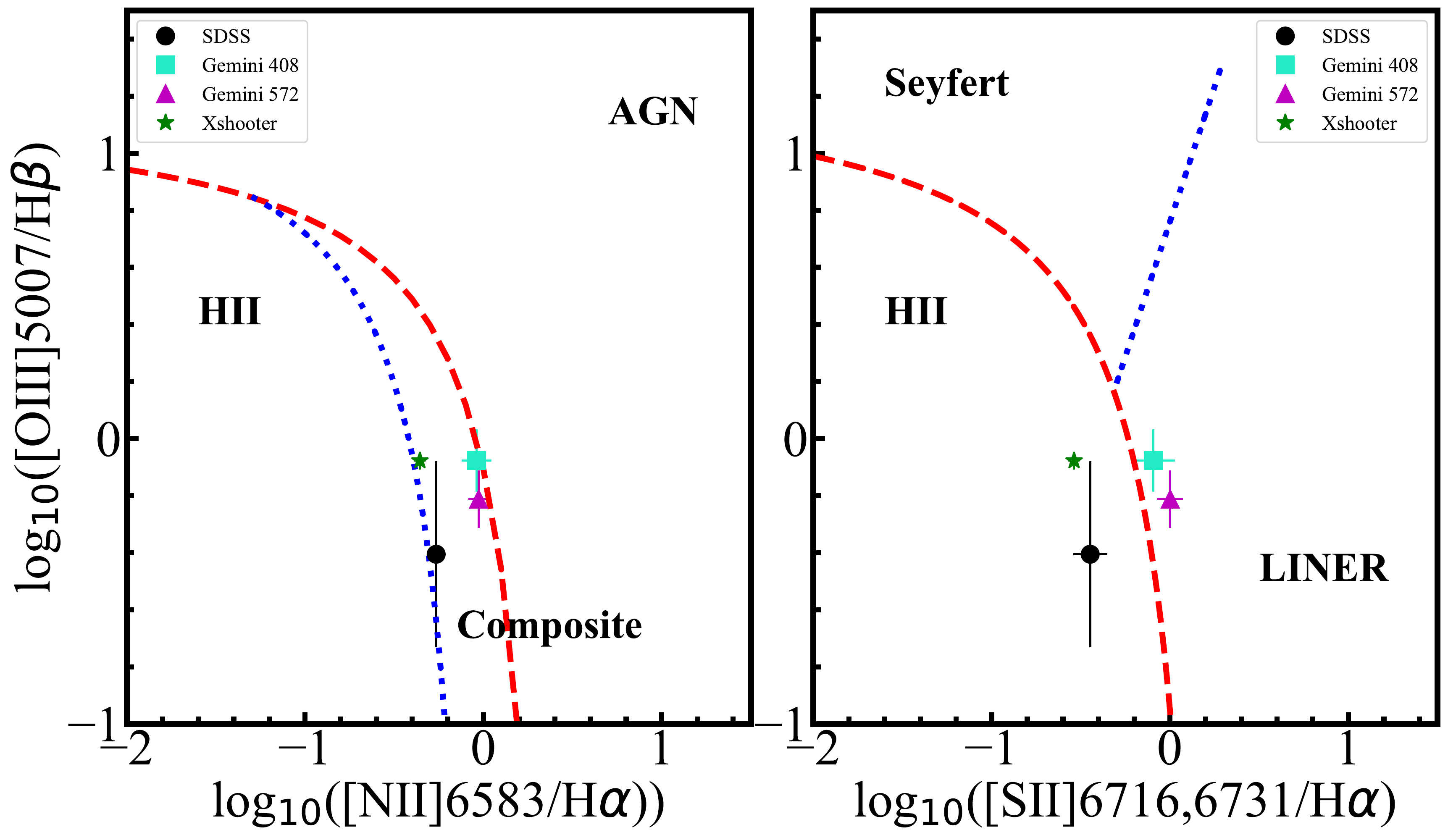}
\caption{BPT diagrams for the host galaxy of AT\,2017gge derived from the equivalent widths of the narrow emission lines detected in the pre-transient SDSS spectrum (black filled point), in the Gemini spectra taken 408 and 572 days from the transient's discovery (cyan filled square and magenta filled triangle, respectively) and in the X-shooter spectrum taken after 1698 days from the transient's discovery. The different activity regions of the diagram are separated by the following lines: red dashed line from \citep{kewley01}, blue dotted line from \citep{kauffmann03} in the left panel and blue dotted line from \citep{kewley06} in the right panel.
}
\label{fig:BPT}
\end{figure}

The detection of AGN-like features in the Gemini optical spectra, their location in the BPT diagnostic diagrams (see Figure \ref{fig:BPT}) and the evolution of the H$\upalpha$/H$\upbeta$ line ratio, suggest an AGN-like behaviour of the AT\,2017gge host galaxy after the TDE occurrence. This led us to use these two spectra and a single epoch (SE) relation to test a different and independent method for the SMBH mass derivation, beside the commonly used $M-\sigma_{\star}$ scaling relations. Recently, \cite[][]{ricci17} derived a SE relation based on the AGN X-ray luminosity in the 2-10 keV band and on the width of \hb\/ and/or \ha\/ broad components. If the broad H emission lines detected in the Gemini spectra are emitted in a virialized photosphere (similar to an AGN-like BLR), then we can use this SE relation to estimate the mass of the black hole. 

To this scope we first have derived the unabsorbed X-ray luminosity in the 2-10 keV band from the {\it Swift}/XRT spectral analysis of $L_{\rm X}$=(6$\pm$3)$\times$10$^{39}$ \unitlum\/. Then we have fitted the H broad emission lines detected in the Gemini spectra. In particular, the broad components detected in the \hb\/ and \ha\/ emission lines of the Gemini spectrum taken at 408 days results to be centered at $\lambda_{c}$ = 4857.30$\pm$1.40 \AA\/ and $\lambda_{c}$= 6559.10$\pm$0.62 \AA\/ and are characterized by a FWHM=2318$\pm$268 \kms\/ and  FWHM=2760$\pm$69 \kms\/, for the \hb\/ and the \ha\/ respectively. In the Gemini spectrum of 572 days only the \ha\/ broad component is detected, which results centered at $\lambda_{c}$=6556.08 $\pm$ 0.49 \AA\/ and with a FWHM=2274$\pm$38 \kms. 

By using the virial estimator presented in Table 4 of \cite[][]{ricci17} with the zero points constants valid for all the classes of host galaxies we derived a black hole mass of $\log M_{\rm BH}$=5.5$\pm$0.3 when using the width of the broad \hb, a $\log M_{\rm BH}$=5.8$\pm$0.3 and $\log M_{\rm BH}$=5.4$\pm$0.3 when using the width of the broad \ha\/ measured in the Gemini spectra of 408 days and 572 days, respectively.  
Finally, we derive the BH mass of the host galaxy by using the $M_{\rm BH}-\sigma_{\star}$ scaling relation of \cite{McConnell13} and \cite{gultekin09}, following the method illustrated in \cite{wevers17,wevers19b} for a sample of TDEs. To this purpose, we measured the stellar velocity dispersion of the host galaxy of AT\,2017gge by applying the \texttt{ppxf} fitting procedure to the X-shooter UVB and VIS spectra. We derive a stellar velocity dispersion of $\sigma_{\star}$=(97$\pm$3) \kms, which correspond to a $\log M_{\rm BH}$=6.55$\pm$0.45 or to a $\log M_{\rm BH}$=6.80$\pm$0.43, when using the $M_{\rm BH}-\sigma_{\star}$ scaling relation of \cite{McConnell13} or the relation of \cite{gultekin09}, respectively.
We note that by using the SE relation of \cite{ricci17} we obtain values for the BH mass that are $\sim$ one order of magnitudes lower than what is derived with the $M_{\rm BH}-\sigma_{\star}$ relation. Moreover, given the bolometric luminosity of $L_{bol}$=4.6$\times$10$^{44}$ at peak, the BH masses derived with the SE relation would imply a super Eddington accretion (with an Eddington ratio $\sim$10). Instead, a Eddington limited accretion would requires a BH mass of $\sim$4$\times$10$^{6}$ \Msun\/, which is consistent with the values obtained with the $M_{\rm BH}-\sigma_{\star}$ relations. Thus we favour the BH mass as derived from the $M_{\rm BH}-\sigma_{\star}$ scaling relation for AT\,2017gge and we suggests that the region emitting the broad lines in this TDE is not virialized as in the case of the AGN BLR. 

\section{Discussion}
The nuclear transient AT\,2017gge was first discovered by the ATLAS survey on MJD 57\,968.35 and was alerted as a possible TDE by the ePESSTO spectroscopic classification \cite[][]{fraser17}. A first claim on the TDE nature for this transient was recently presented by \cite{wang22}, who investigated on its properties following the detection of a MIR flare in the WISE lightcurve \cite[][]{jiang21}.
The optical/UV photometric evolution, the properties of the X-ray emission and the dense spectroscopic sequence as presented in this work, not only confirm the TDE nature of AT\,2017gge, but have also allowed us to accurately investigate on the observational properties and their evolution with time in a multiwavelength approach. This led us to the finding of a strong connection between the TDE flare and the appearance of extreme coronal line emission (ECLEs), to the the first detection of a transient high-ionization coronal NIR line in a TDE and to the suggestion a possible scenario for the emission mechanism and the geometry of the emitting region.

\subsection{AT\,2017gge as a gas-rich TDE surrounded by a dusty environment}

The detection of very broad (FWHM $\sim$ 10$^4$ \kms) H emission lines in the optical spectra place AT\,2017gge among the H-rich TDEs subclass \cite[][]{arcavi14, vanVelzen20, charalampopoulos22}. Notably, in correspondence of a delayed X-ray flare, a broad (FWHM $\sim$10$^{3}$ \kms) \ion{He}{II} $\lambda$4686 emerges after 194 days from the transient discovery, indicating a transition toward a TDE H+He subclass. Together with the \ion{He}{II} $\lambda$4686, also a number of high ionization coronal emission lines appear. These spectral features and the IR echo, strongly indicate the presence of a gas-rich and dust-rich environment surrounding the source, consistent with the starforming nature of the host galaxy. 

The MIR reverberation signal is detected with a delay of $\sim$200 days from the optical peak emission and corresponds to a distance for the dust producing the IR echo of $\sim$0.16 pc, as derived in this work by using two cross correlations methods. This value is similar to the sublimation radius obtained for the case of the TDE PTF-09ge \cite[][]{vanvelzen16} and suggests that AT\,2017gge may have sublimated the pre-existing in-situ dust out to this radius.
Our estimation of the the total (optical/UV+IR) radiated energy of 2$\times$10$^{51}$ erg is significantly lower that the 10$^{53}$ erg that has been theoretically predicted for the case of the disruption of a solar mass star, despite the large covering factor of $\sim$0.2 derived by \cite{wang22}. This could be explained by a large proportion of far-UV emission remaining unobserved, as potentially indicated by the large intrinsic luminosity required to produce the delay time we observe in the IR. This can be compared with TDEs in galaxies with a very high covering factor for the SMBH such as the IR-luminous, highly obscured TDE Arp 299-B AT1, where a direct integration over time of the IR SED yielded a total radiated energy of $>$2$\times$10$^{52}$ erg \cite[][]{mattila18, reynolds22}.

We have presented three NIR spectra taken after the IR echo and we report the detection of a broad feature in the \ion{He}{I} $\lambda$10830 and a intense \ion{Fe}{XIII} coronal emission line, both gradually disappearing in the subsequent spectra. This is the first time that a broad feature and a high-ionization coronal line are detected in the NIR spectra of a TDE taken following an IR echo from surrounding dust \cite[but see the case of the candidate TDE AT\,2017gbl in which NIR spectra have been presented,][]{kool20}.

The presence of the luminous and transient high ionization coronal emission lines in the spectra obtained soon after the detection of the soft X-ray emission implies a strong correlation with the soft X-ray emission and a gas-rich environment of the TDE, as expected from the starforming nature of the host galaxy. These high ionization coronal lines are long-lasting as, together with a narrow \ion{He}{II} $\lambda$4686, are still present in the X-shooter medium resolution spectrum taken after 1698 days from the transient discovery. Together with the TDE AT\,2019qiz (Short et al., in prep.), this is the first time that high ionization coronal emission lines are detected in the optical spectra of TDEs following the X-ray outburst. This strongly indicates a close connection between the two phenomena and support the hypothesis that the extreme coronal emission lines detected in the spectra of a sample of non active galaxies could be a signature of the occurrence of a TDE in the past \cite[as suggested by][]{wang12}. 

\subsection{The broad line emitting region}

We study the evolution of the EW, the FWHM and the line ratios of the broad \hb\/, \ion{He}{II} $\lambda$4686, \ha\/ and \ion{He}{I} $\lambda$5876 emission lines detected in the host-subtracted spectra of AT\,2017gge. With the exception of the \ion{He}{II} emission line, all the other broad lines are detected starting from the very first spectra (at day $\sim$40) with FWHM$\sim$10$^{4}$ \kms and show a declining trend with time. In particular, the \hb\/ have a different behaviour with respect to the \ha\/ and \ion{He}{I} $\lambda$5876, being characterized by a faster decline before becoming undetectable after $\sim$400 days from the transient discovery. Instead, the broad \ha\/ is characterized by a slow declining trend and it is long-lived as it is still detected in the X-shooter spectrum, after 1698 days from the transient discovery, when all the other broad lines are no longer detectable.     
Contrary to what observed in the \ha, the broad \ion{He}{I} $\lambda$5876 keeps constant high values of the FWHM for $\sim$250 days and soon after this phase it starts a very rapid decline, which last only $\sim$50 days. Notably, this almost coincides with the appearing of the broad \ion{He}{II} $\lambda$4686 emission line. This feature it is first detected at 194 days, soon after the delayed X-ray flare, and it does not show a particular declining trend. Instead it keeps an almost constant width of FWHM$\sim$6$\times$10$^{3}$ \kms\/ (one order of magnitude smaller than what measured for the broad H and \ion{He}{I}) until its last detection at 338.6 days. 

The observed different behaviour of the various broad emission lines is indicative of a stratified photosphere where different lines are produced at a different distance from the continuum source. Moreover, the time evolution of the various lines ratios can give us more insight in the properties of the region emitting the broad lines. 
In particular, the observed rising trend of the \ion{He}{II}/\ion{He}{I} line ratio can be ascribe to an increase in the photoionization of the broad Helium emitting region. Given that the \ion{He}{II} line is known to be produced by photoionisation due to (soft) X-ray photons, we suggests that the observed delayed X-ray flare could be responsible for this phenomenon.
Furthermore, the observed evolution of the H$\upalpha$/H$\upbeta$ line ratio, which experiences variation from a values consistent with the Case B recombination to values predicted for homogeneous AGN BLR models, is mainly due to the faster decay of the broad \hb\/ with respect to the \ha.

\subsection{A two-process scenario for the AT\,2017gge emission}

The X-ray peak brightening detected with a delay of $\sim$200 days with respect to the UV/optical peak emission, followed by the development of a broad \ion{He}{II} $\lambda$4686, suggests a different origin for the two signals in AT\,2017gge. A delayed X-ray emission is not new in the TDE field. It has been predicted to eventually occur in some models of TDE emission \cite[i.e. reprocessing envelope model, the stream-stream collision scenario and the TDE unified model,][]{guillochon14, piran15, dai18} and it has been already observed in some TDEs \cite[i.e. ASASSN-14li, ASASSN-15oi, AT\,2019azh and OGLE16aaa,][]{pasham17, gezari17,holoien18, liu22, kajava20,shu20}. Recently, \cite{hayasakijonker21} proposed an interesting two emission scenario in order to explain such a time delay between the two signals, observed in a sub-sample of optically selected TDEs. In particular, in this picture, the optical emission is produced by the collision of intercepting stellar debris streams during the initial phase of the stellar disruption, while the X-ray flare is emitted following the formation of an accretion disk, after the end of the circularization process. 

As suggested also by \cite{wang22}, the AT\,2017gge observed emission can be well described by this scenario. Indeed, the results from the optical photometry, the X-ray spectral analysis and even from the optical spectroscopy, are in line with this picture. In fact, while the optical/UV black body radius results ranging between $R_{\rm BB}$=7-13$\times$10$^{14}$ cm, from the X-ray spectral fitting we derive a much smaller value for the black body radius, $R_{\rm BB}\sim$1.4$\times$10$^{11}$ cm, suggesting a different location for the source of the two signals. Moreover, an accretion disk model is able to represent well the X-ray spectrum. 

The non-detection of the \ion{He}{II} in the optical spectra taken prior to the X-ray flare may indicate the absence of the X-ray emission at these epochs, as the \ion{He}{II} line is known to be correlated with the soft X-ray photons \cite[][]{pakull86,schaerer19,cannizzaro21}. However we note that in the case of the TDE ASASSN-15oi and AT\,2019azh a broad \ion{He}{II} feature is detected well before the onset of a delayed soft X-ray flare \cite[][]{gezari17, vanVelzen21, hinkle21}. Additionally, in the recent work of \cite{nicholl22}, where the light-curves of 32 optically bright TDEs have been thoroughly analyzed, it has been found that the events without \ion{He}{II} were consistent with a stream-crossing origin for the luminosity, while those with \ion{He}{II} were more consistent with forming accretion disks. The late-time developing of the \ion{He}{II} observed in AT\,2017gge could further support this scenario as this TDE seems to transition between these two regimes.
However, we caution that there are also alternative explanations connecting both the optical/UV and X-ray emissions to the accretion phenomena, such as the lowering of the optical depth of an outflowing wind or of a reprocessing region, that cannot be excluded from our study.
In particular, the delayed X-ray brightening observed in AT\,2017gge could still be explained within a reprocessing scenario in which the decreasing of the optical depth of a reprocessing material at later times finally reveals the gas-rich circumnuclear environment from the flare and release the soft X-rays photons, allowing them to power the the narrow He II and iron coronal lines.

\subsection{The awakening of the host galaxy activity}

In order to understand the impact that a TDE can have on the host galaxy nucleus, we have used the pre-transient SDSS and the Gemini and X-shooter late-time optical spectra (taken after $\sim$408, 572 and 1698 days from the AT\,2017gge discovery, respectively) to investigate on the activity status of host galaxy through the BPT diagrams. From this analysis a variation on the host galaxy activity clearly emerges. In particular, while the line ratio measured from the pre-transient SDSS spectrum are consistent with the starforming classification (although its location on the the extreme starburst line suggests the presence of an additional ionising component), the values derived from the Gemini spectra place the galaxy at the border between the composite and the AGN region and well inside the LINER region, suggesting an AGN-like activity between 400 and 600 days from the transient discovery. Finally, the galaxy activity status has come back to the initial values at 1698 days, as derived from the X-shooter spectrum. All this evidence suggests the possibility that the occurrence of the TDE resulted in the enhancement of the activity of the SMBH, which passes from quiescent to an AGN-like, with the formation of a transient stratified photosphere, similar to a BLR, surrounding the source.

In the hypothesis of the formation of a virialized BLR-like photosphere, we test an independent method to derive the SMBH mass by using the SE relation developed by \cite{ricci17} and the broad \ha\/ and \hb\/ as measured in the Gemini spectra (which correspond to the phase of the increased activity of the host galaxy in the BPT diagrams). We derive a SMBH mass which is one order of magnitude smaller than what derived by using the canonical $M_{\rm BH}-\sigma_{\star}$ scaling relations developed by \cite{McConnell13} and \cite{gultekin09} and the stellar velocity dispersion as measured from the X-shooter spectrum ($\log M_{\rm BH}$=5.4-5.7 and $\log M_{\rm BH}$=6.6-6.8, respectively). Given the peak bolometric luminosity of $L_{\rm bol}$=1.4$\times$10$^{44}$ \unitlum, an Eddington limited accretion would requires a BH mass of $\sim$4$\times$10$^{6}$ \Msun, consistent with the values obtained from the $M_{\rm BH}-\sigma_{\star}$ relation. Thus, we consider this method more trustworthy and we suggest that the broad lines emitting region is not virialized. 

\section{Conclusion}
We have presented the results from our dense, long-lasting and multiwavelength follow-up of the nuclear transient AT\,2017gge. Based on the on the detection of very broad \hb\/ and \ha\/ emission lines in the optical spectra and on the results from the bolometric light-curve and the X-ray analysis, we confirm the TDE nature of this transient and we classify this object as a H-rich TDE (TDE H) in transition toward the H+He TDE (TDE H+He) subclass.
The SMBH mass as derived from the $M_{\rm BH}-\sigma_{\star}$ relation results to be $\log M_{\rm BH}$=(6.55$\pm$0.45) or $\log M_{\rm BH}$=(6.80$\pm$0.43)
(depending on the scaling relation used) and is consistent with the typical values derived for the TDEs host galaxies. 

The occurrence of the TDE have an impact on the activity of the host galaxy which passes from a quiescent starforming galaxy to a composite/AGN in the BPT diagram, to come back to quiescence after 1698 days. 
The photoionitazion induced by the delayed X-ray flare is responsible for the production of a number of high ionization coronal emission lines, which also indicate that the TDE occurred in a gas-rich environment.

The picture that we suggest from our analysis is that of a TDE occurred in a dusty and gas-rich environment, in which the UV/optical emission is produced at a distance of $R \sim$10$^{15}$ cm from the SMBH by the collision between intercepting streams of stellar debris during the initial phase of the stellar disruption. After $\sim$200 days from the transient discovery, the circularization process ends and a newly-formed accretion disk released a soft X-ray flare. However we caution that the reprocessing scenarios are still able to explain the delayed X-ray brightening observed in AT\,2017gge. The emitting region of the observed broad lines is consistent with a symmetric and stratified photosphere. Finally, placed at at a distance $R \sim$10$^{17}$ cm, an absorbing dust surrounds the whole system. The covering factor of $\sim$0.2 suggests that a large proportion of UV radiation could still be unobserved in both the optical and IR bands and could be behind the observed total (optical/UV+IR) energy radiated being significantly lower than the energy expected to be released by the disruption of a solar mass star.

Our dense and long lasting monitoring campaign of AT\,2017gge have revealed a remarkable TDE which have induced a variety phenomena in the host environment,
allowing us to build a picture describing both the geometrical and the physical properties of the system. Moreover, its occurrence in a star-forming galaxy suggests a close connection between a gas-rich environment, the soft X-ray flare and the production of long-lasting high ionization coronal emission lines, strongly supporting the idea that extreme coronal line emitter galaxies may have indeed experienced a TDE in the past. This demonstrate the importance of such dense multiwavelength follow-up campaigns of TDEs in the study of accretion processes around quiescent SMBHs.   

\section*{Acknowledgements}
\scriptsize
We thanks the anonymous referee for the useful comments that improved the manuscript.
The research leading to these results has received funding from the European Union's Horizon 2020 Program under the AHEAD project (grant agreement n. 654215). NUTS is supported in part by IDA (The Instrument Centre for Danish Astronomy). 
This work is based [in part] on observations made with the GTC operated by the Instituto de Astrofisica de Canarias, the NOT, owned in collaboration by the University of Turku and Aarhus University, and operated jointly by Aarhus University, the University of Turku and the University of Oslo, representing Denmark, Finland and Norway, the University of Iceland and Stockholm University, the WHT operated by the Isaac Newton Group of Telescopes, the Liverpool Telescope operated by Liverpool John Moores University with financial support from the UK Science and Technology Facilities Council, the TNG (program A42DDT4) operated by the Fundacion Galileo Galilei of the INAF. All these facilities are located at the Spanish Roque de los Muchachos Observatory of the Instituto de Astrofisica de Canarias on the island of La Palma. 
This work was supported by K-GMT Science Program (PID: GEMINI-KR-2018B-007 and GEMINI-KR-2019A-014) of Korea Astronomy and Space Science Institute (KASI). 
ATLAS is primarily funded to search for near earth asteroids through NASA grants NN12AR55G, 80NSSC18K0284, and 80NSSC18K1575. Science products are made possible by grants Kepler/K2 J1944/80NSSC19K0112,HST GO-15889, and STFC grants ST/T000198/1 and ST/S006109/1 and contributions form University of Hawaii Institute for Astronomy, the Queen’s University Belfast, the Space Telescope Science Institute, the South African Astronomical Observatory, and The Millennium Institute of Astrophysics (MAS), Chile. 
Pan-STARRS telescopes are supported by the National Aeronautics and Space Administration under Grants NNX12AR65G NNX14AM74G, from the Near-Earth Object Observations Program. Data are processed at Queen's University Belfast enabled through the STFC grants ST/P000312/1 and ST/T000198/1.
This publication makes use of data products from the Near-Earth Object Wide-field Infrared Survey Explorer (NEOWISE), which is a joint project of the Jet Propulsion Laboratory/California Institute of Technology and the University of Arizona. NEOWISE is funded by the National Aeronautics and Space Administration.
SDSS-IV acknowledges support and resources from the Center for High Performance Computing  at the University of Utah. The SDSS website is www.sdss.org.
Based on observations collected at the European Organisation for Astronomical Research in the Southern Hemisphere, Chile, as part of ePESSTO (the advanced Public ESO Spectroscopic Survey for Transient Objects Survey) and ePESSTO+ (the advanced Public ESO Spectroscopic Survey for Transient Objects Survey).
ePESSTO+ observations were obtained under ESO programs ID 1103.D-0328 and 106.216C
(PI: Inserra), while ePESSTO under ESO ID 199.D-0143 (PI: Smartt).
FO acknowledges support from MIUR, PRIN 2017 (grant 20179ZF5KS) ``The new frontier of the Multi-Messenger Astrophysics: follow-up of electromagnetic transient counterparts of gravitational wave sources'' and the support of AHEAD2020 grant agreement n.871158. MK was supported by a National Research Foundation of Korea (NRF) grant (No.\ 2020R1A2C4001753 and 2022R1A4A3031306) funded by the Korean government (MSIT). MN is supported by the European Research Council (ERC) under the European Union’s Horizon 2020 research and innovation programme (grant agreement No.\ 948381) and by a Fellowship from the Alan Turing Institute. MF is supported by a Royal Society - Science Foundation Ireland University Research Fellowship. MG is supported by the EU Horizon 2020 research and innovation programme under grant agreement No 101004719. NI is partially supported by Polish NCN DAINA grant No. 2017/27/L/ST9/03221. G.L. and M.P are supported by a research grant (19054) from VILLUM FONDEN. TEMB acknowledges financial support from the Spanish Ministerio de Ciencia e Innovaci\'on (MCIN), the Agencia Estatal de Investigaci\'on (AEI) 10.13039/501100011033 under the PID2020-115253GA-I00 HOSTFLOWS project, and from Centro Superior de Investigaciones Cient\'ificas (CSIC) under the PIE project 20215AT016 and the I-LINK 2021 LINKA20409. TEMB was also partially supported by the program Unidad de Excelencia Mar\'ia de Maeztu CEX2020-001058-M.

\section{DATA AVAILABILITY}

The data underlying this article are available in the article and in its online supplementary material. 
The NTT spectra and images are publicity available through the PESSTO SSDR4 ESO Phase 3 Data Release (see the \href{http://archive.eso.org/wdb/wdb/adp/phase3_spectral/form?phase3_collection=PESSTO&release_tag=1}{ESO archive search and retrieve interface}). The NOT, TNG, WHT, GTC data are public and available through the corresponding telescope archives. The processed data underlying this article will be shared on request to the corresponding author.



\bibliographystyle{mnras}
\bibliography{mybib_TDE.bib} 




\appendix

\section{Some extra material}

\begin{table*}
\centering
\begin{minipage}{150mm}
 \caption{\textit{Swift}/UVOT photometric measurements}
 \label{tbl:UVOTphot}
 \begin{center}
 \begin{tabular}{@{}llcccccc}
 \hline
MJD      &  Phase & \textit {UVW2}        &  \textit{UVM2}             & \textit {UVW1}              & \textit {U}              & \textit {B}              & \textit {V}  \\
(1)      & (2)            & (3)            &(4)             & (5)            & (6)            & (7)        & (8)     \\
\hline
58\,031.12 & \phantom{1}62.8 & 18.03$\pm$0.08& 17.86$\pm$0.08& 17.66$\pm$0.09 & 16.51$\pm$0.12& 17.26$\pm$0.14 &16.62$\pm$0.17\\
58\,041.75 & \phantom{1}73.4 & 18.23$\pm$0.09& 18.02$\pm$0.08& 18.00$\pm$0.11 & 16.81$\pm$0.15& 17.77$\pm$0.21&  16.93$^{\ddagger}$\\ 
58\,046.29 & \phantom{1}77.9 & 18.18$\pm$0.08& 18.14$\pm$0.08& 17.98$\pm$0.10 & 16.97$\pm$0.14& 17.83$\pm$0.17& 17.20$\pm$0.22\\
58\,051.78 & \phantom{1}83.4 & 18.42$\pm$0.09& 18.11$\pm$0.08& 17.99$\pm$0.11 & 16.84$\pm$0.16& 17.48$\pm$0.17& 16.84$^{\ddagger}$\\
58\,055.04 & \phantom{1}86.7 & 18.34$\pm$0.09& 18.17$\pm$0.09& 18.18$\pm$0.11 & 16.65$\pm$0.14& 17.84$^{\ddagger}$& 16.85$^{\ddagger}$\\
58\,143.57 & 175.2& 19.09$\pm$0.13& 19.25$\pm$0.14& 19.48$^{\ddagger}$  & 17.35$\pm$0.18& 18.20$^{\ddagger}$&  17.11$^{\ddagger}$ \\
58\,156.93 & 188.6& 19.49$\pm$0.10& 19.57$\pm$0.1&19.22$\pm$0.11&17.87$\pm$0.13&18.33$\pm$0.13 &17.23$\pm$0.13\\
58\,162.98 & 194.6& 19.42$\pm$0.09& 19.55$\pm$0.10&19.22$\pm$0.12& 17.91$\pm$0.14& 18.44$\pm$0.14& 17.53$\pm$0.16\\
58\,168.35 & 200.0& 19.55$\pm$0.11&19.93$\pm$0.13& 19.65$\pm$0.16& 18.07$\pm$0.18& 18.46$\pm$0.17& 17.50$\pm$0.18\\
58\,174.28 & 205.9& 19.40$^{\ddagger}$ & $\cdots$  & 19.50$\pm$0.16 & 17.85$\pm$0.21& 18.07$^{\ddagger}$& $\cdots$\\
58\,177.72 & 209.4& 19.49$\pm$0.12 & 19.76$\pm$0.13& 19.30$\pm$0.15& 18.07$\pm$0.19& 18.34$\pm$0.17& 17.46$\pm$0.20\\
58\,186.15 & 217.8& $\cdots$      & 19.71$\pm$0.09 &$\cdots$ & $\cdots$& $\cdots$& $\cdots$\\
58\,193.72 & 225.4& 19.77$\pm$0.12& $\cdots$      & $\cdots$       & $\cdots$&$\cdots$ & $\cdots$\\
58\,200.76 & 232.4& $\cdots$      & $\cdots$      & $\cdots$       & 18.06$\pm$0.09&$\cdots$ & $\cdots$\\
58\,211.60 & 243.3& $\cdots$      & $\cdots$      & 19.64$\pm$0.10 & $\cdots$&$\cdots$ & $\cdots$\\
58\,213.85 & 245.5& 19.51$\pm$0.09& $\cdots$      & $\cdots$       & $\cdots$&$\cdots$ & $\cdots$\\
58\,215.98 & 247.6& $\cdots$      & $\cdots$ & 19.57$\pm$0.09      & $\cdots$&$\cdots$ & $\cdots$\\
59\,651.84 & 1683.49 & 20.74$\pm$0.12 & 20.82$\pm$0.12& 20.34$\pm$0.14 & 18.43$\pm$0.13& 18.56$\pm$0.11& 17.56$\pm$0.11\\
59\,655.29 & 1686.94 & 20.86$\pm$0.15 & 20.90$\pm$0.16 & 20.44$\pm$0.16 & 18.29$\pm$0.14 & 18.30$\pm$0.10 & 17.62$\pm$0.13\\
\hline
\end{tabular}
\end{center}
Notes: (1) MJD date of observations; (2) Phase (days) with respect to the discovery date MJD 57\,968.35  (3), (4) and (5), apparent magnitudes and uncertainties for the UVOT filters \textit {UVW2}, \textit{UVM2}, \textit {UVW1}, in AB system; (6), (7) and (8), apparent magnitudes and uncertainties for the UVOT filters \textit {U}, \textit {B} and \textit {V}, in the Vega system. The values indicated with $\ddagger$ are upper limits.
All the magnitudes reported are uncorrected for foreground extinction.
With  $\cdots$ we indicate epochs with no data available (no observations). The full table including photometric measurement until MJD 59\,655.29 is available online.
\noindent
\end{minipage}
\end{table*} 

\begin{table*}
\centering
\begin{minipage}{150mm}
 \caption{Optical photometric measurements}
 \label{tbl:Optphot}
 \begin{center}
 \begin{tabular}{@{}llccccccl}
 \hline
MJD      &  Phase &  $c$   & $o$& $u^\prime$        &  $g^\prime$             & $r^\prime$/$w$              & $i^\prime$ & Telescope \\
(1)      & (2)            & (3)            &(4)             & (5)            & (6)            & (7)        & (8)     & (9) \\
\hline
57\,968.35 & \phantom{11}0.00     & $\cdots$  & 18.70$\pm$0.17& $\cdots$ & $\cdots$ & $\cdots$& $\cdots$& ATLAS\\
57\,971.32 & \phantom{11}2.97  & $\cdots$  & 18.48$\pm$0.17& $\cdots$ & $\cdots$ & $\cdots$& $\cdots$ & ATLAS\\
57\,972.30 & \phantom{11}3.95  & $\cdots$  &$\cdots$ & $\cdots$ & $\cdots$ & $\cdots$ &18.57$\pm$0.04 & PS1\\
57\,975.32 & \phantom{11}6.97  & $\cdots$  & 18.38$\pm$0.13& $\cdots$ & $\cdots$ & $\cdots$& $\cdots$ & ATLAS\\
57\,977.30 & \phantom{11}8.95  & $\cdots$  & 18.24$\pm$0.07& $\cdots$ & $\cdots$ & $\cdots$& $\cdots$ & ATLAS\\
57\,979.29 &\phantom{1}10.94  & $\cdots$  & 18.13$\pm$0.10& $\cdots$ & $\cdots$ & $\cdots$& $\cdots$ & ATLAS\\
57\,981.27 &\phantom{1}12.92  & $\cdots$  & 17.89$\pm$0.06& $\cdots$ & $\cdots$ & $\cdots$& $\cdots$ & ATLAS\\
57\,983.30 &\phantom{1}14.95  & 17.60$\pm$0.04 & $\cdots$ & $\cdots$ & $\cdots$ & $\cdots$& $\cdots$ & ATLAS\\
57\,985.26 &\phantom{1}16.91  & $\cdots$  &17.76$\pm$0.05 & $\cdots$ & $\cdots$ & $\cdots$ & $\cdots$& ATLAS\\
57\,987.30 &\phantom{1}18.95  & 17.43$\pm$0.10 & $\cdots$ & $\cdots$ & $\cdots$ & $\cdots$& $\cdots$ &ATLAS\\
57\,989.26 &\phantom{1}20.91  & $\cdots$  &17.56$\pm$0.05 & $\cdots$ & $\cdots$ & $\cdots$ & $\cdots$&ATLAS\\
57\,995.27 &\phantom{1}26.92  & $\cdots$  &18.03$\pm$0.10 & $\cdots$ & $\cdots$ & $\cdots$ & $\cdots$&ATLAS\\
57\,996.27 &\phantom{1}27.92  & $\cdots$  &$\cdots$ & $\cdots$ & $\cdots$ & $\cdots$ &17.42$\pm$0.02 & PS1\\
57\,997.24 &\phantom{1}28.89  & $\cdots$  &18.28$\pm$0.10 & $\cdots$ & $\cdots$ & $\cdots$ & $\cdots$ &ATLAS\\
57\,999.27 &\phantom{1}30.92  & $\cdots$  &18.12$\pm$0.12 & $\cdots$ & $\cdots$ & $\cdots$ & $\cdots$ &ATLAS\\
58\,007.27 &\phantom{1}38.92  & $\cdots$  &18.08$\pm$0.07 & $\cdots$ & $\cdots$ & $\cdots$ & $\cdots$ &ATLAS\\
58\,009.28 &\phantom{1}40.93  & $\cdots$  &18.45$\pm$0.10 & $\cdots$ & $\cdots$ & $\cdots$ & $\cdots$ &ATLAS\\
58\,013.26 &\phantom{1}44.91  & $\cdots$  &18.29$\pm$0.12 & $\cdots$ & $\cdots$ & $\cdots$ & $\cdots$ &ATLAS\\
58\,162.22 &193.87 & $\cdots$  & $\cdots$ & $\cdots$ &19.97$\pm$0.10 & 19.87$\pm$0.10 & $\cdots$ &NOT\\
58\,186.39 &218.04 & $\cdots$  & $\cdots$ & $\cdots$ &19.95$\pm$0.16 & 19.63$\pm$0.21 & $\cdots$ &NTT\\
58\,194.25 &225.90 & $\cdots$  & $\cdots$ & 20.19$\pm$0.14 & 20.13$\pm$0.15 & 20.45$\pm$0.22 & $\cdots$ &TNG\\
58\,198.10 &229.75 & $\cdots$  & $\cdots$ & 20.33$\pm$0.24 & 20.32$\pm$0.11 & 21.10$\pm$0.26 & $\cdots$ &TNG\\
58\,201.19 &232.84 & $\cdots$  & $\cdots$ & $\cdots$ &20.35$\pm$0.09 & 20.44$\pm$0.14 & $\cdots$ &NOT\\
58\,202.37 &234.02 & $\cdots$  & $\cdots$ & $\cdots$ &20.27$\pm$0.230 & 20.08$\pm$0.16 & $\cdots$ &NTT\\
58\,207.15 &238.80 & $\cdots$  & $\cdots$ & 21.47$\pm$0.25 &19.97$\pm$0.18 & 20.64$\pm$0.18 & $\cdots$ &TNG\\
58\,216.40 &248.05 & $\cdots$  & $\cdots$ & $\cdots$ &20.15$\pm$0.20 & 19.42$\pm$0.15 & $\cdots$ &NTT\\
58\,263.10 &294.75 & $\cdots$  & $\cdots$ & $\cdots$ &19.70$\pm$0.20 & 20.87$\pm$0.25 & $\cdots$ &NOT\\
58\,263.17 &294.82 & $\cdots$  &$\cdots$ & $\cdots$ & $\cdots$ & $\cdots$ &20.06$\pm$0.20 &PS1\\
58\,274.52 &306.17 & $\cdots$  &$\cdots$ & $\cdots$ & $\cdots$ & $\cdots$ &20.34$\pm$0.25 &PS1\\
58\,538.89 &570.54 & $\cdots$  & $\cdots$ & 21.41$\pm$0.19 &20.80$\pm$0.23 & 20.45$\pm$0.19 & $\cdots$ &LT\\
58\,610.57 &641.65 & $\cdots$  &$\cdots$ & $\cdots$ & $\cdots$ &20.59$\pm$0.08& $\cdots$ &PS1$^{*}$\\
58\,611.58 &643.23 & $\cdots$  &$\cdots$ & $\cdots$ & $\cdots$ &20.32$\pm$0.05& $\cdots$ &PS1$^{*}$\\ 
58\,633.53 &665.18 & $\cdots$  &$\cdots$ & $\cdots$ & $\cdots$ &20.74$\pm$0.11& $\cdots$ &PS1$^{*}$\\
58\,693.35 &725.00 & $\cdots$  &$\cdots$ & $\cdots$ & $\cdots$ &20.73$\pm$0.10& $\cdots$ &PS1$^{*}$\\
58\,696.32 &727.97 & $\cdots$  &$\cdots$ & $\cdots$ & $\cdots$ &21.08$\pm$0.23& $\cdots$ &PS1$^{*}$\\
\hline
            &        & $J$           & $H$            & $Ks$ & &  & & \\
58\,347.043 & 374.75 &19.96$\pm$0.16 & 17.50$\pm$0.09 & 16.12$\pm$0.09& $\cdots$ & $\cdots$ & $\cdots$&NTT\\
\hline
\end{tabular}
\end{center}
Notes: (1) MJD date of observations; (2) Phase (days) with respect to the discovery date MJD 57\,968.35;  (3) and (4) host-subtracted apparent magnitudes and uncertainties for the ATLAS filters $c$ and $o$, respectively; (5) and (6) host-subtracted apparent magnitudes and uncertainties for the filters $u^\prime$ and $g^\prime$ obtained in the framework o our monitoring campaign; (7) host-subtracted apparent magnitudes and uncertainties for the filters $r^\prime$ obtained in the framework o our monitoring campaign and for the PS1 filter $w$; (8) host-subtracted apparent magnitudes and uncertainties for the filter $i^\prime$, obtained by PS1; (9) telescope used.
NIR magnitude from SofI/NTT observations are reported in the last row of the table. 
All the optical magnitudes are uncorrected for foreground extinction and are in the AB system, while the NIR data are in the Vega system.
With  $\cdots$ we indicate epochs with no data available (no observations).
$^{*}$PS1 data in $w$ band central wavelength correspond to the $r^\prime$ one \cite[][]{smith20}. The full table including photometric measurement until MJD 58\,696.32 is available online.
\noindent
\end{minipage}
\end{table*}

\begin{figure*}
\includegraphics[width=2\columnwidth, angle = 0]{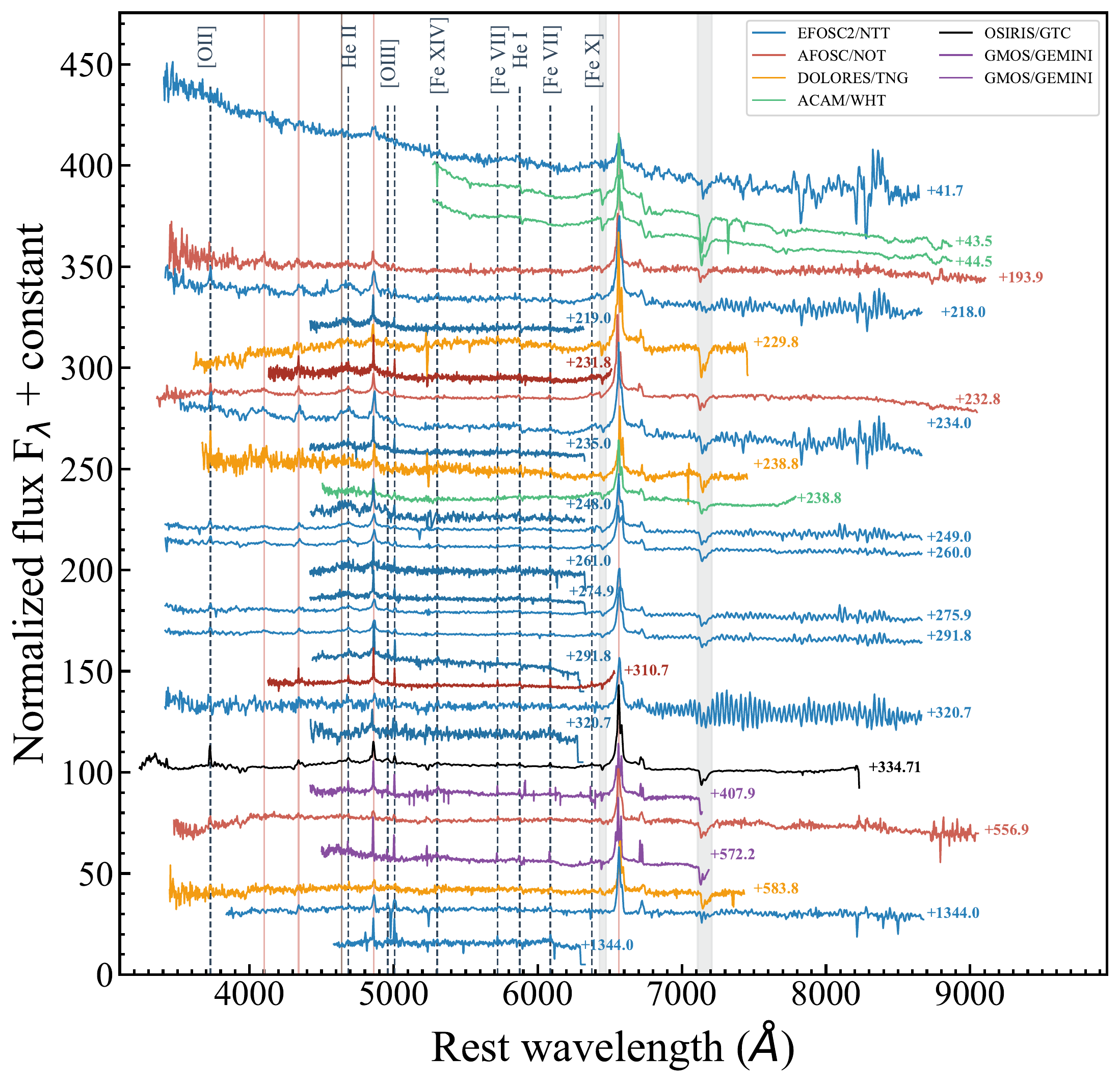}
\caption{Sequence of the rest-frame optical spectra of AT\,2017gge taken with different facilities: EFOSC2/NTT (in blue), ALFOSC/NOT (in red), DOLORES/TNG (in orange), ACAM/WHT (in green), OSIRIS/GTC (in black) and GMOS/Gemini (in purple). All the spectra have been corrected for reddening. The time of the observation in days since the transient discovery and the main emission lines are indicated. The location of telluric absorption lines is indicated by gray bands.}
\label{fig:spec}
\end{figure*}

\begin{figure*}
\includegraphics[scale=0.18, angle=0]{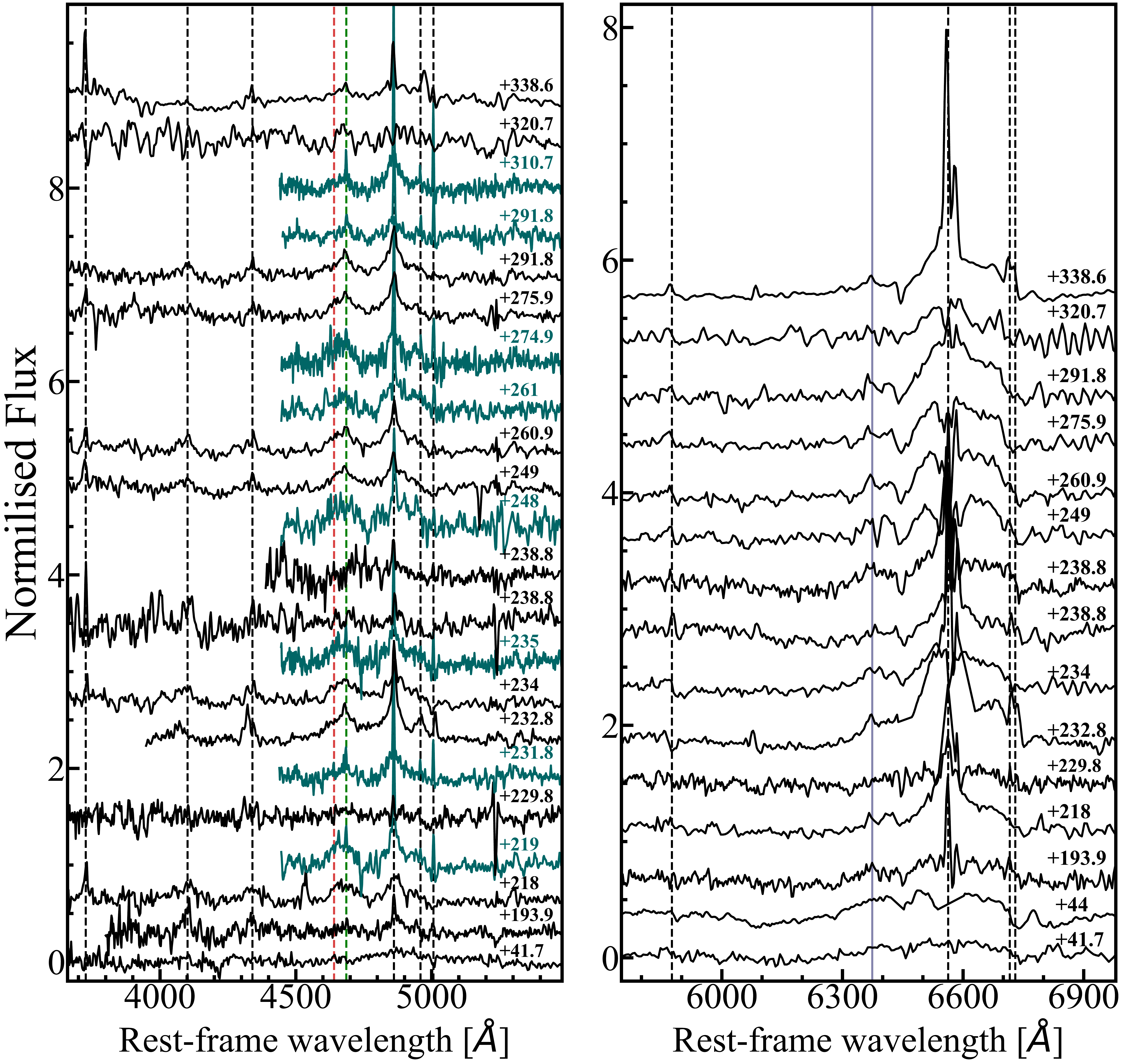}
\caption{Sequence of the host-subtracted spectra of AT\,2017gge. In the left panel the \ion{He}{II}+\hb region is shown, while in the right panel the \ha\/ region. The host-subtracted spectra obtained for the EFOSC2 Gr$\#$18 and the ALFOSC Gr$\#$19 are shown in green in the left panel only (as they don't cover the \ha\/ region). The position of the H lines and of the [\ion{O}{II}] $\lambda$3727 are shown with black vertical dashed lines, while the \ion{He}{II} and the position of the Bowen blend at $\lambda$4640 are indicated by green ad red dashed lines, respectively. The solid gray vertical lines indicates the position of the high ionization coronal lines [\ion{Fe}{X}] $\lambda$6374. For each spectrum, the days from the transient discovery are also indicated.}
\label{fig:host_sub_spec}
\end{figure*}

\begin{figure*}
\includegraphics[width=2\columnwidth, angle = 0]{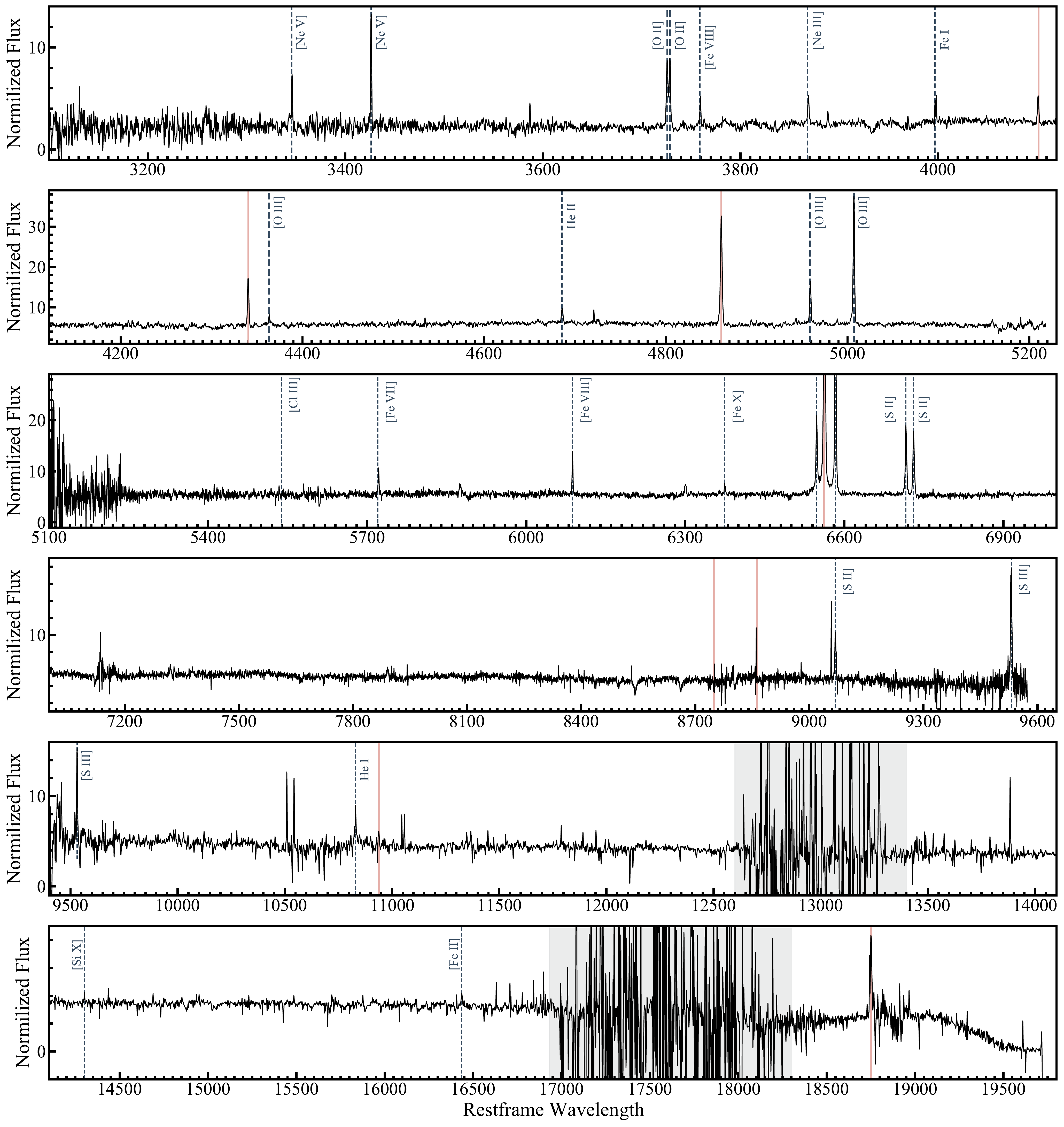}
\caption{Rest-frame X-shooter spectrum (UVB, VIS and NIR) of the AT\,2017gge host galaxy taken 1648 days after the transient discovery. The spectrum has been corrected for reddening. The main emission lines are indicated with dashed vertical lines, while red solid vertical lines indicate the position of the H lines. The region affected by telluric absorption lines is indicated by grey bands.}
\label{fig:Xsh_spec}
\end{figure*}


\bsp	
\label{lastpage}
\end{document}